\documentclass[11pt]{article}
\usepackage{graphicx,verbatim,array,multicol,palatino,algorithm}
\usepackage{amsmath,amssymb,amsfonts,ifthen,setspace,color,
            natbib,epsfig,epstopdf,relsize,accents}
\newboolean{ShowFigures}
\newboolean{UnBlinded}
\newboolean{DoubleSpaced}
\setboolean{ShowFigures}{true}
\setboolean{UnBlinded}{true}
\setboolean{DoubleSpaced}{false}

\definecolor{DarkGreen}{rgb}{0.412,0.545,0.133}
\definecolor{SlateGrey}{rgb}{0.439,0.502,0.565}

\def\myand{\&\ }
\def\ba{\boldsymbol{a}}
\def\bb{\boldsymbol{b}}

\def\bdf{\boldsymbol{f}}
\def\kappatilde{{\widetilde\kappa}}
\def\lambdatilde{{\widetilde\lambda}}
\def\bu{\boldsymbol{u}}

\def\Diff{{\sf D}}
\def\Hess{{\sf H}}
\def\bv{\boldsymbol{v}}

\def\bx{\boldsymbol{x}}
\def\by{\boldsymbol{y}}

\def\bA{\boldsymbol{A}}
\def\bB{\boldsymbol{B}}
\def\bC{\boldsymbol{C}}
\def\bD{\boldsymbol{D}}
\def\bX{\boldsymbol{X}}

\def\bI{\boldsymbol{I}}
\def\bM{\boldsymbol{M}}
\def\threehalves{{\textstyle{3\over2}}}
\def\bmu{\boldsymbol{\mu}}
\def\bSigma{\boldsymbol{\Sigma}}
\def\bQ{\boldsymbol{Q}}
\def\br{\boldsymbol{r}}
\def\eighth{{\textstyle{1\over8}}}
\def\sigeps{\sigma_{\varepsilon}}
\def\bZ{\boldsymbol{Z}}
\def\bzero{\boldsymbol{0}}
\def\bone{\boldsymbol{1}}
\def\btheta{\boldsymbol{\theta}}
\def\vecof{\mbox{vec}}
\def\diag{\mbox{diag}}
\def\bTheta{\boldsymbol{\Theta}}
\def\bLambda{\boldsymbol{\Lambda}}
\def\Csc{{\mathcal C}}
\def\diagarg#1{\diagdum_{#1}}
\def\bO{\boldsymbol{O}}
\def\blockdiagdum{\mathop{\mbox{blockdiag}}}
\def\blockdiag#1{\blockdiagdum_{#1}}
\def\bOmega{\boldsymbol{\Omega}}
\def\maxdum{\mathop{\mbox{max}}}
\def\maxsub#1{\maxdum_{#1}}
\def\tr{\mbox{tr}}
\def\bJ{\boldsymbol{J}}
\def\bU{\boldsymbol{U}}
\def\Aeps{A_{\varepsilon}}

\def\Au{A_u}

\def\real{{\mathbb R}}
\def\bT{\boldsymbol{T}}
\def\Psc{{\mathcal P}}
\def\thickboxit#1{\vbox{{\hrule height 1mm}\hbox{{\vrule width 1mm}\kern6pt
          \vbox{\kern6pt#1\kern6pt}\kern6pt{\vrule width 1mm}}
               {\hrule height 1mm}}}
\def\punder{\underline{p}}
\def\simind{\stackrel{{\tiny \mbox{ind.}}}{\sim}}
\def\diagdum{\mathop{\mbox{diag}}}
\def\subsubsubsection#1{\vskip2mm\noindent$\underline{\mbox{\em #1}}$\vskip2mm}
\def\bxi{\boldsymbol{\xi}}
\def\bXi{\boldsymbol{\Xi}}
\def\bnu{\boldsymbol{\nu}}
\def\bomega{\boldsymbol{\omega}}
\def\diagonal{\mbox{diagonal}}
\def\recipMills{\zeta'}
\def\bibmpa{\vskip1.5pt\par\noindent\hangindent=8 true mm\hangafter=1}

\def\NonEntropy{\mbox{\rm NonEntropy}}
\def\indicB{I^B}
\def\bindicB{\bI^B}
\def\GVMP{G_{\mbox{\tiny VMP}}}
\def\Agbl{A_{\mbox{\tiny gbl}}}
\def\Agrp{A_{\mbox{\tiny grp}}}

\def\bAU{\bA_U}
\def\bZU{\bZ_{U}}
\def\Gdiag{G_{\mbox{\tiny diag}}}
\def\bZgbl{\bZ_{\mbox{\tiny gbl}}}
\def\bZgblW{\bZ_{\mbox{\tiny gbl}}^W}
\def\bZgblB{\bZ_{\mbox{\tiny gbl}}^B}
\def\bZgrp{\bZ_{\mbox{\tiny grp}}}
\def\bugbl{\bu_{\mbox{\tiny gbl}}}
\def\bugblW{\bu_{\mbox{\tiny gbl}}^W}
\def\bugblB{\bu_{\mbox{\tiny gbl}}^B}
\def\bugrp{\bu_{\mbox{\tiny grp}}}
\def\Kgbl{K_{\mbox{\tiny gbl}}}
\def\ugblk{u_{\mbox{\tiny gbl},k}}
\def\zgblk{z_{\mbox{\tiny gbl},k}}
\def\agbl{a_{\mbox{\tiny gbl}}}
\def\agrp{a_{\mbox{\tiny grp}}}
\def\aeps{a_{\varepsilon}}
\def\sigmagbl{\sigma_{\mbox{\tiny gbl}}}
\def\Kgrp{K_{\mbox{\tiny grp}}}
\def\ugrpik{u_{\mbox{\tiny grp},ik}}
\def\zgrpk{z_{\mbox{\tiny grp},k}}
\def\sigmagrp{\sigma_{\mbox{\tiny grp}}}

\def\lambdaInvGau{\lambda}
\def\neighbors{\mbox{neighbors}}
\def\subscriptneighbors{\mbox{\scriptsize neighbors}}

\def\kappaIG{{\widetilde\kappa}}
\def\lambdaIG{{\widetilde\lambda}}
\def\shpOne{\alpha}
\def\shpTwo{\beta}
\def\Entropy{\mbox{Entropy}}
\def\ftrue{f_{\mbox{\tiny true}}}
\def\kappaTheta{\kappa_{\mbox{\tiny $\bTheta$}}}
\def\LambdaTheta{\bLambda_{\mbox{\tiny $\bTheta$}}}
\def\thickarrow{\longleftarrow}

\def\dtheta{d^{\,\mbox{\tiny$\theta$}}}
\def\dTheta{d^{\,\mbox{\tiny$\Theta$}}}
\def\mR{m}

\def\bdeta{{\boldsymbol{\eta}}}
\def\bbeta{{\boldsymbol{\beta}}}

\def\smhalf{{\textstyle{\frac{1}{2}}}}

\def\quarter{{\textstyle{1\over4}}}
\def\jump{\vskip3mm\noindent}

\def\biggerf{\mbox{\large $f$}}
\def\biggerm{\mbox{\large $m$}}
\def\biggermtilde{\mbox{\large $\widetilde{m}$}}

\def\etaSUBpthetaVecThetaVecCONNThetaEll
{\biggerbdeta_{\mbox{\footnotesize$p(\btheta_0,\ldots,\btheta_L|
\,\bTheta_1,\ldots,\bTheta_L)\leftrightarrow\bTheta_{\ell}$}}}

\def\etaSUBpthetaVecThetaVecCONNthetaVec
{\biggerbdeta_{\mbox{\footnotesize$p(\btheta_0,\ldots,\btheta_L|
\,\bTheta_1,\ldots,\bTheta_L)
\leftrightarrow(\btheta_0,\ldots,\btheta_L)$}}}

\def\etaSUBpthetaVecThetaVecTOthetaVec
{\biggerbdeta_{\mbox{\footnotesize$p(\btheta_0,\ldots,\btheta_L|
\,\bTheta_1,\ldots,\bTheta_L)\to(\btheta_0,\ldots,\btheta_L)$}}}

\def\etaSUBthetaVecTOpthetaVecThetaVec
{\biggerbdeta_{\mbox{\footnotesize$(\btheta_0,\ldots,\btheta_L)
\to p(\btheta_0,\ldots,\btheta_L|\,\bTheta_1,\ldots,\bTheta_L)$}}}

\def\etaSUBthetaOneTOpythetaOnethetaTwo
{\biggerbdeta_{\mbox{\footnotesize$\btheta_1\to p(\by|\,\btheta_1,\theta_2)$}}}
\def\etaSUBthetaTwoTOpythetaOnethetaTwo
{\biggerbdeta_{\mbox{\footnotesize$\theta_2\to p(\by|\,\btheta_1,\theta_2)$}}}
\def\etaSUBpythetaOnethetaTwoTOthetaTwo
{\biggerbdeta_{\mbox{\footnotesize$p(\by|\,\btheta_1,\theta_2)\to\theta_2$}}}
\def\etaSUBpythetaOnethetaTwoTOthetaOne
{\biggerbdeta_{\mbox{\footnotesize$p(\by|\,\btheta_1,\theta_2)\to\btheta_1$}}}
\def\etaSUBpthetaVecThetaVecTOThetaEll{\biggerbdeta_{\mbox{\footnotesize$p(\btheta_0,\ldots,\btheta_L|
\,\bTheta_1,\ldots,\bTheta_L)\to\bTheta_{\ell}$}}}
\def\etaSUBThetaEllTOpthetaVecThetaVec{\biggerbdeta_{\mbox{\footnotesize$
\bTheta_{\ell}\to p(\btheta_0,\ldots,\btheta_L|\,\bTheta_1,\ldots,\bTheta_L)$}}}

\def\etaSUBpThetaTOTheta{\biggerbdeta_{\mbox{\footnotesize$p(\bTheta)\to\bTheta$}}}
\def\etaSUBpthetaTOtheta{\biggerbdeta_{\mbox{\footnotesize$p(\btheta)\to\btheta$}}}
\def\etaSUBpbetaTObeta{\biggerbdeta_{\mbox{\footnotesize$p(\bbeta)\to\bbeta$}}}
\def\etaSUBpbetausigsquTObetau{
\biggerbdeta_{\mbox{\footnotesize$p(\bbeta,\bu|\,\sigsq_u)\to(\bbeta,\bu)$}}}
\def\etaSUBpbetausigsquTOsigsqu{
\biggerbdeta_{\mbox{\footnotesize$p(\bbeta,\bu|\,\sigsq_u)\to\sigsq_u$}}}
\def\etaSUBpythetaOnethetaTwoCONNthetaOne
{\biggerbdeta_{\mbox{\footnotesize$p(\by\,|\,\btheta_1,\theta_2)
\leftrightarrow\btheta_1$}}}
\def\etaSUBpythetaOnethetaTwoCONNthetaTwo
{\biggerbdeta_{\mbox{\footnotesize$p(\by\,|\,\btheta_1,\theta_2)
\leftrightarrow\theta_2$}}}

\def\etaSUBpbetaybetasigsqusigsqCONNsigsqu{
\biggerbdeta_{\mbox{\footnotesize$p(\bbeta,\bu|\,\sigsq_u)
\leftrightarrow\sigma_u^2$}}}
\def\etaSUBpbetausigsquCONNbetau{
\biggerbdeta_{\mbox{\footnotesize$p(\bbeta,\bu|\,\sigsq_u)
\leftrightarrow(\bbeta,\bu)$}}}
\def\etaSUBpybetasigsqTObeta{
\biggerbdeta_{\mbox{\footnotesize$p(\by|\bbeta,\sigma^2)\to\bbeta$}}}
\def\etaSUBpybetausigsqepsTObetau{
\biggerbdeta_{\mbox{\footnotesize$p(\by|\bbeta,\bu,\sigma_{\varepsilon}^2)
          \to(\bbeta,\bu)$}}}
\def\etaSUBpybetausigsqepsTOsigsqeps{
\biggerbdeta_{\mbox{\footnotesize$p(\by|\,\bbeta,\bu,\sigma_{\varepsilon}^2)
          \to\sigma^2_{\varepsilon}$}}}
\def\etaSUBpybetasigsqTOsigsq{\biggerbdeta_
     {\mbox{\footnotesize$p(\by|\bbeta,\sigma^2)\to\sigma^2$}}}
\def\etaSUBpsigsqaTOsigsq{
      \biggerbdeta_{\mbox{\footnotesize$p(\sigma^2|a)\to\sigma^2$}}}
\def\etaSUBpsigsqepsaepsTOsigsqeps{
      \biggerbdeta_{\mbox{\footnotesize$p(\sigeps^2|a_{\varepsilon})\to\sigeps^2$}}}
\def\etaSUBpsigsquauTOsigsqu{
      \biggerbdeta_{\mbox{\footnotesize$p(\sigma_u^2|a_u)\to\sigma_u^2$}}}
\def\etaSUBpsigsquauTOau{
      \biggerbdeta_{\mbox{\footnotesize$p(\sigma_u^2|a_u)\to a_u$}}}
\def\etaSUBpsigsqaTOa{
      \biggerbdeta_{\mbox{\footnotesize$p(\sigma^2|a)\to a$}}}
\def\etaSUBbetaTOpbeta{
      \biggerbdeta_{\mbox{\footnotesize$\bbeta\to p(\bbeta)$}}}
\def\etaSUBaTOpa{
     \biggerbdeta_{\mbox{\footnotesize$a\to p(a)$}}}

\def\etaSUBbetaTOpybetasigsq{\biggerbdeta_{\mbox{\footnotesize
          $\bbeta\to p(\by|\bbeta,\sigma^2)$}}}
\def\etaSUBsigsqTOpybetasigsq{\biggerbdeta_
     {\mbox{\footnotesize$\sigma^2\to p(\by|\bbeta,\sigma^2)$}}}
\def\etaSUBpybetasigsqCONNsigsq{\biggerbdeta_
     {\mbox{\footnotesize$p(\by|\bbeta,\sigma^2)\leftrightarrow\sigma^2$}}}
\def\etaSUBpybetasigsqCONNbeta{\biggerbdeta_
     {\mbox{\footnotesize$p(\by|\bbeta,\sigma^2)\leftrightarrow\bbeta$}}}
\def\etaSUBsigsqTOpsigsqa{
     \biggerbdeta_{\mbox{\footnotesize$\sigma^2\to p(\sigma^2|a)$}}}
\def\etaSUBaTOpsigsqa{\biggerbdeta_{\mbox{\footnotesize$a\to p(\sigma^2|a)$}}}
\def\etaSUBpsigsqaCONNa{\biggerbdeta_{
          \mbox{\footnotesize$p(\sigma^2|\,a)\leftrightarrow a$}}}
\def\etaSUBpsigsqaCONNsigsq{\biggerbdeta_{\mbox{\footnotesize$p(\sigma^2|\,a)
          \leftrightarrow\sigsq$}}}

\def\mSUBpbetausigsquTOsigsqu{
\biggerm_{\mbox{\footnotesize$p(\bbeta,\bu|\,\sigma_u^2)\to\sigma_u^2$}}
(\sigma_u^2)}
\def\mSUBpbetausigsquTObetau{
\biggerm_{\mbox{\footnotesize$p(\bbeta,\bu|\,\sigma_u^2)\to(\bbeta,\bu)$}}
(\bbeta,\bu)}
\def\mSUBpybetasigsqTObeta{
\biggerm_{\mbox{\footnotesize$p(\by|\bbeta,\sigma^2)\to\bbeta$}}(\bbeta)}

\def\mSUBpsigsqaTOa{
      \biggerm_{\mbox{\footnotesize$p(\sigma^2|a)\to a$}}(a)}
\def\mSUBpsigsqaTOsigsq{
      \biggerm_{\mbox{\footnotesize$p(\sigma^2|a)\to\sigma^2$}}(\sigma^2)}
\def\mSUBsigsqTOpsigsqa{
      \biggerm_{\mbox{\footnotesize$\sigma^2\to p(\sigma^2|a)$}}(\sigma^2)}
\def\mSUBpthetaTOtheta{
      \biggerm_{\mbox{\footnotesize$p(\btheta)\to\btheta$}}(\btheta)}
\def\mSUBpythetaTOtheta{
      \biggerm_{\mbox{\footnotesize$p(\by|\btheta)\to\btheta$}}(\btheta)}
\def\munderSUBpythetaTOthetaxi{
   \biggermunder_{\,\mbox{\footnotesize$p(\by|\btheta)\to\btheta$}}(\btheta;\bxi)}
\def\mSUBpThetaTOTheta{
      \biggerm_{\mbox{\footnotesize$p(\bTheta)\to\bTheta$}}(\bTheta)}

\def\biggerbdeta{\mbox{\Large $\bdeta$}}

\def\biggerlambda{\mbox{\Large $\lambda$}}
\def\biggerm{\mbox{\Large $m$}}
\def\munder{\underline{m}}
\def\biggermunder{\mbox{\Large $\munder$}}

\def\biggermu{\mbox{\Large $\mu$}}
\def\biggerE{\mbox{\Large $E$}}
\def\etaSUBpaTOa{\biggerbdeta_{\mbox{\footnotesize$p(a)\to a$}}}
\def\etaSUBpauTOau{\biggerbdeta_{\mbox{\footnotesize$p(a_u)\to a_u$}}}

\def\etaSUBpsigsqaTOsigsq{
      \biggerbdeta_{\mbox{\footnotesize$p(\sigma^2|\,a)\to\sigma^2$}}}
\def\etaSUBpauTOau{
      \biggerbdeta_{\mbox{\footnotesize$p(a_u)\to a_u$}}}
\def\etaSUBpaepsTOaeps{
      \biggerbdeta_{\mbox{\footnotesize$p(\aeps)\to a_{\varepsilon}$}}}
\def\etaSUBpsigsquauTOau{
      \biggerbdeta_{\mbox{\footnotesize$p(\sigma_u^2|\,a_u)\to a_u$}}}
\def\etaSUBpsigsqepsaepsTOaeps{
      \biggerbdeta_{\mbox{\footnotesize$p(\sigeps^2|\,\aeps)\to a_{\varepsilon}$}}}
\def\etaSUBpsigsqaTOa{
      \biggerbdeta_{\mbox{\footnotesize$p(\sigma^2|\,a)\to a$}}}

\def\etaSUBaTOpa{
     \biggerbdeta_{\mbox{\footnotesize$a\to p(a)$}}}

\def\etaSUBsigsqTOpsigsqa{
     \biggerbdeta_{\mbox{\footnotesize$\sigma^2\to p(\sigma^2|\,a)$}}}
\def\etaSUBaTOpsigsqa{\biggerbdeta_{\mbox{\footnotesize$a\to p(\sigma^2|\,a)$}}}
\def\mSUBpThetaOneThetaTwoTOThetaOne{
\biggerm_{\mbox{\footnotesize$p(\bTheta_1|\bTheta_2)\to\bTheta_1$}}(\bTheta_1)}
\def\etaSUBpThetaOneThetaTwoTOThetaOne{
\biggerbdeta_{\mbox{\footnotesize$p(\bTheta_1|\bTheta_2)\to\bTheta_1$}}}
\def\mSUBThetaTwoTOpThetaOneThetaTwo{
\biggerm_{\mbox{\footnotesize$\bTheta_2\to p(\bTheta_1|\bTheta_2)$}}(\bTheta_2)}
\def\mSUBpThetaOneThetaTwoTOThetaTwo{
\biggerm_{\mbox{\footnotesize$p(\bTheta_1|\bTheta_2)\to\bTheta_2$}}(\bTheta_2)}
\def\etaSUBpThetaOneThetaTwoTOThetaTwo{
\biggerbdeta_{\mbox{\footnotesize$p(\bTheta_1|\bTheta_2)\to\bTheta_2$}}}

\def\etaSUBpThetaOneThetaTwoCONNThetaOne{
\biggerbdeta_{\mbox{\footnotesize$p(\bTheta_1|\bTheta_2)\leftrightarrow\bTheta_1$}}}
\def\etaSUBpThetaOneThetaTwoCONNThetaTwo{
\biggerbdeta_{\mbox{\footnotesize$p(\bTheta_1|\bTheta_2)\leftrightarrow\bTheta_2$}}}

\def\mSUBpaTOa{
      \biggerm_{\mbox{\footnotesize$p(a)\to a$}}(a)}
\def\mSUBpsigsqaTOsigsq{
      \biggerm_{\mbox{\footnotesize$p(\sigma^2|\,a)\to\sigma^2$}}(\sigma^2)}
\def\mSUBsigsqTOpsigsqa{
      \biggerm_{\mbox{\footnotesize$\sigma^2\to p(\sigma^2|\,a)$}}(\sigma^2)}

\def\sigsq{\sigma^2}
\def\etaSUBpythetaCONNtheta
{\biggerbdeta_{\mbox{\footnotesize$p(\by|\,\btheta)\leftrightarrow\btheta$}}}
\def\etaSUBpythetaTOtheta
{\biggerbdeta_{\mbox{\footnotesize$p(\by|\,\btheta)\to\btheta$}}}
\def\etaSUBpathetaCONNtheta
{\biggerbdeta_{\mbox{\footnotesize$p(\ba|\,\btheta)\leftrightarrow\btheta$}}}
\def\muSUBpathetaCONNai
{\biggermu_{\mbox{\footnotesize$p(\ba|\,\btheta)\leftrightarrow a_i$}}}
\def\bmuSUBpathetaCONNa
{\biggermu_{\mbox{\footnotesize$p(\ba|\,\btheta)\leftrightarrow\ba$}}}
\def\etaSUBpathetaTOtheta
{\biggerbdeta_{\mbox{\footnotesize$p(\ba|\,\btheta)\to\btheta$}}}
\def\etaSUBthetaTOpatheta
{\biggerbdeta_{\mbox{\footnotesize$\btheta\to p(\ba|\,\btheta)$}}}
\def\etaSUBpythetaTOtheta
{\biggerbdeta_{\mbox{\footnotesize$p(\by|\,\btheta)\to\btheta$}}}
\def\etaSUBthetaTOpytheta
{\biggerbdeta_{\mbox{\footnotesize$\btheta\to p(\by|\,\btheta)$}}}
\def\etaSUBpythetaCONNtheta
{\biggerbdeta_{\mbox{\footnotesize$p(\by|\,\btheta)\leftrightarrow\btheta$}}}

\def\mSUBpyaTOai
{\biggerm_{\mbox{\footnotesize$p(\by|\,\ba)\to a_i$}}(a_i)}
\def\mSUBaiTOpatheta
{\biggerm_{\mbox{\footnotesize$a_i\to p(\ba|\,\btheta)$}}(a_i)}

\def\mSUBpathetaTOai
{\biggerm_{\mbox{\footnotesize$p(\ba|\,\btheta)\to a_i$}}(a_i)}
\def\mSUBpathetaTOtheta
{\biggerm_{\mbox{\footnotesize$p(\ba|\,\btheta)\to\btheta$}}}

\def\mtildeSUBpythetaTOtheta
{\biggermtilde_{\mbox{\footnotesize$p(\by|\,\btheta)\to\btheta$}}(\btheta)}

\def\mSUBthetaTOpytheta
{\biggerm_{\mbox{\footnotesize$\btheta\to p(\by|\,\btheta)$}}(\btheta)}

\def\mSUBpbetaTObeta
{\biggerm_{\mbox{\footnotesize$p(\bbeta)\to\bbeta$}}(\bbeta)}

\def\mSUBpybetasigsqTOsigsq
{\biggerm_{\mbox{\footnotesize$p(\by|\bbeta,\sigma^2)\to\sigma^2$}}(\sigma^2)}

\begin{document}
\ifthenelse{\boolean{DoubleSpaced}}{\setstretch{1.5}}{}
%

\begin{center}
{\Large\bf Fast Approximate Inference for Arbitrarily Large}
\vskip0.5mm
{\Large\bf Semiparametric Regression Models via Message Passing}

\vskip1.5mm
\ifthenelse{\boolean{UnBlinded}}{
{\sc By M.P. Wand}
\footnote{M.P. Wand is Distinguished Professor, 
School of Mathematical and Physical Sciences,
University of Technology Sydney, 
P.O. Box 123, Broadway 2007, Australia, and
Chief Investigator, Australian Research Council Centre of Excellence
for Mathematical and Statistical Frontiers.
\textcolor{SlateGrey}{Date of this version: 05 APR 2016.}}}
{\null}
\end{center}

We show how the notion of \emph{message passing} can be used
to streamline the algebra and computer coding for fast
approximate inference in large Bayesian semiparametric 
regression models. In particular, this approach is amenable to handling 
\emph{arbitrarily large} models of particular types
once a set of primitive operations is established.
The approach is founded upon a message passing formulation
of mean field variational Bayes that utilizes 
\emph{factor graph} representations of statistical
models. The underlying principles apply to general Bayesian
hierarchical models although we focus on semiparametric
regression. The notion of factor graph fragments is introduced
and is shown to facilitate compartmentalization of the
required algebra and coding. The resultant algorithms have
ready-to-implement closed form expressions and allow 
a broad class of arbitrarily large semiparametric
regression models to be handled. Ongoing software projects 
such as \textsf{Infer.NET} and \textsf{Stan} support 
variational-type inference for particular model classes. 
This article is not concerned with software packages
\emph{per se} and focuses on the underlying tenets of scalable
variational inference algorithms.


\section{Introduction}\label{sec:intro}

We derive algorithmic primitives that afford fast approximate
inference for arbitrarily large semiparametric regression models.
The fit updating steps required for fitting a simple semiparametric 
regression model, such as Gaussian response nonparametric regression,
can also be used for a much larger model involving, for example,
multiple predictors, group-specific curves and non-Gaussian responses.
Such update formulae only need to be derived and implemented once, 
representing enormous savings in terms of algebra and computing coding.

Semiparametric regression extends classical statistical
models, such as generalized linear models and linear mixed
models, to accommodate non-linear predictor effects. 
The essence of the extension is penalization of basis functions such 
as B-splines and Debauchies wavelets. Such penalization can be
achieved through random effects models that have the same
form as those used traditionally in longitudinal and multilevel
data analysis. Generalized additive models, group-specific curve models
and varying coefficient models are some of the families of models
that are included within semiparametric regression.
If a Bayesian approach is adopted then semiparametric
regression can be couched within the directed acyclic graphical models
infrastructure and, for example, Markov chain Monte Carlo (MCMC) and
mean field variational Bayes (MFVB) algorithms and software can be used 
for fitting and inference. The MFVB approach has the advantage of
being scalable to very large models and big data-sets.
Recent articles by the author that describe MCMC and MFVB 
approaches to semiparametric regression analysis include 
Ruppert, Wand \myand Carroll (2009), Wand (2009),
Marley \myand Wand (2010), Wand \myand Ormerod (2011)
and Luts, Broderick \myand Wand (2014). 

In this article we revisit MFVB for semiparametric regression but 
instead work with an approach known as \emph{variational message
passing} (VMP) (Winn \myand Bishop, 2005). The MFVB and VMP 
approaches each lead to ostensibly different iterative algorithms 
but, in a wide range of models, converge to the identical posterior 
density function approximations since they are 
each founded upon the same optimization problem.
VMP has the advantage that its iterative updates are 
amenable to modularization, and extension to arbitrarily large
models, via the notion of \emph{factor graph fragments}. 
Factor graphs (Frey \textit{et al.} 1998), described in 
Section \ref{sec:facGraphs}, is a relatively new graphical concept.
As explained in Minka (2005), mean field variational 
approximation iterative updates can be expressed as 
\emph{messages} passed between nodes on a
suitable factor graph. \emph{Message passing} is a general principle
in software engineering for efficient computing within so-called 
distributed systems (e.g. Ghosh, 2015). 
In the contemporary statistics literature,
Jordan (2004) explains how message passing can be used 
to streamline the computation of marginal probability mass 
functions of the nodes on large discrete random variable
probabilistic undirected trees as a pedagogical special 
case of the factor graph treatment given in Kschischang 
\textit{et al.} (2001). This particular message passing strategy
is known as the \emph{sum-product algorithm}. 
Despite its appeal for efficient and modular computation on 
large graphical models, message passing on factor graphs is 
not well-known in mainstream statistics. The thrust of this article 
is an explanation of how it benefits semiparametric regression 
analysis. Even though we concentrate on semiparametric regression, the principles 
apply quite generally and can be transferred to other classes 
of statistical models such as those involving, for example, 
missing data, time series correlation structures and 
classification-oriented loss functions.

The efficiencies afforded by VMP also apply to another message
passing algorithm known as \emph{expectation propagation} 
(e.g. Minka, 2005), although here we focus on the simpler VMP
approach. The high-quality software package \textsf{Infer.NET} 
(Minka \textit{et al.},2014) supports expectation propagation 
and VMP fitting of various Bayesian hierarchical models. 
However, the nature of MFVB/VMP is such that coverage of various
arbitrary scenarios in a general purpose software package
is virtually impossible. The current release of \textsf{Infer.NET}
has limitations in that many important semiparametric 
regression scenarios are not supported and self-implementation 
is the only option. Therefore it is important to understand 
the message passing paradigm and how it can be used to build 
both general purpose and special purpose approximate inference 
engines. This article is a launch pad for the algebra and computing 
required for fitting arbitrary semiparametric regression models, 
and other statistical models, regardless of support by \textsf{Infer.NET}. 
At first glance, the algebra of VMP is foreign-looking
for readers who work in statistics. Section \ref{sec:linReg}
provides the details of VMP for a Bayesian linear regression model
and working through it carefully is recommended for digestion of 
the concept. 

Recently Kucukelbir \textit{et al.} (2016) announced support of Gaussian 
variational approximations in the \textsf{Stan} package 
(Stan Development Team, 2016). This is a different type of approximation
used by \textsf{Infer.NET} and this article.

Mean field restrictions, upon which MFVB/VMP is based, 
often lead to much simpler approximate Bayesian inference algorithms
compared with the unrestricted exact case. The
accuracy of the inference is typically very good 
(e.g. Faes \textit{et al.} 2011, Luts \myand Wand, 2015).
Nevertheless, mean field variational inference is
prone to varying degrees of inaccuracy and, for
classes of models of interest, benchmarking against
Markov chain Monte Carlo fitting is recommended
to see if the accuracy of MFVB/VMP is acceptable
for the intended application. In Sections \ref{sec:GaussResp}
and \ref{sec:generalized} we show how a wide variety 
of Gaussian, Bernoulli and Poisson response semiparametric 
models can be accommodated via a few updating rules. 
Moreover, the updates involve purely matrix algebraic 
manipulations and can be readily implemented, and 
compartmentalized into a small number of functions, in
the analyst's computing environment of choice.

As explained in Section 3.5 of Winn \myand Bishop (2005),
the messages required for VMP fitting can be passed according
to a flexible schedule with convergence occurring, under
mild conditions, regardless of the order in which the 
messages are updated. This entails straightforward 
parallelizability of VMP algorithms,
meaning that for large models the computing can be 
distributed across several cores. Luts (2015) contains
details on parallelization of variational semiparametric
regression analysis for distributed data sets. In a similar vein,
VMP can achieve real-time fitting and inference for
semiparametric regression by analogy with the MFVB approaches
described by Luts, Broderick \myand Wand (2014).

Section \ref{sec:background} provides background material 
relevant to VMP. In Section \ref{sec:linReg}
we use a Bayesian linear regression setting to convey the
main ideas of VMP and then describe the ease of 
extension to larger models. Sections \ref{sec:GaussResp} 
and \ref{sec:generalized} form the centerpiece
of this article. They describe eight factor graph fragments
that are the building blocks of a wide range of arbitrarily
large semiparametric regression models. The more straightforward
Gaussian response case is treated first in Section \ref{sec:GaussResp}
and then, in Section \ref{sec:generalized}, we show how Bernoulli
and Poisson response models can also be accommodated via the
addition of only a handful of algebraic rules. Speed considerations
are briefly discussed in Section \ref{sec:speed} before some concluding
remarks in Section \ref{sec:conclusion}.
An online supplement to this article provides technicalities
such as detailed derivations.

\section{Background Material}\label{sec:background}

Here we provide some notation and coverage of background
material required for our treatment of VMP for semiparametric
regression in upcoming sections.

\subsection{Density Function Notation}

In keeping with the MFVB and VMP literature we let $p$ be
the generic symbol for a density function
when describing models and exact posterior density functions.
Approximate posterior density functions according to MFVB/VMP
restrictions are denoted generically by $q$. 

As an example, consider a model 
having observed data vector $\by$ and parameter vectors $\btheta_1$ 
and $\btheta_2$. The joint posterior density function of $\btheta_1$
and $\btheta_2$ is 
$$p(\btheta_1,\btheta_2|\by)=\frac{p(\btheta_1,\btheta_2,\by)}{p(\by)}.$$
A mean field approximation to $p(\btheta_1,\btheta_2|\by)$, based
on the restriction that $\btheta_1$ and $\btheta_2$ have
posterior independence, is denoted by $q(\btheta_1)q(\btheta_2)$
with the dependence on $\by$ suppressed. The essence of mean field
approximation and references to more detailed descriptions
are given in Section \ref{sec:MFVB}.

\subsection{Matrix Definitions and Results}\label{sec:matrix}

If $\bv$ is a column vector then $\Vert\bv\Vert\equiv\sqrt{\bv^T\bv}$.
For a $d\times d$ matrix $\bA$ we let $\vecof(\bA)$ denote
the $d^2\times 1$ vector obtained by stacking the columns
of $\bA$ underneath each other in order from left to right.
For a $d^2\times1$ vector $\ba$ we let $\vecof^{-1}(\ba)$ denote
the $d\times d$ matrix formed from listing the entries of $\ba$
in a column-wise fashion in order from left to right.
Note that $\vecof^{-1}$ is the usual function inverse
when the domain of $\vecof$ is restricted to square matrices.
In particular, $\vecof^{-1}\{\vecof(\bA)\}=\bA$ 
for $d\times d$ matrices $\bA$ and $\vecof\{\vecof^{-1}(\ba)\}=\ba$
for $d^2\times 1$ vectors $\ba$.
The following identity links $\vecof$ and the matrix trace:
$\tr(\bA^T\bB)=\vecof(\bA)^T\vecof(\bB)$
for any two matrices $\bA$ and $\bB$ such that $\bA^T\bB$ is
defined and square.
If $\ba$ and $\bb$ are both $d\times 1$ vectors then 
$\ba\odot\bb$ denotes their element-wise product and
$\ba/\bb$ denotes their element-wise quotient.
Lastly, we use the convention that function evaluation
is element-wise when applied to vectors. For example,
if $s:\real\to\real$ then $s(\ba)$ denotes the 
$d\times 1$ vector with $i$th entry equal to $s(a_i)$.

\subsection{Exponential Family Distributions}\label{sec:facGraphs}

Univariate exponential family density and 
probability mass functions are those that can be written 
in the form
\begin{equation}
p(x)=\exp\{\bT(x)^T\bdeta-A(\bdeta)\}h(x)
\label{eq:univExponFamForm}
\end{equation}
where $\bT(x)$ is the \emph{sufficient statistic}, 
$\bdeta$ is the \emph{natural parameter}, 
$A(\bdeta)$ is the \emph{log-partition function}
and $h(x)$ is the \emph{base measure}. 
Note that the sufficient statistic is not unique.
However, it is common to take $\bT(x)$ to
be the simplest possible algebraic form given $p(x)$.

An exponential family density function that arises several
times in this article is that corresponding to an
\emph{Inverse Chi-Squared} random variable. The density function
has general form
\begin{equation}
p(x)=\{(\lambda/2)^{\kappa/2}/\Gamma(\kappa/2)\}\,x^{-(\kappa/2)-1}
\exp\{-(\lambda/2)/x\},\quad x>0,
\label{eq:InvWishComm}
\end{equation}
where $\kappa>0$ and $\lambda>0$ are, respectively, shape and
scale parameters. Simple algebraic manipulations show that
(\ref{eq:InvWishComm}) is a special case of (\ref{eq:univExponFamForm})
with 
$$
\bT(x)=\left[
\begin{array}{c}
\log(x)\\
1/x
\end{array}
\right],
\quad
\bdeta=
\left[
\begin{array}{c}
\eta_1\\
\eta_2
\end{array}
\right]=\left[
\begin{array}{c}
-\smhalf(\kappa+2)\\[1ex]
-\smhalf\lambda
\end{array}
\right]
\quad\mbox{and}\quad
h(x)=I(x>0)
$$
where $I(\Psc)=1$ if $\Psc$ is true and
$I(\Psc)=0$ if $\Psc$ is false. The log-partition 
function is $A(\bdeta)=(\eta_1+1)\log(-\eta_2)+\log\Gamma(-\eta_1-1)$.

Section \ref{sec:expFam} of the online supplement chronicles 
the sufficient statistics and natural parameter vectors, and other 
relevant relationships, for several exponential family distributions 
arising in semiparametric regression. Included is extension to
multivariate density functions for random vectors and matrices.

\subsection{Factor Graphs}

A \emph{factor graph} is a graphical representation of the 
factor/argument dependencies of a real-valued function. 
Consider, for example, the function $h$ defined on $\real^5$ as
follows:
\begin{equation}
{\setlength\arraycolsep{1pt}
\begin{array}{rcl}
h(x_1,x_2,x_3,x_4,x_5)&\equiv&(x_1+x_2)\,\sin(x_2+3^{x_3x_4})\,
\,\sqrt{\frac{x_3}{x_3^3+1}}\,\tan^8(x_4^2-x_5)\,
\coth\left(\displaystyle{\frac{x_5+9}{7x_1+1}}\right)\\[2ex]
&=&f_1(x_1,x_2)\,f_2(x_2,x_3,x_4)\,f_3(x_3)\,f_4(x_4,x_5)\,f_5(x_1,x_5)
\end{array}
}
\label{eq:hfunction}
\end{equation}
where, for example, $f_1(x_1,x_2)\equiv x_1+x_2$ and $f_2,\ldots,f_5$
are defined similarly. Then Figure \ref{fig:simpIntroFacGraph}
is a factor graph corresponding to $h$. The circular nodes 
match the arguments of $h$ and the square nodes coincide with the
factors in (\ref{eq:hfunction}). Edges are drawn between each
factor node and arguments of that factor. Factor graphs of functions
are not unique since, for example, $f_1$ and $f_2$ could be
combined into a single factor and a different factor graph
would result.

\ifthenelse{\boolean{ShowFigures}}
{
\begin{figure}[!ht]
\centering
{\includegraphics[width=0.45\textwidth]{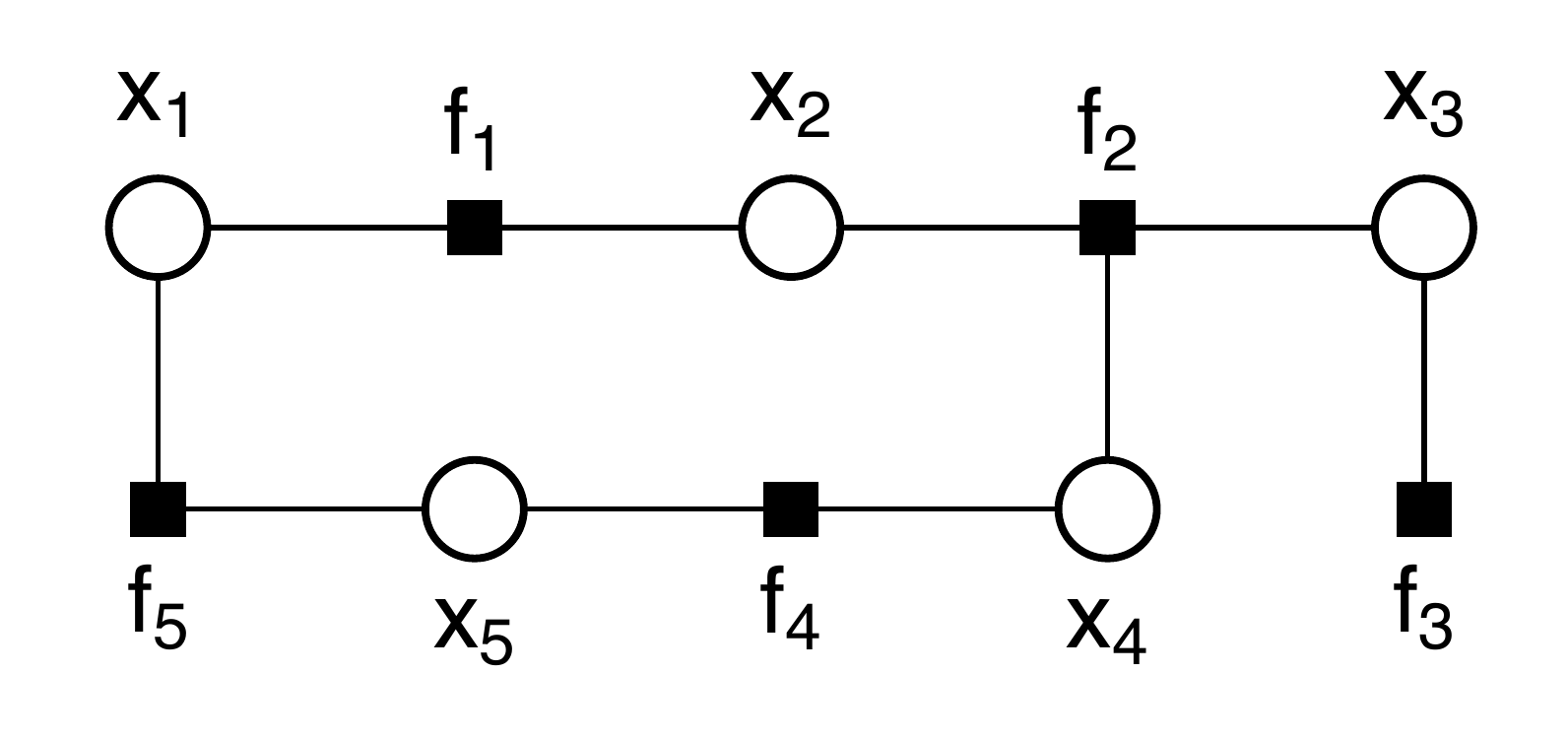}}
\caption{\it A factor graph corresponding to the function 
$h(x_1,x_2,x_3,x_4,x_5)$
defined by (\ref{eq:hfunction}).
}
\label{fig:simpIntroFacGraph} 
\end{figure}
}
{\vskip3mm
\thickboxit{\bf \centerline{Figure here.}}
\vskip3mm
}

All of the factor graphs in the remainder of this article
are such that the circular nodes correspond to random variables, 
random vectors and random matrices. Hence, we use the phrase
\emph{stochastic node} to describe a circular node. A square node
is simply called a \emph{factor}. We use the word \emph{node}
to describe either a stochastic node or a factor. If two nodes
on a factor graph are joined by an edge then we say that
the nodes are \emph{neighbors} of each other.

\subsection{Variational Message Passing}\label{sec:VMP}

Consider a Bayesian statistical model with observed data $\bD$ and 
parameter vector $\btheta$. A mean field variational approximation to the 
posterior density function $p(\btheta|\bD)$ is 
$$p(\btheta|\bD)\approx q^*(\btheta)$$
where $q^*(\btheta)$ is the minimizer of the Kullback-Leibler
divergence
$\int q(\btheta)\log\left\{\frac{q(\btheta)}{p(\btheta|\bD)}\right\}\,d\btheta$
subject to the product density restriction 
$q(\btheta)=\prod_{i=1}^M\,q(\btheta_i)$
and
\begin{equation}
\{\btheta_1,\ldots,\btheta_{M}\}
\label{eq:thetaPartition}
\end{equation}
is some partition of $\btheta$. A useful notation for any
subset $S$ of $\{1,\ldots,M\}$ is
$\btheta_{S}\equiv\{\btheta_i:i\in S\}$.
Given the partition (\ref{eq:thetaPartition}), 
the joint density function of $\btheta$ and $\bD$ is expressible as 
\begin{equation}
p(\btheta,\bD)=\prod_{j=1}^{N}\,f_j\big(\btheta_{S_j}\big)
\ \ \mbox{for subsets $S_j$ of $\{1,\ldots,M\}$ and \emph{factors} $f_j$,\ 
$1\le j\le N$.}
\label{eq:generalFactors}
\end{equation}
For example, if 
$p(\btheta,\bD)$ is a directed acyclic graphical model
with nodes $\btheta_1,\ldots,\btheta_{M}$ and $\bD$ then 
\begin{equation}
p(\btheta,\bD)=
\left\{\prod_{i=1}^M\,p(\btheta_i|\mbox{parents of $\btheta_i$})\right\}
p(\bD|\,\mbox{parents of $\bD$})
\label{eq:DAGfactors}
\end{equation}
is an $N=M+1$ example of (\ref{eq:generalFactors}) with $f_j$, $1\le j\le M$,
corresponding to the density function of $\theta_j$ conditional on its
parents and $f_{M+1}$ corresponding to the likelihood. Each factor
is a function of the subset of (\ref{eq:thetaPartition}) corresponding
to parental relationships in the directed acyclic graph. Further
factorization of (\ref{eq:DAGfactors}) may be possible. 

The factor graph in Figure \ref{fig:generFacGraphBW} shows an
$M=9$, $N=11$ example of (\ref{eq:generalFactors}).
The edges link each factor to the stochastic nodes on 
which the factor depends.

\ifthenelse{\boolean{ShowFigures}}
{
\begin{figure}[!ht]
\centering
{\includegraphics[width=0.6\textwidth]{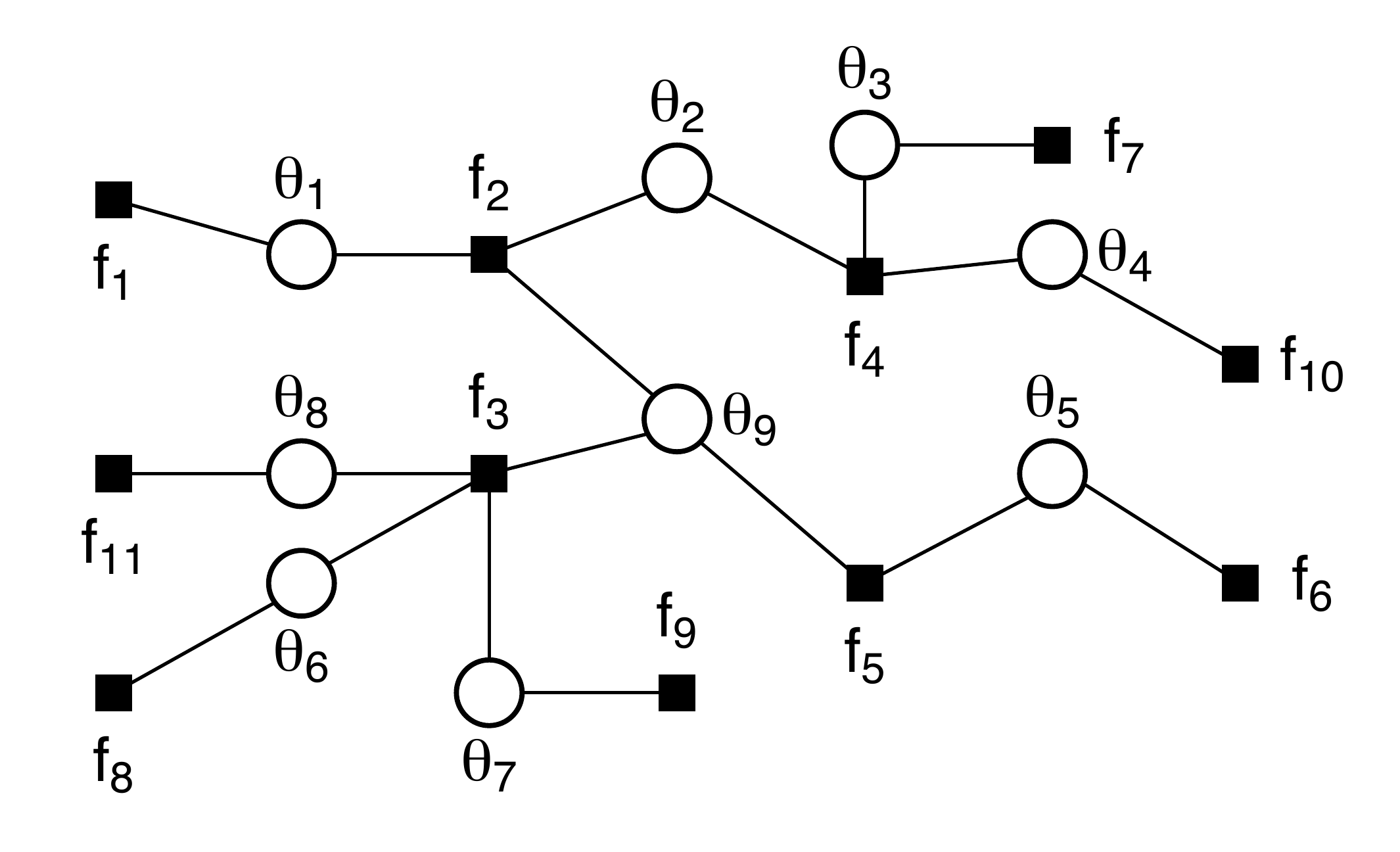}}
\caption{\it A factor graph corresponding to a Bayesian
model with stochastic nodes $\theta_1,\ldots,\theta_9$ and 
factors $f_1,\ldots,f_{11}$.}
\label{fig:generFacGraphBW} 
\end{figure}
}
{\vskip3mm
\thickboxit{\bf \centerline{Figure here.}}
\vskip3mm
}

VMP can be expressed in terms of updating
messages passed between nodes on the factor graph,
and its description benefits from the notation:
$$\neighbors(j)\equiv\{1\le i\le M:\mbox{$\btheta_i$ is a neighbor of $f_j$}\}.$$
Examples of this notation for the Figure \ref{fig:generFacGraphBW}
factor graph are 
$$\neighbors(1)=\{1\},\quad \neighbors(2)=\{1,2,9\}\quad\mbox{and}\quad\
\neighbors(3)=\{6,7,8,9\}.$$
Hence, according to this notation,
$p(\btheta,\bD)=\prod_{j=1}^{N}\,f_j(\btheta_{\subscriptneighbors(j)})$.
For each $1\le i\le M$ and $1\le j\le N$, the VMP stochastic node to 
factor message updates are
\begin{equation}
\biggerm_{\btheta_i\to\ f_j}(\btheta_i)\thickarrow\,\propto
\prod_{j'\ne j:\,i\in\subscriptneighbors(j')}
\biggerm_{f_{j'}\to\btheta_i}(\btheta_i)
\label{eq:stochToFac}
\end{equation}
and the factor to stochastic node message updates are
\begin{equation}
\biggerm_{f_j\to\btheta_i}(\btheta_i)\thickarrow\,\propto\exp\Big[E_{f_j\to\btheta_i}
\Big\{\log\biggerf_j(\btheta_{\subscriptneighbors(j)}\,)\Big\}\Big]
\label{eq:facToStochOne}
\end{equation}
where $E_{f_j\to\btheta_i}$ denotes expectation with respect to the density 
function
\begin{equation}
\frac{\displaystyle{\prod_{i'\in\subscriptneighbors(j)\backslash\{i\}}}
\biggerm_{f_j\to\btheta_{i'}}(\btheta_{i'})\biggerm_{\btheta_{i'}\to f_j}(\btheta_{i'})}
{\displaystyle{\prod_{i'\in\subscriptneighbors(j)\backslash\{i\}}}\int
\biggerm_{f_j\to\btheta_{i'}}(\btheta_{i'})\biggerm_{\btheta_{i'}\to f_j}(\btheta_{i'})
\,d\btheta_{i'}}.
\label{eq:facToStochTwo}
\end{equation}
In (\ref{eq:stochToFac}) and (\ref{eq:facToStochOne})
the $\thickarrow\,\propto$ symbol means that the function of 
$\btheta_i$ on the left-hand side is updated according
to the expression on the right-hand side but that multiplicative 
factors not depending on $\btheta_i$ can be ignored.
For common statistical models, the messages arising from
(\ref{eq:facToStochOne}) are proportional to exponential 
family density functions and some simple examples are
given in Section \ref{sec:altVMP}. 
If $\neighbors(j)\backslash\{i\}=\emptyset$ then 
the expectation in (\ref{eq:facToStochOne}) can be dispensed with
and the right-hand side of (\ref{eq:facToStochOne}) is proportional to  
$\biggerf_j(\btheta_{\subscriptneighbors(j)}\,)$.
The normalizing factor in (\ref{eq:facToStochTwo}) involves
summation if some of the $\btheta_{i'}$ have discrete components.
Upon convergence of the messages, the Kullback-Leibler optimal 
$q$-densities are obtained via
\begin{equation}
q^*(\btheta_i)\propto\prod_{j:i\in\subscriptneighbors(j)}
\biggerm_{f_j\to\btheta_i}(\btheta_i).
\label{eq:qStarVMP}
\end{equation}

The genesis of (\ref{eq:stochToFac})--(\ref{eq:qStarVMP}) 
is given in Minka (2005) where a factor graph-based
approach to VMP is described. Winn \myand Bishop (2005)
develop an alternative version of VMP based on
directed acyclic graphs. Yet another version of VMP is
given in Appendix A of Minka \myand Winn (2008)
which is similar, but not identical to, that
given in Minka (2005). All three versions, as well
as MFVB, converge to the same posterior density function
approximations.

Section 3.6 of Winn \myand Bishop (2005) and Appendix A
of Minka \myand Winn (2008) also describe calculation of the
marginal log-likelihood lower bound
\begin{equation}
\log\punder(\bD;q)\equiv\int q(\btheta)\,
\log\left\{\frac{p(\bD,\btheta)}{q(\btheta)}\right\}\,d\btheta
\label{eq:logMLfirst}
\end{equation}
which satisfies $\log\punder(\bD;q)\le\log\,p(\bD)$ regardless
of $q$. In Winn \myand Bishop (2005) $\log\punder(\bD;q)$
is referred to as the \emph{log evidence}. 
Section \ref{sec:logMLderiv} in the online supplement
describes streamlined computation of this quantity within
the VMP framework.

\subsection{Bayesian Semiparametric Regression}\label{sec:BayeSemiPar}

Detailed descriptions of Bayesian semiparametric regression  
are given in, for example, Chapter 16 of Ruppert \textit{et al.} (2003),
Gurrin \textit{et al.} (2005) and Wand (2009). Here we provide a very
brief account of the topic.

A fundamental ingredient, which facilitates the incorporation of non-linear
predictor effects, is that of \emph{mixed model-based penalized splines}.
If $x$ is a continuous predictor variable then the most common
form of a mixed model-based penalized spline in $x$ is 
\begin{equation}
f(x)=\beta_0+\beta_1\,x+\sum_{k=1}^K\,u_k\,z_k(x),\quad u_k|\sigma_u\simind N(0,\sigma_u^2),\quad
1\le k\le K,
\label{eq:penSplineFirst}
\end{equation}
where $\{z_k:1\le k\le K\}$ is a suitable spline basis. A good default choice
for the $z_k$s are canonical cubic O'Sullivan splines as described in 
Section 4 of Wand \myand Ormerod (2008), although any scatterplot smoother
with a linear basis expansion and a single quadratic penalty can
be re-parametrized to have form (\ref{eq:penSplineFirst}).

In Bayesian semiparametric regression $\beta_0$, $\beta_1$ and $\sigma_u$ are
random variables which require prior distributions to be imposed upon them.
A common choice for $(\beta_0,\beta_1)$ is a Bivariate Normal distribution
prior, which allows straightforward approximate noninformativity to be imposed. As explained
in Gelman (2006), approximate noninformativity of $\sigma_u$ can be achieved via 
Uniform distribution and Half $t$ distribution priors. 
The illustrations given in the current article
use Half Cauchy priors for standard deviation parameters such as $\sigma_u$. 
This entails setting $p(\sigma_u)=2/[\pi\,A_u\{1+(\sigma_u/A_u)^2\}]$,
$\sigma_u>0$, where the scale parameter $A_u>0$ is a user-specified hyperparameter.
We denote this by $\sigma_u\sim\mbox{Half-Cauchy}(A)$.
MFVB and VMP benefit from the following auxiliary variable result:
\begin{equation}
\begin{array}{l}
\mbox{if}\ \sigma_u^2|\,a_u\sim\mbox{Inverse-$\chi^2$}(1,1/a_u)\ \mbox{and}\     
a_u\sim\mbox{Inverse-$\chi^2$}(1,1/A_u^2)\\[1ex] 
\mbox{then}\ \sigma_u\sim\mbox{Half-Cauchy}(A_u).
\end{array}
\label{eq:HalfCauRes}
\end{equation}
A covariance matrix extension of (\ref{eq:HalfCauRes}) is described
in Huang \myand Wand (2013) and is given by (\ref{eq:HuangWand})
in the upcoming Section \ref{sec:IterInvGWish}.

The presence of penalized univariate or multivariate splines and, occasionally, 
penalized versions of other types of basis functions such as wavelets 
(e.g. Wand \myand Ormerod, 2011) is the distinguishing feature of 
semiparametric regression compared with parametric regression. 
We advocate a broad view of the latter with linear models, 
linear mixed models and their various generalized 
response extensions included. According to this viewpoint, Bayesian
versions of many of the models used in longitudinal and multilevel data analysis
(e.g. Diggle \textit{et. al}, 2002; Fitzmaurice \textit{et. al}, 2008; 
Gelman \myand Hill, 2007; Goldstein, 2010) lie within the
realm of Bayesian semiparametric regression.

\subsection{A Central Function: $\GVMP$}\label{sec:GVMP}

For a $d\times 1$ vector $\bv_1$ and 
a $d^2\times 1$ vector $\bv_2$ such that $\vecof^{-1}(\bv_2)$ is symmetric,
the following function is central to VMP for semiparametric 
regression:
\begin{equation}
\begin{array}{rcl}
\GVMP\left(\left[\begin{array}{c}    
\bv_1\\
\bv_2
\end{array}
\right];\bQ,\br,s\right)
&\equiv& -\eighth\,\mbox{tr}\Big(\bQ\{\vecof^{-1}(\bv_2)\}^{-1}
[\bv_1\bv_1^T\{\vecof^{-1}(\bv_2)\}^{-1}-2\bI]\Big)\\
&&\quad-\smhalf\br^T\{\vecof^{-1}(\bv_2)\}^{-1}\bv_1-\smhalf s.
\end{array}
\label{eq:GVMPdefn}
\end{equation}
The secondary arguments of $\GVMP$ are a $d\times d$ matrix $\bQ$,
a $d\times1$ vector $\br$ and $s\in\real$. The function $\GVMP$
arises from the following fact: if if $\btheta$ is a $d\times 1$ 
Multivariate Normal random vector with natural parameter vector 
$\bdeta$ as defined by (\ref{eq:MVNnatVScomm}) in the online supplement 
then
$$E_{\btheta}\big\{-\smhalf\big(\btheta^T\bQ\,\btheta-2\br^T\btheta+s\big)\big\}
=E_{\btheta}\big(-\smhalf\btheta^T\bQ\,\btheta+\br^T\btheta\big)-\smhalf\,s
=\GVMP(\bdeta\,;\bQ,\br,s).
$$
%
%
%
For example, if $\ba$ is an $m\times 1$ vector and $\bA$ is an 
$m\times d$ matrix then
$$E_{\btheta}\left\{-\smhalf\Vert\ba-\bA\btheta\Vert^2\right\}
=\GVMP\Big(\bdeta;\bA^T\bA,\,\bA^T\ba,\,\ba^T\ba\Big).$$

\section{Linear Regression Illustrative Example}\label{sec:linReg}

Consider the Bayesian regression model
\begin{equation}
\begin{array}{c}
\by|\,\bbeta,\sigma^2\sim N(\bX\bbeta,\sigma^2\,\bI),\quad
\bbeta\sim N(\bmu_{\bbeta},\bSigma_{\bbeta}),\\[2ex]
\sigma^2|a\sim\mbox{Inverse-$\chi^2$}(1,1/a),\quad
a\sim\mbox{Inverse-$\chi^2$}(1,1/A^2)
\end{array}
\label{eq:linRegMod}
\end{equation}
where $\by$ is an $n\times 1$ vector of response data and  $\bX$ is an 
$n\times d$ design matrix. The $d\times1$ vector $\bmu_{\bbeta}$, 
the $d\times d$ covariance matrix $\bSigma_{\bbeta}$ and $A>0$
are user-specified hyperparameters that remain fixed throughout
any approximate Bayesian inference procedure for (\ref{eq:linRegMod}).
As explained in Section \ref{sec:BayeSemiPar}, the marginal 
prior distribution on $\sigma$ in (\ref{eq:linRegMod}) is 
$\mbox{Half-Cauchy}(A)$. The joint posterior density function
of the model parameters and auxiliary variable $a$ is 
\begin{equation}
p(\bbeta,\sigma^2,a\,|\,\by)=\frac{p(\bbeta,\sigma^2,a,\by)}{p(\by)}
\label{eq:linModPoster}
\end{equation}
but is analytically intractable and numerically challenging. 
MCMC (e.g. Chapters 11-12, Gelman \textit{et al.}, 2014) is the most common tool 
for making approximate Bayesian inference for $\bbeta$ and $\sigma^2$.
The computationally intensive nature of MCMC entails that, whilst
its speed will be acceptable for some applications, there are others
where faster approximations are desirable or necessary. 
We next describe MFVB as one such fast alternative.

\subsection{Mean Field Variational Bayes Approach}\label{sec:MFVB}

MFVB is a prescription for approximation
of posterior density functions in a graphical model.
References on MFVB for general graphical models include Bishop (2006),
Wainwright \myand Jordan (2008) and Ormerod \myand Wand (2010).
In this section we focus on MFVB for approximation of (\ref{eq:linModPoster}).
This is founded upon $p(\bbeta,\sigma^2,a\,|\,\by)$ being restricted 
to have the product form 
\begin{equation}
q(\bbeta)\,q(\sigma^2)\,q(a)
\label{eq:prodRestrict}
\end{equation}
for density functions $q(\bbeta)$, $q(\sigma^2)$ and $q(a)$.
These \emph{$q$-density functions} are then chosen
to minimize the Kullback-Leibler distance between
$p(\bbeta,\sigma^2,a|\by)$ and $q(\bbeta)\,q(\sigma^2)\,q(a)$:
$$\int q(\bbeta)\,q(\sigma^2)\,q(a)\,
\log\left\{\frac{q(\bbeta)\,q(\sigma^2)\,q(a)}
{p(\bbeta,a,\sigma^2|\by)}\right\}\,d\bbeta\,d\sigma^2\,da.
$$
One can then prove by variational calculus that the optimal $q$-densities satisfy:
$$
\begin{array}{l}
q^*(\bbeta)\  \mbox{is a}\ N(\bmu_{q(\bbeta)},\bSigma_{q(\bbeta)})\ \mbox{density function},\\[2ex]
q^*(\sigma^2)\  \mbox{is an}\ \mbox{Inverse-$\chi^2$}\Big(n+1,\lambda_{q(\sigma^2)}\Big)
\ \mbox{density function, and}\\[2ex]
q^*(a)\  \mbox{is an}\ \mbox{Inverse-$\chi^2$}\Big(2,\lambda_{q(a)}\Big)\ \mbox{density function}
\end{array}
$$
for some $d\times1$ vector $\bmu_{q(\bbeta)}$, $d\times d$ covariance matrix 
$\bSigma_{q(\bbeta)}$ and positive scalars $\lambda_{q(\sigma^2)}$ and $\lambda_{q(a)}$.
These $q$-density parameters do not have closed form solutions but, instead,
can be determined iteratively via coordinate ascent as explained in
Section 10.1.1 of Bishop (2006) and Section 2.2 of Ormerod \myand Wand (2010).
For the model at hand, the coordinate ascent updates reduce to
Algorithm \ref{alg:lmMFVB}. Here and elsewhere ``$\thickarrow$'' indicates
that the quantity on the left-hand side is updated according to the
expression on the right-hand side.

\begin{algorithm}
\begin{center}
\begin{minipage}[t]{150mm}
\hrule
\begin{itemize}
\item[] Initialize:\ $\lambda_{q(\sigma^2)}>0$.
\item[] Cycle:
\begin{itemize}
\item[] $\bSigma_{q(\bbeta)}\thickarrow
\Bigg\{\Bigg(\displaystyle{\frac{n+1}{\lambda_{q(\sigma^2)}}}\Bigg)
\,\bX^T\bX+\bSigma_{\bbeta}^{-1}\Bigg\}^{-1}$
\item[] $\bmu_{q(\bbeta)}\thickarrow 
\Bigg(\displaystyle{\frac{n+1}{\lambda_{q(\sigma^2)}}}\Bigg)
\,\bSigma_{q(\bbeta)}\,
\big(\bX^T\by+\bSigma_{\bbeta}^{-1}\bmu_{\bbeta}\big)$
\ \ \ \ ;\ \ \ \  
$\lambda_{q(a)}\thickarrow 
2\Bigg\{
\Bigg(\displaystyle{\frac{n+1}{\lambda_{q(\sigma^2)}}}\Bigg)
+A^{-2}\Bigg\}$
\item[]
$\lambda_{q(\sigma^2)}\thickarrow 
\by^T\by-2\bmu_{q(\bbeta)}^T\bX^T\by
+\mbox{tr}[(\bX^T\bX)\{\bSigma_{q(\bbeta)}+\bmu_{q(\bbeta)}
\bmu_{q(\bbeta)}^T\}]+2/\lambda_{q(a)}$
\end{itemize}
\item[] until the changes in all $q$-density parameters are negligible.
\end{itemize}
\hrule
\end{minipage}
\end{center}
\caption{\it Mean field variational Bayes algorithm for
approximate inference in the Gaussian response linear regression
model (\ref{eq:linRegMod}).}
\label{alg:lmMFVB} 
\end{algorithm}

\subsection{Alternative Approach Based on Variational Message Passing}\label{sec:altVMP}

We now explain the VMP alternative for the Bayesian linear regression
example. Firstly, note that the joint distribution of all random
variables in model (\ref{eq:linRegMod}) admits the 
factorization 
\begin{equation}
p(\by,\bbeta,\sigma^2,a)=p(\by|\bbeta,\sigma^2)p(\bbeta)p(\sigma^2|a)p(a).
\label{eq:linModDens}
\end{equation}
Treating (\ref{eq:linModDens}) as a function of parameters
corresponding to each factor in the mean field restriction
(\ref{eq:prodRestrict}) we arrive at the factor graph 
shown in Figure \ref{fig:linRegFacGraphMsgs}. 

VMP iteration for fitting model (\ref{eq:linRegMod}) 
involves the updating of \emph{messages} passed from 
each node on the Figure \ref{fig:linRegFacGraphMsgs} 
factor graph to its neighboring nodes. 
Each message is a function of the stochastic node that
receives or sends the message. 
For example, the nodes $\sigma^2$ and $p(\sigma^2|a)$ are neighbors of
each other in the Figure \ref{fig:linRegFacGraphMsgs}
factor graph. The messages passed between these
two nodes are both functions of the stochastic node
$\sigma^2$ and are denoted by $\mSUBsigsqTOpsigsqa$
and $\mSUBpsigsqaTOsigsq$.
The subscripts of $\biggerm$ designates the nodes involved in the message
passing and the direction in which the message is passed.
Figure \ref{fig:linRegFacGraphMsgs} shows all 12 of the messages
between neighboring nodes on the factor graph.

\ifthenelse{\boolean{ShowFigures}}
{
\begin{figure}[!ht]
\centering
{\includegraphics[width=0.85\textwidth]{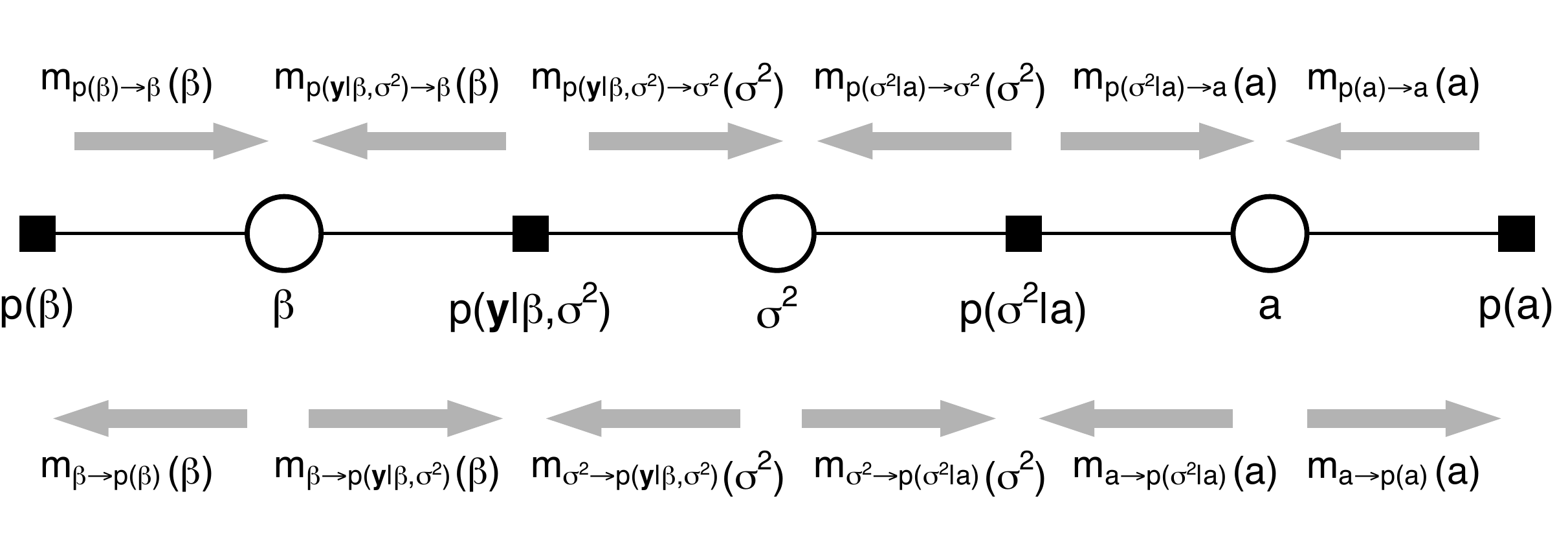}}
\caption{\it A factor graph for the function 
$p(\by,\bbeta,\sigma^2,a)$ with stochastic nodes $\bbeta$,
$\sigma^2$ and $a$, corresponding to the factors in 
product restriction (\ref{eq:prodRestrict}). Also
shown are each of the messages between neighboring nodes
on the factor graph. The gray arrows depict the directions 
in which the messages are passed.}
\label{fig:linRegFacGraphMsgs} 
\end{figure}
}
{\vskip3mm
\thickboxit{\bf \centerline{Figure here.}}
\vskip3mm
}

Based on the VMP updating equations given in Section \ref{sec:VMP},
and with details given in Section \ref{sec:msgFormsDeriv} of the online supplement,
the factor to stochastic node messages have the following functional 
forms after the first iteration:
\begin{equation}
{\setlength\arraycolsep{3pt}
\begin{array}{rcl}
\mSUBpbetaTObeta&=&\exp\left\{
\left[
\begin{array}{c}   
\bbeta\\
\vecof(\bbeta\bbeta^T)
\end{array}
\right]^T\etaSUBpbetaTObeta\right\},\\[1ex]
\mSUBpybetasigsqTObeta&=&\exp\left\{
\left[
\begin{array}{c}   
\bbeta\\
\vecof(\bbeta\bbeta^T)
\end{array}
\right]^T\etaSUBpybetasigsqTObeta\right\},\\[1ex]
\mSUBpybetasigsqTOsigsq&=&\exp\left\{
\left[
\begin{array}{c}   
\log(\sigma^2)\\
1/\sigma^2
\end{array}
\right]^T\etaSUBpybetasigsqTOsigsq\right\},\\[1ex]
\mSUBpsigsqaTOsigsq&=&\exp\left\{
\left[
\begin{array}{c}   
\log(\sigma^2)\\
1/\sigma^2
\end{array}
\right]^T\etaSUBpsigsqaTOsigsq\right\},\\[1ex]
\mSUBpsigsqaTOa&=&\exp\left\{
\left[
\begin{array}{c}   
\log(a)\\
1/a
\end{array}
\right]^T\etaSUBpsigsqaTOa\right\}\\[1ex]
\mbox{and}\quad
\mSUBpaTOa&=&\exp\left\{
\left[
\begin{array}{c}   
\log(a)\\
1/a
\end{array}
\right]^T\etaSUBpaTOa\right\}
\end{array}
}
\label{eq:linModMsgForms}
\end{equation}
for $(d+d^2)\times 1$ vectors $\etaSUBpbetaTObeta$ 
and $\etaSUBpybetasigsqTObeta$ and $2\times 1$ vectors
$\etaSUBpybetasigsqTOsigsq$, $\etaSUBpsigsqaTOsigsq$,
$\etaSUBpsigsqaTOa$ and $\etaSUBpaTOa$.
The fixed form of the messages means that, for the remaining iterations, 
the message updates (\ref{eq:stochToFac}) and (\ref{eq:facToStochOne}) simply 
involve updates for the natural parameter vectors of the messages.
Note that the last four of these 
messages are proportional to Inverse Chi-Squared density functions. 
The first two are proportional to $d$-dimensional Multivariate Normal
distributions, but expressed in exponential family form 
as explained in Section \ref{sec:MultNormDistn}.
Therefore, normalizing factors aside, 
each of the subscripted $\bdeta$ vectors
are natural parameters for a particular exponential
family density function.
The stochastic node to factor messages have the same functional
forms as their reverse messages. For example 
$$
\mSUBsigsqTOpsigsqa=\exp\left\{
\left[
\begin{array}{c}   
\log(\sigma^2)\\
1/\sigma^2
\end{array}
\right]^T\etaSUBsigsqTOpsigsqa\right\}
$$
for some $2\times1$ vector $\etaSUBsigsqTOpsigsqa$.

Once the functional forms of the messages have been determined, 
the VMP iteration loop has the following generic steps:
$$
\begin{array}{lll}
&\mbox{1. Choose a factor.}\\
&\mbox{2. Update the parameter vectors of the messages passed from}\\
&\mbox{\ \ \ \ the factor's neighboring stochastic nodes to the factor.}\\
&\mbox{3. Update the parameter vectors of the messages passed}\\
&\mbox{\ \ \ \ from the factor to its neighboring stochastic nodes.}
\end{array}
$$
For typical semiparametric regression models
the order in which factors are chosen does not matter although
all factors should eventually be chosen as the iterations proceed.
There are some classes of models, outside those treated in this
article, for which local optima exist and the update order may affect
which optimum is attained.

The updates of the stochastic node to factor natural parameter vectors 
have simple forms based on (\ref{eq:stochToFac}) and are updated
as follows:
\begin{equation}
\begin{array}{ll}
\etaSUBbetaTOpbeta\thickarrow\etaSUBpybetasigsqTObeta,\ \ \ \ \ \ 
\etaSUBbetaTOpybetasigsq\thickarrow\etaSUBpbetaTObeta\\[2ex]
\etaSUBsigsqTOpybetasigsq\thickarrow\etaSUBpsigsqaTOsigsq,\ \ \ \ \ \ 
\etaSUBsigsqTOpsigsqa\thickarrow\etaSUBpybetasigsqTOsigsq\\[2ex]
\etaSUBaTOpsigsqa\thickarrow\etaSUBpaTOa\ \ \ \mbox{and}\ \ \ 
\etaSUBaTOpa\thickarrow \etaSUBpsigsqaTOa.
\end{array}
\label{eq:stochToFacLinReg}
\end{equation}
Based on (\ref{eq:facToStochOne}) and (\ref{eq:facToStochTwo})
$\mSUBpbetaTObeta\propto p(\bbeta)$ and
$\mSUBpaTOa\propto p(a)$
so the natural parameter updates for these two messages are simply
\begin{equation}
\etaSUBpbetaTObeta\thickarrow
\left[\begin{array}{c}   
\bSigma_{\bbeta}^{-1}\bmu_{\bbeta}\\
-\smhalf\vecof(\bSigma_{\bbeta}^{-1})
\end{array}
\right]\quad\mbox{and}\quad
\etaSUBpaTOa\to
\left[\begin{array}{c}   
-3/2\\
-1/A^2
\end{array}
\right]
\label{eq:priorUpdates}
\end{equation}
and remain constant throughout the iterations. 
The updates corresponding to the messages sent from
$p(\by|\bbeta,\sigma^2)$ to its neighboring stochastic
nodes can be obtained from (\ref{eq:facToStochOne}) and 
(\ref{eq:facToStochTwo}). The expectation 
in (\ref{eq:facToStochOne}) reduces to a linear combination of  
expected sufficient statistics. Table \ref{tab:ETx} in the online
supplement gives the required expressions. Simple algebra then leads to 
\begin{equation}
\begin{array}{rcl}
\etaSUBpybetasigsqTObeta&\thickarrow&
\left[\begin{array}{c}
\bX^T\by\\[1ex]
-\smhalf\vecof(\bX^T\bX)
\end{array}
\right]\displaystyle{\frac{\Big(\etaSUBpybetasigsqCONNsigsq\Big)_1+1}{
\Big(\etaSUBpybetasigsqCONNsigsq\Big)_2}}\\[6ex]
\mbox{and}\quad\etaSUBpybetasigsqTOsigsq&\thickarrow&
\left[\begin{array}{c}
-\smhalf\,n\\[2ex]
\GVMP\Big(\etaSUBpybetasigsqCONNbeta;\bX^T\bX,\bX^T\by,\by^T\by\Big)
\end{array}
\right]
\end{array}
\label{eq:pLikUpdates}
\end{equation}
where
\begin{equation}
{\setlength\arraycolsep{2pt}
\begin{array}{rcl}
\etaSUBpybetasigsqCONNsigsq&\equiv&\etaSUBpybetasigsqTOsigsq+\etaSUBsigsqTOpybetasigsq,\\[1ex]
\etaSUBpybetasigsqCONNbeta&\equiv&\etaSUBpybetasigsqTObeta+\etaSUBbetaTOpybetasigsq,
\end{array}
}
\label{eq:doubleArrowFirst}
\end{equation}
$\Big(\etaSUBpybetasigsqCONNsigsq\Big)_i$ denotes the $i$th entry of
$\etaSUBpybetasigsqCONNsigsq$ and $\GVMP$ is explained
in Section \ref{sec:GVMP}.
The parameter updates for the messages passed from
$p(\sigma^2|a)$ to its neighbors are
\begin{equation}
\begin{array}{rcl}
\etaSUBpsigsqaTOsigsq&\thickarrow&
\left[
{\setlength\arraycolsep{0.2pt}
\begin{array}{c}
-\frac{3}{2} \\[1ex]
\displaystyle{
\frac{-\Big(\etaSUBpsigsqaCONNa\Big)_1-1}
{2\Big(\etaSUBpsigsqaCONNa\Big)_2}}
\end{array}
}
\right]\\[6ex]
\quad \mbox{and}\quad
\etaSUBpsigsqaTOa&\thickarrow&
\left[
{\setlength\arraycolsep{0.2pt}
\begin{array}{c}
-\smhalf \\[1ex]
\displaystyle{
\frac{-\Big(\etaSUBpsigsqaCONNsigsq\Big)_1-1}
{2\Big(\etaSUBpsigsqaCONNsigsq\Big)_2}}
\end{array}
}
\right]
\end{array}
\label{eq:psigsqaUpdates}
\end{equation}
where the definitions of $\etaSUBpybetasigsqCONNsigsq$ and 
$\etaSUBpsigsqaCONNa$ are analogous to those given in 
(\ref{eq:doubleArrowFirst}).

After initializing the stochastic node to factor natural
parameters, updates (\ref{eq:stochToFacLinReg}),
(\ref{eq:priorUpdates}), (\ref{eq:pLikUpdates}) and (\ref{eq:psigsqaUpdates}) 
form an iterative scheme in the message natural parameter space. 
Once convergence of the messages has been attained, the $q$-density 
natural parameters can be obtained from (\ref{eq:qStarVMP}) as:
\begin{equation}
\begin{array}{rcl}
\bdeta_{q(\bbeta)}&\thickarrow&\etaSUBpbetaTObeta+\etaSUBpybetasigsqTObeta,\\[2ex]
\bdeta_{q(\sigma^2)}&\thickarrow&\etaSUBpybetasigsqTOsigsq+\etaSUBpsigsqaTOsigsq\\[2ex]
\mbox{and}\quad\bdeta_{q(a)}&\thickarrow&\etaSUBpsigsqaTOa+\etaSUBpaTOa.
\end{array}
\label{eq:qetaupdates}
\end{equation}
Updates (\ref{eq:qetaupdates}) show that the
natural parameters of a $q$-density of a stochastic node 
and incoming messages to that node have simple linear relationships.
This, together with (\ref{eq:stochToFac}), 
motivates working with natural parameters in VMP.

The $q$-density common parameters can be obtained from (\ref{eq:qetaupdates}) 
using (\ref{eq:InvChiSqNatVsComm}) and (\ref{eq:MVNnatVScomm}) and lead to
\begin{equation}
\begin{array}{l}
\bmu_{q(\bbeta)}=-\smhalf\{\vecof^{-1}\big((\bdeta_{q(\bbeta)})_2\big)\}^{-1}
(\bdeta_{q(\bbeta)})_1,\ \ 
\bSigma_{q(\bbeta)}=-\smhalf\{\vecof^{-1}\big((\bdeta_{q(\bbeta)})_2\big)\}^{-1},\\[2ex]
\lambda_{q(\sigma^2)}= -2\big(\bdeta_{q(\sigma^2)}\big)_2\quad\mbox{and}\quad
\lambda_{q(a)}= -2\big(\bdeta_{q(a)}\big)_2
\end{array}
\end{equation}
where $(\bdeta_{q(\bbeta)})_1$ contains the first $d$ entries of $\bdeta_{q(\bbeta)}$
and $(\bdeta_{q(\bbeta)})_2$ contains the remaining $d^2$ entries of $\bdeta_{q(\bbeta)}$.
The values of $\bmu_{q(\bbeta)}$, $\bSigma_{q(\bbeta)}$, $\lambda_{q(\sigma^2)}$
and $\lambda_{q(a)}$ are the same regardless of whether one uses the MFVB
approach encapsulated in Algorithm \ref{alg:lmMFVB} or the VMP approach described
in this section. On face value, it would appear that the MFVB approach
is superior due to its succinctness. However this ranking of MFVB over VMP
is within the confines of approximate inference for model (\ref{eq:linRegMod}).
As we now explain in Section \ref{sec:arbit}, VMP is a more attractive
proposition when semiparametric regression models are extended arbitrarily.

\subsubsection{Arbitrarily Large Model Viewpoint}\label{sec:arbit}

We now turn attention to variational inference for
arbitrarily large semiparametric regression models and 
how the message passing approach allows streamlining of 
the required calculations.

Figure \ref{fig:carExamp} shows both simple linear regression
and nonparametric regression fits to data on 93 passenger
car models on sale in U.S.A. in 1993 (source: Lock, 1993).
The $i$th response observation ($y_i$) is fuel efficiency on city roads 
(miles/gallon) and the $i$th predictor observation ($x_i$) is weight of 
the car (pounds).

\ifthenelse{\boolean{ShowFigures}}
{
\begin{figure}[!ht]
\centering
{\includegraphics[width=0.85\textwidth]{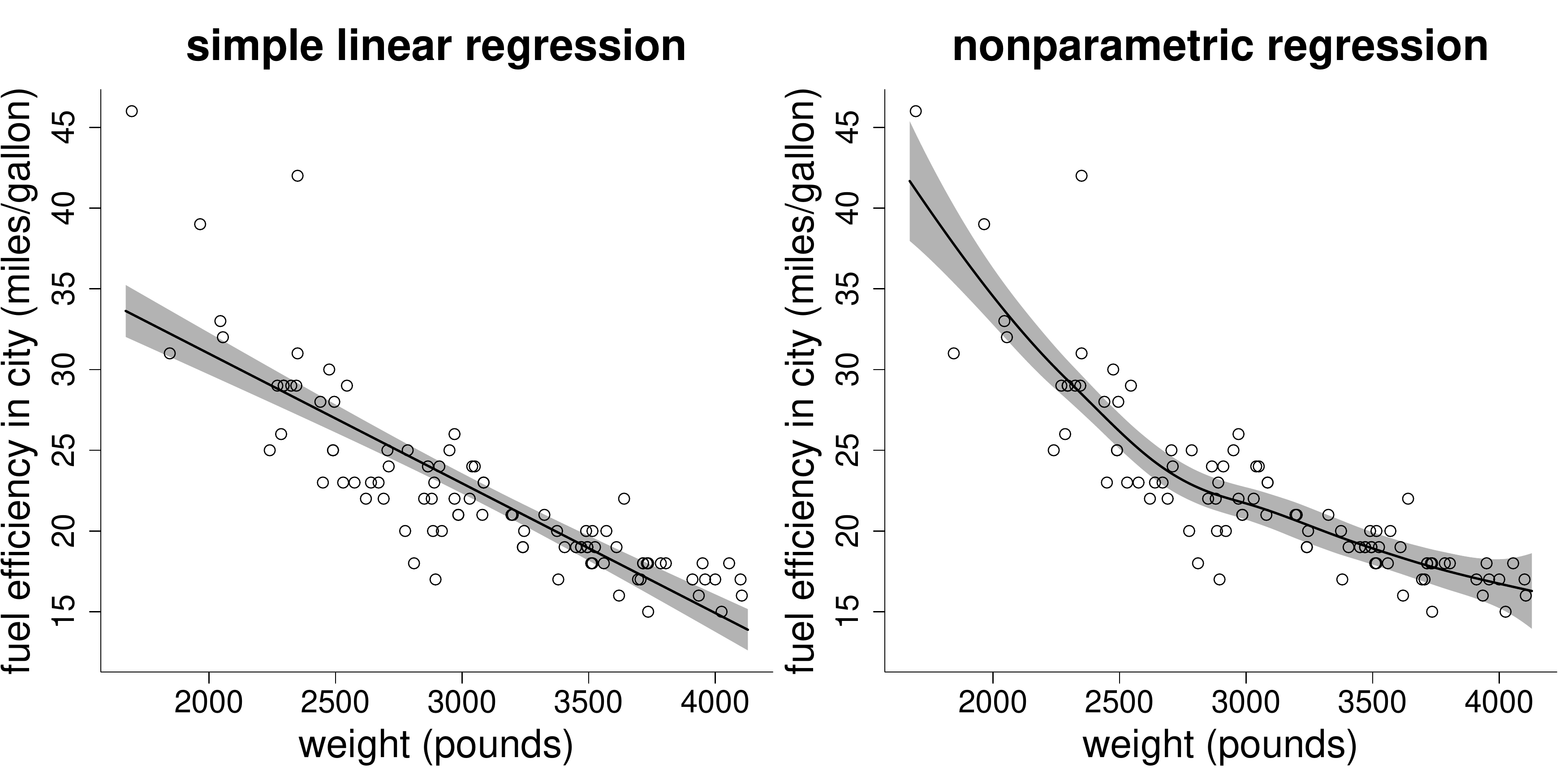}}
\caption{\it Left panel: VMP-based simple linear regression fit
to data on fuel efficiency and  weight of 93 passenger
car models on sale in U.S.A. in 1993 (source: Lock, 1993). 
The fitted line is the posterior mean and the shaded region
shows pointwise 95\% credible sets according to the 
mean field approximation (\ref{eq:prodRestrict}).
Right panel: Similar to the left panel but for 
nonparametric regression according to the mixed model-based
penalized spline extension (\ref{eq:penSplRegMod}) with
mean field approximation (\ref{eq:penSplProdForm}).
}
\label{fig:carExamp} 
\end{figure}
}
{\vskip3mm
\thickboxit{\bf \centerline{Figure here.}}
\vskip3mm
}

The simple linear regression fit is obtained using VMP
applied to the special case of (\ref{eq:linRegMod})
with $\bX=[1\ x_i]_{1\le i\le n}$. The nonparametric
regression fit in Figure \ref{fig:carExamp} is according 
to mixed model-based penalized spline model
\begin{equation}
\begin{array}{c}
\by|\,\bbeta,\bu,\sigeps^2\sim N(\bX\bbeta+\bZ\bu,\sigeps^2\,\bI),\quad
\bu|\,\sigma_u^2\sim N(0,\sigma_u^2),\quad
\bbeta\sim N(\bmu_{\bbeta},\bSigma_{\bbeta}),\\[2ex]
\sigma_u^2|a_u\sim\mbox{Inverse-$\chi^2$}(1,1/a_u),\quad
a_u\sim\mbox{Inverse-$\chi^2$}(1,1/A_u^2),\\[2ex]
\sigeps^2|\,a_{\varepsilon}\sim\mbox{Inverse-$\chi^2$}(1,1/a_{\varepsilon}),\quad
a_{\varepsilon}\sim\mbox{Inverse-$\chi^2$}(1,1/A_{\varepsilon}^2),
\end{array}
\label{eq:penSplRegMod}
\end{equation}
where
$$\bZ\equiv\left[
\begin{array}{ccc}
z_1(x_1)&\cdots&z_K(x_1)\\
\vdots  & \ddots&\vdots \\ 
z_1(x_n)&\cdots&z_K(x_n)
\end{array}
\right]
$$
for a spline basis $\{z_k:1\le k\le K\}$ as defined adjacent to 
(\ref{eq:penSplineFirst}). The mean field approximation being
used here is 
\begin{equation}
p(\bbeta,\bu,\sigma_u^2,a_u,\sigeps^2,a_{\varepsilon}|\by)
\approx q(\bbeta,\bu,a_u,a_{\varepsilon})q(\sigma_u^2,\sigeps^2).
\label{eq:penSplProdFormMin}
\end{equation}
However, further product density forms arise due to conditional
independencies in the model (e.g. Section 10.2.5 of Bishop, 2006)
and it can be established that (\ref{eq:penSplProdFormMin}) 
is equivalent to 
\begin{equation}
p(\bbeta,\bu,\sigma_u^2,a_u,\sigeps^2,a_{\varepsilon}|\by)
\approx q(\bbeta,\bu)q(\sigma_u^2)q(a_u)\,q(\sigeps^2)q(a_{\varepsilon}).
\label{eq:penSplProdForm}
\end{equation}

The extension of the VMP updates when transitioning from the linear regression
model (\ref{eq:linRegMod}) to (\ref{eq:penSplRegMod}) benefits from:
\jump

{\sl Definition.}
A \emph{factor graph fragment}, or \emph{fragment} for short, is a sub-graph 
of a factor graph consisting of a single factor and each of the stochastic nodes 
that are neighbors of the factor.

\jump
Figure \ref{fig:penSplFacGraph} shows the factor graph corresponding to 
(\ref{eq:penSplRegMod}) with mean field approximation (\ref{eq:penSplProdForm}).
This factor graph has six factors and therefore six fragments. Five of them
have the same form as the fragments of the factors of 
Figure \ref{fig:linRegFacGraphMsgs} and are colored gray. 
The black-colored fragment corresponds to the following
penalization of the coefficient vector:
$$
\left[
\begin{array}{c}
\bbeta\\
\bu
\end{array}
\right]\Big|\sigma_u^2 \sim
N\left(
\left[
\begin{array}{cc}
\sigma_{\bbeta}^2\,\bI_2 & \bzero\\
\bzero                 & \sigma_{u}^2\,\bI_K
\end{array}
\right]
\right)
$$
and is a distributional form that does not appear in 
the linear regression model.

\ifthenelse{\boolean{ShowFigures}}
{
\begin{figure}[!ht]
\centering
{\includegraphics[width=0.85\textwidth]{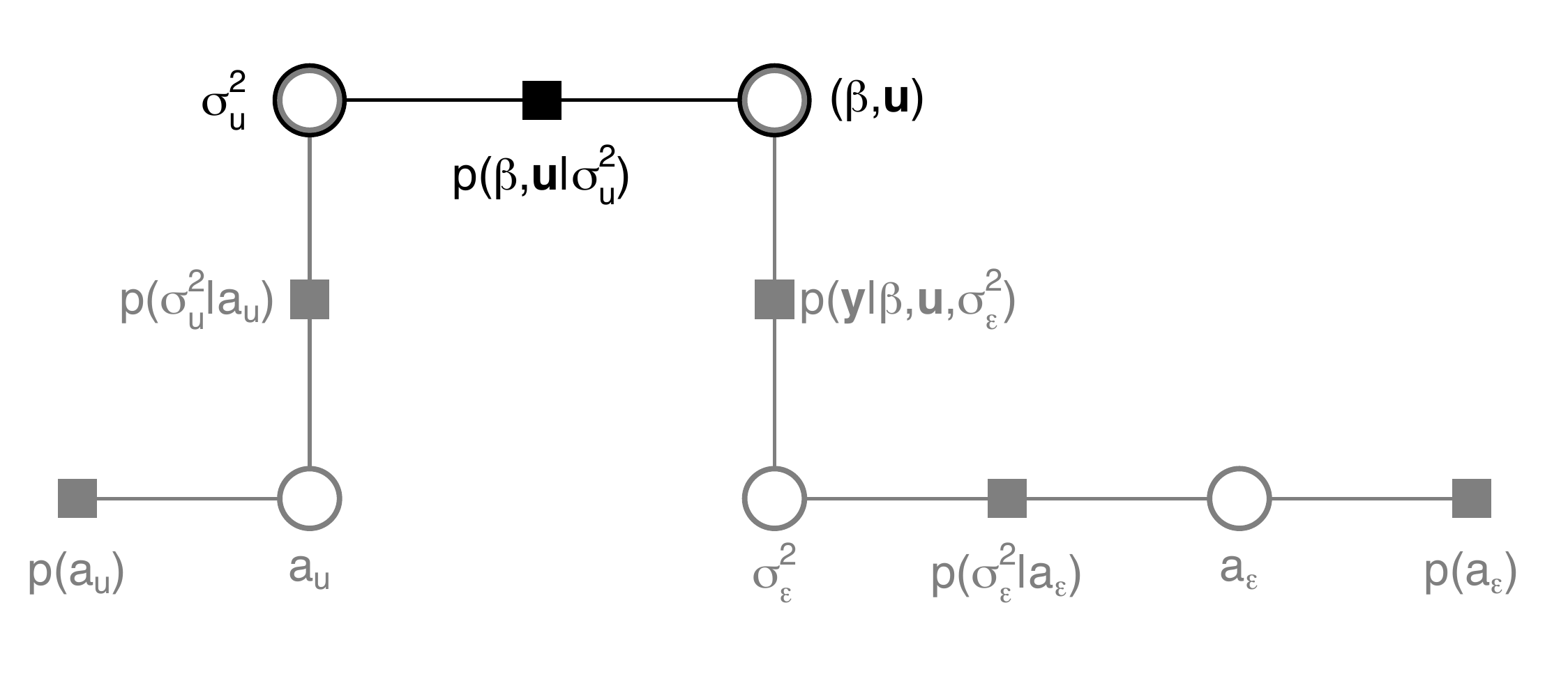}}
\caption{\it Diagrammatic depiction of the extension 
from simple linear regression to penalized spline regression.
The fragment shown in black is the only one that
is of a different type compared with the fragments in the
Bayesian linear regression model. The fragments
shown in gray are present in the linear model factor
graph shown in Figure \ref{fig:linRegFacGraphMsgs}.}
\label{fig:penSplFacGraph} 
\end{figure}
}
{\vskip3mm
\thickboxit{\bf \centerline{Figure here.}}
\vskip3mm
}

The stochastic node to factor messages in Figure \ref{fig:penSplFacGraph} 
have trivial updates analogous to those given in 
(\ref{eq:stochToFacLinReg}). The factor to stochastic messages
are more complicated, but for the five fragments shown 
in gray in Figure \ref{fig:penSplFacGraph} they 
are identical or very similar to analogous updates 
on the Figure \ref{fig:linRegFacGraphMsgs} factor graph,
as we now explain:

\begin{enumerate}
\item\label{it:firstComm} The message passed from $p(a_u)$ to $a_u$ 
has the same form as that passed from $p(a)$ to $a$
for  model (\ref{eq:linRegMod}). The natural parameter updates
$\etaSUBpauTOau$ takes the same form as that for $\etaSUBpaTOa$ 
in (\ref{eq:priorUpdates}). 
\item Comments similar to those given in \ref{it:firstComm}. apply to the 
messages passed from $p(a_{\varepsilon})$ to $a_{\varepsilon}$.
\item\label{it:thirdComm} 
The messages passed from the factor $p(\sigma_u^2|\,a_u)$ to
its neighboring stochastic nodes $\sigma_u^2$ and $a_u$ have the
same form as those passed from $p(\sigma^2|a)$ to $\sigma^2$ and $a$
for model (\ref{eq:linRegMod}). The natural parameter updates 
$\etaSUBpsigsquauTOsigsqu$ and $\etaSUBpsigsquauTOau$
take the same forms as those for $\etaSUBpsigsqaTOsigsq$ and 
$\etaSUBpsigsqaTOa$ in (\ref{eq:psigsqaUpdates}). 
\item Comments similar to those given in \ref{it:thirdComm}. apply to the 
messages passed from $p(\sigma_{\varepsilon}^2|a_{\varepsilon})$ 
to its neighboring stochastic nodes.
\item The messages passed from the factor 
$p(\by|\bbeta,\bu,\sigma_{\epsilon}^2)$ to its neighboring
stochastic nodes $(\bbeta,\bu)$ and $\sigma_{\epsilon}^2$
have a similar form to those passed from 
$p(\by|\bbeta,\sigma^2)$ to $\beta$ and $\sigma^2$ for
model (\ref{eq:linRegMod}). The natural parameter updates 
$\etaSUBpybetasigsqTObeta$ and $\etaSUBpybetasigsqTOsigsq$
take the same forms as those for $\etaSUBpybetausigsqepsTObetau$ and 
$\etaSUBpybetausigsqepsTOsigsqeps$ in (\ref{eq:pLikUpdates})
but with $\bbeta$ replaced by $(\bbeta,\bu)$, $\sigma^2$ replaced by
$\sigma_{\varepsilon}^2$  and $\bX$ replaced by
$\bC\equiv[\bX\ \bZ]$.
\end{enumerate}

It remains to take care of the black-colored fragment of
Figure \ref{fig:penSplFacGraph}.
The message passed from $p(\bbeta,\bu|\sigma_u^2)$ to 
$\sigma_u^2$ is 
$$\mSUBpbetausigsquTOsigsqu=
\exp\left\{
\left[
\begin{array}{c}
\log(\sigma_u^2)\\
1/\sigma_u^2
\end{array}
\right]^T\etaSUBpbetausigsquTOsigsqu\right\}
$$
where
$$\etaSUBpbetausigsquTOsigsqu\thickarrow
\left[
\begin{array}{c}
-K/2 \\[1ex]
\GVMP\Big(\etaSUBpbetausigsquCONNbetau;\bD,\bzero,0\Big)
\end{array}
\right]
$$
where $\bD\equiv\diag(\bzero_2,\bone_K)$ and the function $\GVMP$ is
defined by (\ref{eq:GVMPdefn}).
The message passed from $p(\bbeta,\bu|\sigma_u^2)$ to 
$(\bbeta,\bu)$ will be shown (Section \ref{sec:GaussPenFrag}) to equal 
\def\bbetabu{
\left[   
\begin{array}{c}
\bbeta\\
\bu
\end{array}
\right]}
$$\mSUBpbetausigsquTObetau=
\exp\left\{
\left[
\begin{array}{c}    
\bbeta\\
\bu\\
\vecof\left(
\bbetabu\bbetabu^T\right)
\end{array}
\right]^T\etaSUBpbetausigsquTObetau\right\}
$$
with natural parameter update
$$\etaSUBpbetausigsquTObetau\thickarrow
\left[
\begin{array}{c}
\bSigma_{\bbeta}^{-1}\bmu_{\bbeta} \\[1ex]
\bzero_K\\[1ex]
-\smhalf\vecof\left(\mbox{blockdiag}\left(\bSigma_{\bbeta}^{-1},
\left\{\displaystyle{\frac{\Big(\etaSUBpbetaybetasigsqusigsqCONNsigsqu\Big)_1+1}{
\Big(\etaSUBpbetaybetasigsqusigsqCONNsigsqu\Big)_2}}\right\}\bI_K\right)\right)
\end{array}
\right].
$$

In Section \ref{sec:GaussResp} we catalog fragment types and identify five 
that are fundamental to semiparametric regression analysis via MFVB/VMP.
The form of the factor to stochastic node updates for these fragments
only needs to be derived and implemented once if developing a suite
of programs for VMP-based semiparametric regression. Such cataloging
allows for arbitrarily large models to be handled without an onerous 
algebraic and computational overhead.

\subsubsection{Conjugate Factor Graphs}\label{sec:conj}

We will say that a factor graph corresponding to a variational message 
passing scheme is \emph{conjugate} if, for each stochastic node, 
the messages passed to the node are in the same exponential family.
The two factor graphs of this section, shown in 
Figures \ref{fig:linRegFacGraphMsgs} and \ref{fig:penSplFacGraph},
are conjugate factor graphs. For example it is apparent from
(\ref{eq:linModMsgForms}) that, in Figure \ref{fig:linRegFacGraphMsgs},
the two messages passed to $\sigma^2$ are both proportional 
to Inverse Chi-Squared density functions. However, some of
the exponential forms do not correspond to proper density
functions. In Figure \ref{fig:linRegFacGraphMsgs}, the 
convergent form of $\mSUBpsigsqaTOa$ is
$$\mSUBpsigsqaTOa=\exp\left\{
\left[
\begin{array}{c}   
\log(a)\\
1/a
\end{array}
\right]^T\left[
\begin{array}{c}   
-\smhalf\\
-\biggerlambda_{p(\sigma^2|\,a)\to a}
\end{array}
\right]\right\}\quad\mbox{for some}\quad 
\biggerlambda_{p(\sigma^2|\,a)\to a}>0
$$
which is not proportional to a proper density function.

The concept of a conjugate factor graph can be extended
to sub-graphs of the factor graph at hand, in that
conjugacy holds in some parts of a factor graph but
not necessarily in other parts. 

\section{Gaussian Response Semiparametric Regression}\label{sec:GaussResp}

Since many popular Gaussian response semiparametric regression models 
admit conjugate factor graphs, we first focus on
their fitting via VMP. Generalized response models are more challenging 
and their treatment is postponed until Section \ref{sec:generalized}.
We start by identifying five fundamental fragments.

\subsection{Five Fundamental Fragments}\label{sec:fiveFundFrags}

Table \ref{tab:fiveFrags} shows five factor graph
fragments that are fundamental to VMP-based semiparametric
regression. We use generic notation, such as $\btheta$
for a random vector and $\bA$ for a design matrix, rather
than notation that matches specific semiparametric regression 
models. This is in keeping with update formulae within fragments 
being the building blocks for the handling of arbitrarily large 
models.

\ifthenelse{\boolean{ShowFigures}}
{
\begin{table}[!ht]
\begin{center}
\begin{footnotesize}
\begin{tabular}{lcl}
\hline\\[-1.3ex]
Fragment name         & Diagram   & Distributional statement    \\[0.1ex]
\hline\\[-0.9ex]
1. Gaussian prior             &   \includegraphics[width=0.2\textwidth, trim=0 8mm 0 0]{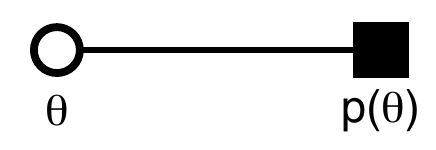}  &  
$\btheta\sim N(\bmu_{\btheta},\bSigma_{\btheta})$\\[3ex]
2. Inverse Wishart          & \includegraphics[width=0.2\textwidth, trim=0 8mm 0 0]{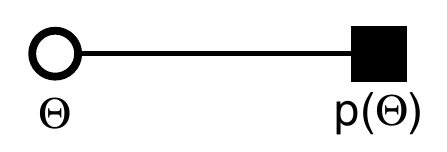}          &  
$\bTheta\sim\mbox{Inverse-Wishart}(\kappa_{\bTheta},\bLambda_{\bTheta})$  \\
\ \ \ \ prior                     &            &    \\[3ex]
3. Iterated Inverse        &    \includegraphics[width=0.2\textwidth, trim=0 8mm 0 0]{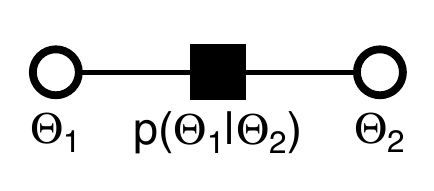}       & 
$\bTheta_1|\bTheta_2\sim\mbox{Inverse-G-Wishart}(G,\kappa,\bTheta_2^{-1})$  \\
\ \ \ \ G-Wishart               &        &                                                \\[3ex] 
4. Gaussian penalization  & \includegraphics[width=0.3\textwidth, trim=0 24mm 0 0]{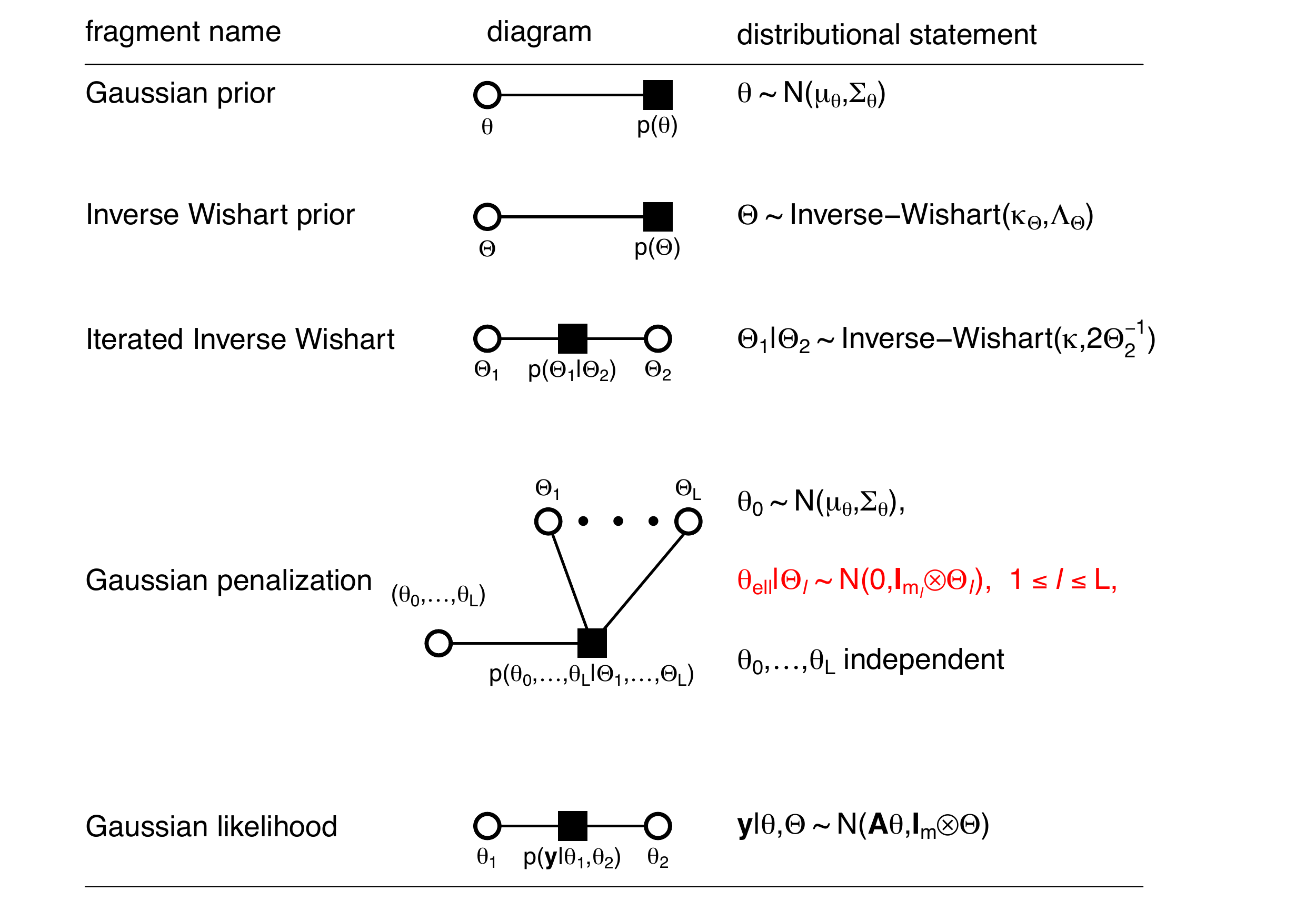}         &  
$\left[
\begin{array}{c}
\btheta_0\\
\vdots\\
\btheta_L
\end{array}
\right]
\Bigg|
\bTheta_1,\ldots,\bTheta_L\,\sim$\\
& & $
N\left(
\left[
\begin{array}{c}
\bmu_{\btheta_0}\\
\bzero
\end{array}
\right],
\left[
\begin{array}{cc}
\bSigma_{\btheta_0} & \bO^T \\
\bO         &\displaystyle{\blockdiag{1\le\ell\le L}}
              (\bI_{\mR_{\ell}}\otimes\bTheta_{\ell})
\end{array}
\right]
\right)$                       \\
5. Gaussian likelihood       &  \includegraphics[width=0.2\textwidth, trim=0 8mm 0 0]{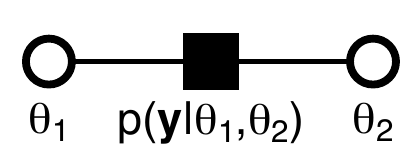}     &    
$\by|\,\btheta_1,\theta_2\sim N(\bA\btheta_1,\theta_2\,\bI)$   \\[5ex]
\hline
\end{tabular}
\end{footnotesize}
\end{center}
\caption{\it Five fundamental factor graph fragments 
for Gaussian response semiparametric regression.}
\label{tab:fiveFrags} 
\end{table}
}
{\vskip3mm
\thickboxit{\bf \centerline{Figure here.}}
\vskip3mm
}

\subsubsection{Gaussian Prior Fragment}\label{sec:GaussPriorFrag}

The \emph{Gaussian prior fragment} corresponds to the following prior 
specification of the $\dtheta\times 1$ random vector $\btheta$:
$$\btheta\sim N(\bmu_{\btheta},\bSigma_{\btheta}).$$
The $\dtheta\times1$ vector $\bmu_{\btheta}$ and
$\dtheta\times\dtheta$ covariance matrix 
$\bSigma_{\btheta}$ are user-specified hyperparameters. The fragment is shown 
in Table \ref{tab:fiveFrags} and has factor
$$p(\btheta)=(2\pi)^{-\dtheta/2}|\bSigma_{\btheta}|^{-1/2}
\exp\big\{-\smhalf(\btheta-\bmu_{\btheta})^T\bSigma_{\btheta}^{-1}
(\btheta-\bmu_{\btheta})\big\}
$$
and the single stochastic node $\btheta$.
Using the natural form of the Multivariate Normal
distribution described in Section \ref{sec:MultNormDistn}, 
the factor to stochastic node message is proportional to 
$p(\btheta)$ and has natural parameter form:
$$\mSUBpthetaTOtheta=
\exp\left\{
\left[\begin{array}{c}\btheta\\[1ex]
\vecof(\btheta\btheta^T)\end{array}
\right]^T\etaSUBpthetaTOtheta\right\}.
$$
The natural parameter vector is a fixed vector depending only 
on the hyperparameters:
$$\etaSUBpthetaTOtheta\thickarrow\left[
\begin{array}{c}
\bSigma_{\btheta}^{-1}\bmu_{\btheta}\\[1ex]   
-\smhalf\vecof(\bSigma_{\btheta}^{-1})
\end{array}
\right].
$$

\subsubsection{The Inverse Wishart Prior Fragment}\label{sec:InvWishPriorFrag}

We define the \emph{Inverse Wishart prior} fragment to correspond to the
$\dTheta\times\dTheta$ random matrix $\bTheta$ satisfying
$$\bTheta\sim\mbox{Inverse-Wishart}(\kappaTheta,\LambdaTheta)$$
where $\kappaTheta>0$ and $\LambdaTheta$ is a $\dTheta\times\dTheta$ 
symmetric positive definite matrix. This fragment, also shown in Table 
\ref{tab:fiveFrags}, has factor
$$
p(\bTheta)=\Csc_{\dTheta,\kappaTheta}^{-1}|\LambdaTheta|^{\kappaTheta/2}\,
|\bTheta|^{-(\kappaTheta+\dTheta+1)/2}
\exp\{-\smhalf\tr(\LambdaTheta\bTheta^{-1} )\},
$$
where 
\begin{equation}
\Csc_{d,\kappa}\equiv2^{d\kappa/2}\pi^{d(d-1)/4}
\prod_{j=1}^d\Gamma\left(\frac{\kappa+1-j}{2}\right),
\label{eq:CscDefn}
\end{equation}
and the single stochastic node $\bTheta$, a symmetric and positive definite
$\dTheta\times\dTheta$ matrix. 
From the natural form of the Inverse Wishart distribution 
given in Section \ref{sec:invWishDist}, the factor to stochastic 
node message is proportional to $p(\bTheta)$ and has natural 
parameter form:
$$\mSUBpThetaTOTheta
=\exp\left\{
\left[\begin{array}{c}\log|\bTheta|\\[1ex]
\vecof(\bTheta^{-1})\end{array}
\right]^T\etaSUBpThetaTOTheta\right\}.
$$
The natural parameter vector is a fixed vector that depends 
only on the hyperparameters:
$$\etaSUBpThetaTOTheta\thickarrow\left[
\begin{array}{c}
-\smhalf(\kappaTheta+\dTheta+1)\\[1ex]   
-\smhalf\vecof(\LambdaTheta)
\end{array}
\right].
$$

\subsubsection{Iterated Inverse G-Wishart Fragment}\label{sec:IterInvGWish}

The \emph{iterated Inverse G-Wishart fragment} is shown in 
Table \ref{tab:fiveFrags} and corresponds to the conditional
distributional specification
$$\bTheta_1|\bTheta_2\sim\mbox{Inverse-G-Wishart}(G,\kappa,\bTheta_2^{-1}),$$
where $\bTheta_1$ and $\bTheta_2$  are $\dTheta\times\dTheta$ 
random matrices, $\kappa>\dTheta-1$ is deterministic and $G$ is
a $\dTheta$-node undirected graph. See Section \ref{sec:GinvWishDist} 
in the online supplement for the definition 
of the Inverse G-Wishart distribution.

The rationale for this fragment for Bayesian semiparametric regression
stems from the family of marginally noninformative covariance matrix 
priors given in Huang \myand Wand (2013). In particular, for 
a $d\times d$ covariance matrix $\bSigma$, their equation (2)
is equivalent to 
\begin{equation}
{\setlength\arraycolsep{3pt}
\begin{array}{rcl}
\bSigma|\bA &\sim& \mbox{Inverse-Wishart}(\nu+d-1,\bA^{-1}),\\[1ex]
\bA         &\sim& \mbox{Inverse-G-Wishart}\Big(\Gdiag,1,
                   \frac{1}{\nu}\displaystyle{\diagarg{1\le k\le K}}(1/A_k^2)\Big)
\end{array}
}
\label{eq:HuangWand}
\end{equation}
where $\nu,A_1,\ldots,A_K>0$ are hyperparameters and $\Gdiag$ is
defined in Section \ref{sec:GinvWishDist} of the online supplement.
Setting $d=\nu=1$ leads to the variance 
parameter result (\ref{eq:HalfCauRes}). 
For $d>1$ setting $\nu=2$ has the attraction of
imposing $\mbox{Uniform}(-1,1)$ priors on the correlation 
parameters in $\bSigma$ (Huang \myand Wand, 2013).

The fragment factor is of the form 
$$p(\bTheta_1|\,\bTheta_2)\propto|\bTheta_2|^{-\kappa/2}
|\bTheta_1|^{-(\kappa+\dTheta+1)/2}
\exp\{-\smhalf\tr(\bTheta_1^{-1}\bTheta_2^{-1})\}.
$$
From (\ref{eq:facToStochOne}) and (\ref{eq:facToStochTwo}), 
the message that $p(\bTheta_1|\bTheta_2)$ passes to $\bTheta_1$ is
$$\mSUBpThetaOneThetaTwoTOThetaOne=\exp\left\{
\left[\begin{array}{c}\log|\bTheta_1|\\[1ex]
\vecof(\bTheta_1^{-1})\end{array}
\right]^T\etaSUBpThetaOneThetaTwoTOThetaOne\right\}
$$
where 
\begin{equation}
\etaSUBpThetaOneThetaTwoTOThetaOne\thickarrow
\left[
\begin{array}{c}
-(\kappa+\dTheta+1)/2\\[1ex]
-\smhalf\vecof\Big(
E_{\mbox{\footnotesize$p(\bTheta_1|\bTheta_2)\to\bTheta_1$}}(\bTheta_2^{-1})\Big)
\end{array}
\right]
\label{eq:goatHeadSoup}
\end{equation}
and $E_{\mbox{\footnotesize$p(\bTheta_1|\bTheta_2)\to\bTheta_1$}}$
denotes expectation with respect to the density function formed by
normalizing the message product
$\mSUBpThetaOneThetaTwoTOThetaTwo\,\mSUBThetaTwoTOpThetaOneThetaTwo$.
Under the conjugacy assumption that messages passed to $\bTheta_2$
from its other neighboring factors are also within the Inverse-G-Wishart
family the expectation in (\ref{eq:goatHeadSoup}) is a special case of
\begin{equation}
E(\bX^{-1})\quad\mbox{where}\quad\bX\sim\mbox{Inverse-G-Wishart}(G,\kappa,\bLambda)
\label{eq:genericMeanInvGWish}
\end{equation}
or, equivalently, the mean of a G-Wishart random matrix. Similarly
$$\mSUBpThetaOneThetaTwoTOThetaTwo=\exp\left\{
\left[\begin{array}{c}\log|\bTheta_2|\\[1ex]
\vecof(\bTheta_2^{-1})\end{array}
\right]^T\etaSUBpThetaOneThetaTwoTOThetaTwo\right\}.
$$
where
$$
\etaSUBpThetaOneThetaTwoTOThetaTwo\thickarrow
\left[
\begin{array}{c}
-\kappa/2\\[1ex]
-\smhalf\vecof\Big(
E_{\mbox{\footnotesize$p(\bTheta_1|\bTheta_2)\to\bTheta_2$}}
(\bTheta_1^{-1})\Big)
\end{array}
\right]
$$
and, assuming that all other messages passed to $\bTheta_2$ are within the
Inverse G-Wishart family, the natural parameter update
is also a special case of (\ref{eq:genericMeanInvGWish}).

For general undirected graphs (\ref{eq:genericMeanInvGWish})
can be very complicated (Uhler \textit{et al.}, 2014).
However, for important special cases the required
expectation admits a simple closed form expression.
These cases are discussed next.

\subsubsubsection{The Case of $\dTheta=1$}

If $\dTheta=1$ then $\bTheta_1$ and $\bTheta_2$ reduce 
to variance parameters and results concerning Inverse Chi-Squared 
random variables apply. The updates become
\begin{equation}
\etaSUBpThetaOneThetaTwoTOThetaOne\thickarrow
\left[
{\setlength\arraycolsep{0.1pt}
\begin{array}{c}
-\smhalf(\kappa+2)  \\[2ex]
-\smhalf\Big(\big(\etaSUBpThetaOneThetaTwoCONNThetaTwo\big)_1+1\Big)\Big/
(\etaSUBpThetaOneThetaTwoCONNThetaTwo\big)_2
\end{array}
}
\right]
\label{eq:Th1Th2TOTh1Simp}
\end{equation}
and
\begin{equation}
\etaSUBpThetaOneThetaTwoTOThetaTwo\thickarrow
\left[
{\setlength\arraycolsep{0.1pt}
\begin{array}{c}
-\smhalf\kappa  \\[2ex]
-\smhalf\Big(\big(\etaSUBpThetaOneThetaTwoCONNThetaOne\big)_1+1\Big)\Big/
\big(\etaSUBpThetaOneThetaTwoCONNThetaOne\big)_2
\end{array}
}
\right].
\label{eq:Th1Th2TOTh2Simp}
\end{equation}
Note use of the notation first used at (\ref{eq:doubleArrowFirst}).
%

\subsubsubsection{The Case of $\dTheta>1$ and $G$ Totally Connected or Totally Disconnected}

If $G$ is a totally connected $d$-node graph, meaning that there is an edge
between each pair of nodes, then the Inverse G-Wishart distribution
coincides with the ordinary Inverse Wishart distribution and the well-known result 
\begin{equation}
\bX\sim\mbox{Inverse-Wishart}(\kappa,\bLambda)\quad\mbox{implies}\quad
E(\bX^{-1})=\kappa\bLambda^{-1}
\label{eq:ErecipXInvWish}
\end{equation}
applies. Suppose instead that $G$ is totally disconnected, meaning that it has no edges. 
Then $G=\Gdiag$ in the notation of (\ref{sec:GinvWishDist}) and $\bLambda$ is
a diagonal matrix. It is easily established that (\ref{eq:ErecipXInvWish})
also applies in the totally disconnected case. Switching to
natural parameters via (\ref{eq:natParmInvWish}) we obtain the update expressions
\begin{equation}
\etaSUBpThetaOneThetaTwoTOThetaOne\thickarrow
\left[
{\setlength\arraycolsep{0.1pt}
\begin{array}{c}
-\smhalf(\kappa+\dTheta+1)  \\[2ex]
-\smhalf\Big\{\big(\etaSUBpThetaOneThetaTwoCONNThetaTwo\big)_1
+\frac{\dTheta+1}{2}\Big\}\\[1ex]
\times
\vecof\Big[\Big\{\vecof^{-1}
\Big(\big(\etaSUBpThetaOneThetaTwoCONNThetaTwo\big)_2\Big)\Big\}^{-1}\Big]
\end{array}
}
\right]
\label{eq:Th1Th2TOTh1}
\end{equation}
and 
\begin{equation}
\etaSUBpThetaOneThetaTwoTOThetaTwo\thickarrow
\left[
{\setlength\arraycolsep{0.1pt}
\begin{array}{c}
-\smhalf\kappa  \\[2ex]
-\smhalf\Big\{\big(\etaSUBpThetaOneThetaTwoCONNThetaOne\big)_1
+\frac{\dTheta+1}{2}\Big\}\\[1ex]
\times
\vecof\Big[\Big\{\vecof^{-1}
\Big(\big(\etaSUBpThetaOneThetaTwoCONNThetaOne\big)_2\Big)\Big\}^{-1}\Big]
\end{array}
}
\right]
\label{eq:Th1Th2TOTh2}
\end{equation}
where
$$
\left[
\begin{array}{c}
\big(\etaSUBpThetaOneThetaTwoCONNThetaOne\big)_1\\[2ex]
\big(\etaSUBpThetaOneThetaTwoCONNThetaOne\big)_2
\end{array}
\right]
$$
is the partition of $\etaSUBpThetaOneThetaTwoCONNThetaOne$
for which $\big(\etaSUBpThetaOneThetaTwoCONNThetaOne\big)_1$
is the first entry of the vector and 
$\big(\etaSUBpThetaOneThetaTwoCONNThetaOne\big)_2$
contains the remaining entries. Similar partitional notation
applies to $\etaSUBpThetaOneThetaTwoCONNThetaTwo$.

\subsubsubsection{The Case of $\dTheta>1$ and $G$ Partially Connected}

This case suffers from the fact that (\ref{eq:genericMeanInvGWish}) does
not have a simple expression for general partially connected $G$. 
However, the Inverse G-Wishart forms that commonly arise in 
Bayesian semiparametric regression analysis are covered by the 
previous cases. Hence, this case can be left aside for common models.

\subsubsection{Gaussian Penalization Fragment}\label{sec:GaussPenFrag}

The fourth fragment in Table \ref{tab:fiveFrags} is 
the \emph{Gaussian penalization fragment} since it
imposes Gaussian distributional penalties on random effects
parameters. The corresponding conditional distributional
specification is
$$
\left[
\begin{array}{c}
\btheta_0\\
\vdots\\
\btheta_L
\end{array}
\right]
\Bigg|
\bTheta_1,\ldots,\bTheta_L
\sim
N\left(
\left[
\begin{array}{c}
\bmu_{\btheta_0}\\
\bzero
\end{array}
\right],
\left[
\begin{array}{cc}
\bSigma_{\btheta_0} & \bO^T \\
\bO         &\displaystyle{\blockdiag{1\le\ell\le L}}
              (\bI_{\mR_{\ell}}\otimes\bTheta_{\ell})
\end{array}
\right]
\right)
$$
where $\bO$ is an appropriately-sized matrix of zeroes.
The $\dtheta_0\times1$ vector $\btheta_0$ is a fixed effects
parameter and has a $\dtheta_0\times1$ deterministic mean $\bmu_{\btheta_0}$ and
$\dtheta_0\times\dtheta_0$ deterministic covariance matrix 
$\bSigma_{\btheta_0}$. The covariance matrices $\bTheta_{\ell}$
are stochastic and have dimension $\dTheta_{\ell}\times\dTheta_{\ell}$, $1\le\ell\le L$.
The random effects vectors $\btheta_{\ell}$ are also stochastic and
have dimension $(\mR_{\ell}\dTheta_{\ell})\times 1$, $1\le\ell\le L$.

The fragment factor is 
\begin{eqnarray*}
&&p(\btheta_0,\ldots,\btheta_L|\bTheta_1,\ldots,\bTheta_L)
=(2\pi)^{-\dtheta_0/2}|\bSigma_{\btheta_0}|^{-1/2}
\exp\left\{-\smhalf(\btheta_0-\bmu_{\btheta_0})^T
\bSigma_{\btheta_0}^{-1}(\btheta_0-\bmu_{\btheta_0})\right\}\\[1ex]
&&\qquad\qquad\times\prod_{\ell=1}^L
(2\pi)^{-\mR_{\ell}\dTheta_{\ell}/2}|\bI_{\mR_{\ell}}\otimes\bTheta_{\ell}|^{-1/2}
\exp\{-\smhalf\btheta_{\ell}^T
(\bI_{\mR_{\ell}}\otimes\bTheta_{\ell}^{-1})\btheta_{\ell}\}.
\end{eqnarray*}
The structure of the fragment is depicted in its diagram in 
Table \ref{tab:fiveFrags}. We assume that each of 
$(\btheta_0,\ldots,\btheta_L)$ and $\bTheta_1,\ldots,\bTheta_L$ 
receive messages from outside the fragment that are conjugate
with the message it receives from 
$p(\btheta_0,\ldots,\btheta_L|\bTheta_1,\ldots,\bTheta_L)$.
Update (\ref{eq:stochToFac}) implies that the message from 
$(\btheta_0,\ldots,\btheta_L)$ to the fragment factor
is proportional to a Multivariate Normal density function 
with natural parameter vector
$\etaSUBthetaVecTOpthetaVecThetaVec$
and the message from each $\bTheta_{\ell}$ is proportional
to an Inverse-G-Wishart density function with natural parameter
vector $\etaSUBThetaEllTOpthetaVecThetaVec$. It follows
that the inputs for the Gaussian penalization fragment are
$$\etaSUBthetaVecTOpthetaVecThetaVec,\quad\etaSUBpthetaVecThetaVecTOthetaVec$$
and 
$$\etaSUBThetaEllTOpthetaVecThetaVec\quad
\etaSUBpthetaVecThetaVecTOThetaEll,\quad 1\le\ell\le L.$$
Using (\ref{eq:facToStochOne}), (\ref{eq:facToStochTwo})
and Table \ref{tab:ETx} in the online supplement,
the message from this factor to the coefficient vector 
$(\btheta_0,\cdots,\btheta_L)$ has natural parameter update
$$
\etaSUBpthetaVecThetaVecTOthetaVec\thickarrow
\left[
\begin{array}{c}
\bSigma_{\btheta_0}^{-1}\bmu_{\btheta_0}\\[2ex]
\bzero\\[2ex]
-\smhalf\vecof\Big(\mbox{blockdiag}\big(
\bSigma_{\btheta_0}^{-1},
\displaystyle{\blockdiag{1\le\ell\le L}}(\bI_{\mR_{\ell}}\otimes\bOmega_{\ell})\big)
\Big)
\end{array}
\right]
$$
where $\bzero$ is the $\sum_{\ell=1}^L m_{\ell}\,\dtheta_{\ell}\times1$ 
vector of zeroes and
$$\bOmega_{\ell}\equiv\Big(\big(\etaSUBpthetaVecThetaVecCONNThetaEll\big)_1
+\textstyle{\frac{\dTheta_{\ell}+1}{2}}\Big)
\left\{\vecof^{-1}\Big(\big(\etaSUBpthetaVecThetaVecCONNThetaEll\big)_2\Big)\right\}^{-1}.$$
Similarly, the message from 
$p(\btheta_0,\ldots,\btheta_L|\bTheta_1,\ldots,\bTheta_L)$ to 
each $\bTheta_{\ell}$, $1\le\ell\le L$, has natural parameter update
$$\etaSUBpthetaVecThetaVecTOThetaEll\thickarrow
\left[
\begin{array}{c}
-\mR_{\ell}/2\\[2ex]
\GVMP\left(\etaSUBpthetaVecThetaVecCONNthetaVec;\bD_{\ell},\bzero,0\right)
\end{array}
\right]
$$
where $\GVMP$ is defined by (\ref{eq:GVMPdefn}) and
$$\bD_{\ell}\equiv\mbox{blockdiag}\Big(\bO_{\dTheta_0},\bO_{m_1\dTheta_1},\ldots,
\blockdiag{1\le k\le m_{\ell}}\big(\bJ_{\dTheta_{\ell}}\big),\ldots,\bO_{m_L\dTheta_L}\Big)$$
with $\bJ_d$ denoting the $d\times d$ matrix with each entry equal to $1$ and 
$\bO_d$ denoting the $d\times d$ matrix with each entry equal to $0$.

Whilst the formulae given in this section cover a wide range of penalization
scenarios arising in semiparametric regression we have, with
succinctness in mind, left out multilevel models with the number of levels 
exceeding two. The extension to arbitrarily high levels would take
significantly more algebra and obscure the main message regarding
the fragment approach. In the same vein, we are not using matrix algebraic 
streamlining as described in Lee \myand Wand (2016a, 2016b) for MFVB.
Matrix algebraic streamlining is concerned with matters such as avoiding large
indicator matrices and redundant calculations. Its extension to
VMP would also require significantly more algebra and is left for 
future research.

\subsubsection{Gaussian Likelihood Fragment}\label{sec:gaussLikFrag}

The \emph{Gaussian likelihood fragment} corresponds to the form
$$\by|\,\btheta_1,\theta_2\sim N(\bA\btheta_1,\theta_2\,\bI)$$
where $\by$ is an $n\times1$ vector of observed data values,
and $\bA$ is an $n\times\dtheta$ design matrix. The stochastic nodes are the
$\dtheta\times1$ coefficient vector $\btheta_1$ and the variance
parameter $\theta_2>0$. The factor is
$$p(\by|\,\btheta_1,\theta_2)=(2\pi\theta_2)^{-n/2}
\exp\{-(2\theta_2)^{-1}\Vert\by-\bA\btheta_1\Vert^2\}.
$$

For this fragment, shown in Table \ref{tab:fiveFrags}, we assume that
each of the stochastic nodes, $\btheta_1$ and $\theta_2$, 
receive messages from factors outside of the fragment that 
are conjugate with the message it receives from 
$p(\by|\btheta_1,\theta_2)$. Because of (\ref{eq:stochToFac}) 
this implies that the message from $\btheta_1$ to $p(\by|\btheta_1,\theta_2)$
is proportional to a Multivariate Normal density
function with natural parameter $\etaSUBthetaOneTOpythetaOnethetaTwo$
and that from $\theta_2$ to $p(\by|\btheta_1,\theta_2)$ is
proportional to an Inverse Chi-Squared density function 
with natural parameter $\etaSUBthetaTwoTOpythetaOnethetaTwo$.
It follows that the inputs for the Gaussian likelihood fragment are
$$\etaSUBthetaOneTOpythetaOnethetaTwo,\quad\etaSUBpythetaOnethetaTwoTOthetaOne,\quad
\etaSUBthetaTwoTOpythetaOnethetaTwo\quad\mbox{and}\quad\etaSUBpythetaOnethetaTwoTOthetaTwo.$$
%
%
The outputs are the following updated natural parameters
of the messages passed from $p(\by|\,\btheta_1,\theta_2)$ 
to $\btheta_1$ and $\theta_2$:
\begin{equation}
\etaSUBpythetaOnethetaTwoTOthetaOne\thickarrow
\left[
\begin{array}{c}
\bA^T\by \\
-\smhalf\vecof(\bA^T\bA)
\end{array}
\right]
\frac{\big(\etaSUBpythetaOnethetaTwoCONNthetaTwo\big)_1+1}
{\big(\etaSUBpythetaOnethetaTwoCONNthetaTwo\big)_2}
\label{eq:likTOthetaSimp}
\end{equation}
and
%
%
\begin{equation}
\etaSUBpythetaOnethetaTwoTOthetaTwo\thickarrow
\left[
\begin{array}{c}
-n/2 \\[1ex]
\GVMP\left(\etaSUBpythetaOnethetaTwoCONNthetaOne;\bA^T\bA,\bA^T\by,\by^T\by\right)
\end{array}
\right]
\label{eq:likTOTheta}
\end{equation}
where the notational convention of (\ref{eq:pLikUpdates}) is 
followed and $\GVMP$ is defined by (\ref{eq:GVMPdefn}).

\subsection{Models Accommodated by the Five Fundamental Fragments}

The five fragments covered in Section \ref{sec:fiveFundFrags} 
are fundamental to VMP-based Bayesian semiparametric regression 
and accommodate a wide range of models. Table \ref{tab:laundryList} 
lists the types of models that can be handled via the Bayesian mixed model-based
penalized splines approach to semiparametric regression laid out in
Section \ref{sec:BayeSemiPar}.

\begin{table}
\begin{center}
\begin{tabular}{ll}
\hline\\[-0.9ex]
Linear regression        & Factor-by-curve interactions          \\        
Linear mixed             & Varying coefficients  \\ 
Nonparametric regression & Multivariate nonparametric regression    \\
Additive                 & Geoadditive   \\
Additive mixed           &  Group-specific curves                 \\[1ex]
\hline
\end{tabular}
\end{center}
\caption{\it Types of Gaussian response semiparametric regression models that
are accommodated by VMP with factor to stochastic node updates as given
in Section \ref{sec:fiveFundFrags} for the five fundamental fragments.}
\label{tab:laundryList} 
\end{table}

The models in the left column of Table \ref{tab:laundryList} are part
of mainstream semiparametric regression analysis for 
cross-sectional and grouped data as summarized in, 
for example, Wood (2006) and Hodges (2013). 
Chapters 2--9 of Ruppert \textit{et al.} (2003) 
summarize the specific approach taken in the current article. 
Factor-by-curve interactions are detailed in Coull \textit{et al.} (2001)
whilst Kammann \myand Wand (2003) describe multivariate nonparametric
regression and geoadditive models that are in accordance with 
the VMP fragment set-up of Section \ref{sec:fiveFundFrags}.
Similarly, the group-specific curves model of 
Durban \textit{et al.} (2005) is accommodated by 
the Section \ref{sec:fiveFundFrags} fragments and 
is illustrated in Section \ref{sec:growInd}.
Group-specific curve models have a number of alternative formulations
(e.g. Donnelly \textit{et al.}, 1995; Verbyla \textit{et al.}, 1999).

Of the five fragments, only the last is specific to Gaussian
response semiparametric regression. The other four are
applicable to non-Gaussian response models and, when combined
with the fragments of Section \ref{sec:generalized}, 
facilitate handling of a wider range of models such as 
generalized additive models and generalized linear mixed models.

\subsection{Coding Issues}

According to the VMP approach with fragment identification, the
updates of the natural parameters for factor to stochastic 
node messages only need to be coded once and can be then
compartmentalized into functions. Once this is achieved
for all fragments present in a particular class of models then
the factor to stochastic node messages for a
specific model within that class can be handled with 
calls to these functions. The stochastic node to 
factor messages are trivial and require only a few lines
of code.

A more ambitious software project 
is one that allows the user to specify a  model using 
either syntactic coding rules or a directed acyclic graph
drawing interface, such as those used by the 
\textsf{BUGS} (Lunn \textit{et al.}, 2012), 
\textsf{Infer.NET} (Minka \textit{et al.}, 2014),
\textsf{VIBES} (Bishop \textit{et al.}, 2003) and \textsf{Stan} 
(Stan Development Team, 2015) Bayesian inference engines, 
and then internally construct an appropriate factor graph and perform 
VMP message updates.

As already discussed, \textsf{Infer.NET} is the main software platform
providing support for VMP-based inference for general classes of 
Bayesian models and its interior architecture makes use of fragment-type
rules such as those treated in (\ref{sec:fiveFundFrags}) to handle
arbitrarily large models that are accommodated by these rules.
In Wang \myand Wand (2011) and Luts {\it et al.} (2015) 
we show that versions of \textsf{Infer.NET} can handle various 
semiparametric regression models provided that particular `tricks' are used. 
For example, the conjugacy rules of \textsf{Infer.NET 2.5, Beta 2} do not
allow for the standard auxiliary variable representation of the 
Laplace distribution (e.g. equation (4) of Park \myand Casella, 2008) 
and an alternative approximate representation is used in Section 8 of 
Luts {\it et al.} (2015). More complicated semiparametric regression
scenarios such as interactions handled using tensor product splines 
(e.g.\ Wood \textit{et al.} 2013), nonparametric variance function estimation
(e.g.\ Menictas \myand Wand, 2015), streamlined variational
inference for longitudinal and multilevel models (e.g. Lee \myand Wand, 2016)
and missing data (e.g.\ Faes {\it et al.}, 2011) require 
self-implementation and the development of new fragments.

This article is not concerned primarily with coding issues
but rather the mathematics of VMP aimed at facilitating personal 
coding of VMP and development of updating formulae for more elaborate 
semiparametric regression and other statistical models.

\subsection{Illustration for Group-Specific Curves Semiparametric Regression}\label{sec:growInd}

The five fundamental fragments of Section \ref{sec:fiveFundFrags} can handle
quite complicated models as we now demonstrate for data from a longitudinal 
study on adolescent somatic growth, described in detail by Pratt \textit{et al.} (1989).
The main variables are
{\setlength\arraycolsep{3pt}
\begin{eqnarray*}
y_{ij}&=&\mbox{$j$th height measurement (centimetres) of subject $i$},\\[1ex]
\mbox{and}\ x_{ij}&=&\mbox{age (years) of subject $i$ when $y_{ij}$ is recorded,}
\end{eqnarray*}
}
for $1\le j\le n_i$ and $1\le i\le m$. We restrict attention to 
the males in the study, which results in $m=116$ subjects. The subjects
are categorized into black ethnicity (28 subjects) and 
white ethnicity (88 subjects) and comparison of 
mean height between the two populations is of interest. The group-specific
curve model takes the form
$$y_{ij}=\left\{\begin{array}{cc}
                 f_B(x_{ij})+g_i(x_{ij})+\varepsilon_{ij} &  \mbox{for black subjects}\\[1ex]
                 f_W(x_{ij})+g_i(x_{ij})+\varepsilon_{ij} &  \mbox{for white subjects}
               \end{array}\right.
$$
where $f_B$ is the mean height function for the black population,
$f_W$ is the mean height function for the white population, 
the functions $g_i$, $1\le i\le m$, represent the 
deviations from $i$th subject's mean function and $\varepsilon_{ij}$ 
is the within-subject random error. The penalized spline models are 
of the form
{\setlength\arraycolsep{3pt}
\begin{eqnarray*}
f_W(x)&=&\beta_0^{\mbox{\tiny W}}
+\beta_1^{\mbox{\tiny W}}\,x+\sum_{k=1}^{\Kgbl}\,\ugblk^W\zgblk(x),\\
f_B(x)&=&\beta_0^{\mbox{\tiny W}}+\beta_0^{\mbox{\tiny BvsW}}+
(\beta_1^{\mbox{\tiny W}}+\beta_1^{\mbox{\tiny BvsW}})\,x
+\sum_{k=1}^{\Kgbl}\,\ugblk^B\zgblk(x)\\
\mbox{and}\quad
g_i(x)&=&U_{0i}+U_{1i}\,x+\sum_{k=1}^{\Kgrp}\,\ugrpik\zgrpk(x),
\end{eqnarray*}
}
where $\{\zgblk:1\le k\le \Kgbl\}$ and $\{\zgrpk:1\le k\le \Kgrp\}$
are spline bases of size $\Kgbl$ and $\Kgrp$. The contrast function is
\begin{equation}
c(x)\equiv f_B(x) - f_W(x)
=\beta_0^{\mbox{\tiny BvsW}}+\beta_1^{\mbox{\tiny BvsW}}\,x
+\sum_{k=1}^{\Kgbl}(\ugblk^B-\ugblk^W)\zgblk(x).
\label{eq:contrastCurve}
\end{equation}
Following the mixed model formulation of Durban \textit{et al.} (2005) and
adopting a Bayesian approach leads to the model
\begin{equation}
\begin{array}{c}
\by|\,\bbeta,\bugblW,\bugblB,\bU,\bugrp,\sigeps\sim
N(\bX\bbeta+\bZgblW\bugblW+\bZgblB\bugblB+\bZU\bU+\bZgrp\bugrp,\sigeps^2\bI),\\[2ex]
\bugbl^W|\,\sigmagbl^W\sim N(\bzero,(\sigmagbl^W)^2),
\qquad
\bugbl^B|\,\sigmagbl^B\sim N(\bzero,(\sigmagbl^B)^2),
\\[2ex]
\bU|\,\bSigma\sim N(\bzero,\bI_m\otimes\bSigma),\quad
\bugrp|\,\sigmagrp\sim N(\bzero,\sigmagrp^2),
\quad
\bbeta\sim N(\bzero,\sigma_{\bbeta}^2\,\bI),
\\[2ex]
(\sigmagbl^W)^2|\agbl^W\sim\mbox{Inverse-$\chi^2$}(1,1/\agbl^W),\quad
(\sigmagbl^B)^2|\agbl^B\sim\mbox{Inverse-$\chi^2$}(1,1/\agbl^B),\\[2ex]
\sigmagrp^2|\agrp\sim\mbox{Inverse-$\chi^2$}(1,1/\agrp),\quad 
\bSigma|\bA\sim\mbox{Inverse-Wishart}\big(3,\bA^{-1}\big),\\[2ex]
\agbl^W\sim\mbox{Inverse-$\chi^2$}(1,1/\Agbl^2),\quad
\agbl^B\sim\mbox{Inverse-$\chi^2$}(1,1/\Agbl^2),\\[2ex]
\agrp\sim\mbox{Inverse-$\chi^2$}(1,1/\Agrp^2),
\quad\bA\sim\mbox{Inverse-G-Wishart}(\Gdiag,1,\smhalf\bAU^{-2}),\\[2ex]
\sigeps^2|\aeps\sim\mbox{Inverse-$\chi^2$}(1,1/\aeps),\quad
\aeps\sim\mbox{Inverse-$\chi^2$}(1,1/\Aeps^2)
\end{array}
\label{eq:somaticModel}
\end{equation}
for hyperparameters $\sigma_{\bbeta}^2$, $\Agbl$, $\Agrp$ $\Aeps$
all positive scalars, $\bAU$ a $2\times2$ positive definite 
diagonal matrix and $\Gdiag$ is a two-node graph without edges,
so that $\bA$ has off-diagonal entries equaling zero.
All distributional notation is given in Section \ref{sec:expFam}.
The coefficient vectors in (\ref{eq:somaticModel}) are
$$\bbeta\equiv
\left[
\begin{array}{c}
\beta_0^W\\
\beta_1^W\\
\beta_0^{\mbox{\tiny BvsW}}\\
\beta_1^{\mbox{\tiny BvsW}}\\
\end{array}
\right],
\quad
\bugbl^W\equiv
\left[
\begin{array}{c}
u_{\mbox{\tiny gbl},1}^W\\
\vdots\\
u_{\mbox{\tiny gbl},\Kgbl}^W
\end{array}
\right],
\quad
\bU\equiv
\left[
\begin{array}{c}
U_{01}\\
U_{11}\\
\vdots\\
U_{0m}\\
U_{1m}
\end{array}
\right]\quad\mbox{and}\quad
\bugrp\equiv
\left[
\begin{array}{c}
u_{\mbox{\tiny grp},11}\\
\vdots\\
u_{\mbox{\tiny grp},1\Kgrp}\\
\vdots\\
u_{\mbox{\tiny grp},m1}\\
\vdots\\
u_{\mbox{\tiny grp},m\Kgrp}\\
\end{array}
\right]
$$
with $\bugbl^B$ defined analogously to $\bugbl^W$.
The design matrices $\bX$, $\bZgbl^B$ and $\bZ_U$ are 
$$
\bX\equiv
\left[
\begin{array}{cccc}
\bone      & \bx_1 & \bindicB_1 & \bindicB_1\odot \bx_1\\
\vdots & \vdots& \vdots & \vdots\\ 
\bone      & \bx_m & \bindicB_m&  \bindicB_m\odot \bx_m\\
\end{array}
\right],
\quad
\bZgbl^B\equiv 
\left[
\begin{array}{ccc}
\bindicB_1\odot\,z_{\mbox{\tiny gbl},1}(\bx_1)&\cdots& 
\bindicB_1\odot\,z_{\mbox{\tiny gbl},\Kgbl}(\bx_1)\\
\vdots & \ddots & \vdots\\ 
\bindicB_m\odot\,z_{\mbox{\tiny gbl},1}(\bx_m)&\cdots& 
\bindicB_m\odot\,z_{\mbox{\tiny gbl},\Kgbl}(\bx_m)\\
\end{array}
\right]
$$
$$\mbox{and}\quad\bZU\equiv\blockdiag{1\le i\le m}[\bone\ \bx_i]$$
with $\bx_i$ equaling the $n_i\times 1$ vector containing
the $x_{ij}$, $1\le j\le n_i$, and $\bindicB_i$ equaling the
$n_i\times 1$ vector with each entry set to $\indicB_i$ with
$\indicB_{i}=1$ if the $i$th subject is black and 
$\indicB_{i}=0$ if the $i$th subject is white.
The matrix $\bZgbl^W$ is defined in a similar manner to 
$\bZgbl^B$, but with $\indicB_i$ replaced by $1-\indicB_i$.
The design matrix $\bZgrp$ has block diagonal structure similar
to $\bZU$ with blocks analogous to $\bZgbl^B$ and $\bZgbl^W$
but there is allowance for a different, typically smaller,
spline basks of size $\Kgrp$.
The prior on $\bSigma$, in terms of
the auxiliary variable $\bA$, has entries that
are marginally noninformative as explained in Huang \myand Wand (2013).

Figure \ref{fig:grpSpecCurvContrFacGraph} shows the factor graph
of (\ref{eq:somaticModel}) according to the $q$-density product 
restriction
\begin{eqnarray*}
&&q\big(\bbeta,\bu,\aeps,\agbl^B,\agbl^W,\agrp,\bA,
\sigeps^2,(\sigmagbl^W)^2,(\sigmagbl^B)^2,\bSigma,\sigmagrp^2\big)\\[1ex]
&&\qquad=
q\big(\bbeta,\bu\big)\,q\big(\aeps,\agbl^B,\agbl^W,\agrp,\bA,
\sigeps^2,(\sigmagbl^W)^2,(\sigmagbl^B)^2,\bSigma,\sigmagrp^2\big)\\[1ex]
&&\qquad=
q(\bbeta,\bu)q(\aeps)q\big(\agbl^B\big)q\big(\agbl^W\big)q(\agrp)
q(\bA)q(\sigeps^2)q\big((\sigmagbl^W)^2\big)
q\big((\sigmagbl^B)^2\big)q(\bSigma)q(\sigmagrp^2).
\end{eqnarray*}
with the second equality justified by 
induced factorization theory (e.g. Section 10.2.5 of Bishop, 2006).

\ifthenelse{\boolean{ShowFigures}}
{
\begin{figure}[!ht]
\centering
{\includegraphics[width=0.85\textwidth]{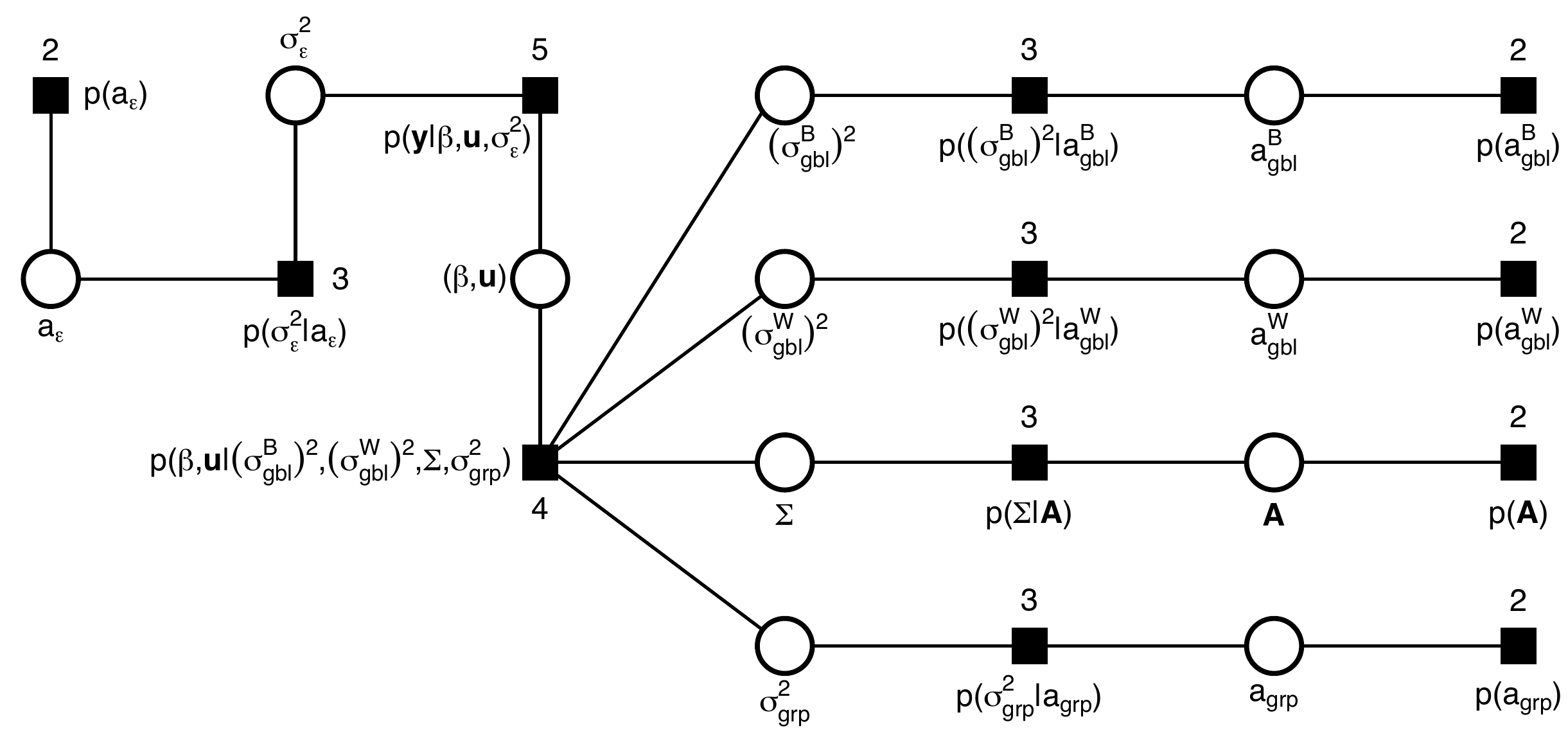}}
\caption{\it Factor graph corresponding to the group specific curve
model (\ref{eq:somaticModel}). The number adjacent to each factor 
signifies the fragment number in Table \ref{tab:fiveFrags}.}
\label{fig:grpSpecCurvContrFacGraph} 
\end{figure}
}
{\vskip3mm
\thickboxit{\bf \centerline{Figure here.}}
\vskip3mm
}

\noindent
Notwithstanding the complexity of Figure \ref{fig:grpSpecCurvContrFacGraph},
it is simply a conglomeration of four of the fundamental
fragments of Table \ref{tab:fiveFrags}, indicated by the number adjacent
to each factor. Therefore the factor to stochastic node messages 
for VMP-based inference are special cases of the messages given 
in Section \ref{sec:fiveFundFrags} and can be updated using the
formulae given there. The stochastic node to factor messages 
have trivial updates based on (\ref{eq:stochToFac}). Running 100
iterations of these updates leads to the fitted group-specific
curves for 35 randomly chosen subjects and 
contrast curve shown in Figure \ref{fig:maleGrowIndContVMP}.
MCMC-based fits, obtained using the \textsf{R} package \textsf{rstan} 
(Stan Development Team, 2016), are also shown for comparison. VMP is seen to 
be in very good agreement with MCMC. The right panel of 
Figure \ref{fig:maleGrowIndContVMP} shows the estimated height 
gap between black male adolescents and white male adolescents 
as a function of age. It is highest and (marginally) statistically 
significant up to about 14 years of age, peaking at 13 years of age. 
Between 17 and 20 years old there is no discernible height difference 
between the two populations.

\ifthenelse{\boolean{ShowFigures}}
{
\begin{figure}[!ht]
\centering
\begin{tabular}{cc}
\null
{\includegraphics[width=0.45\textwidth]{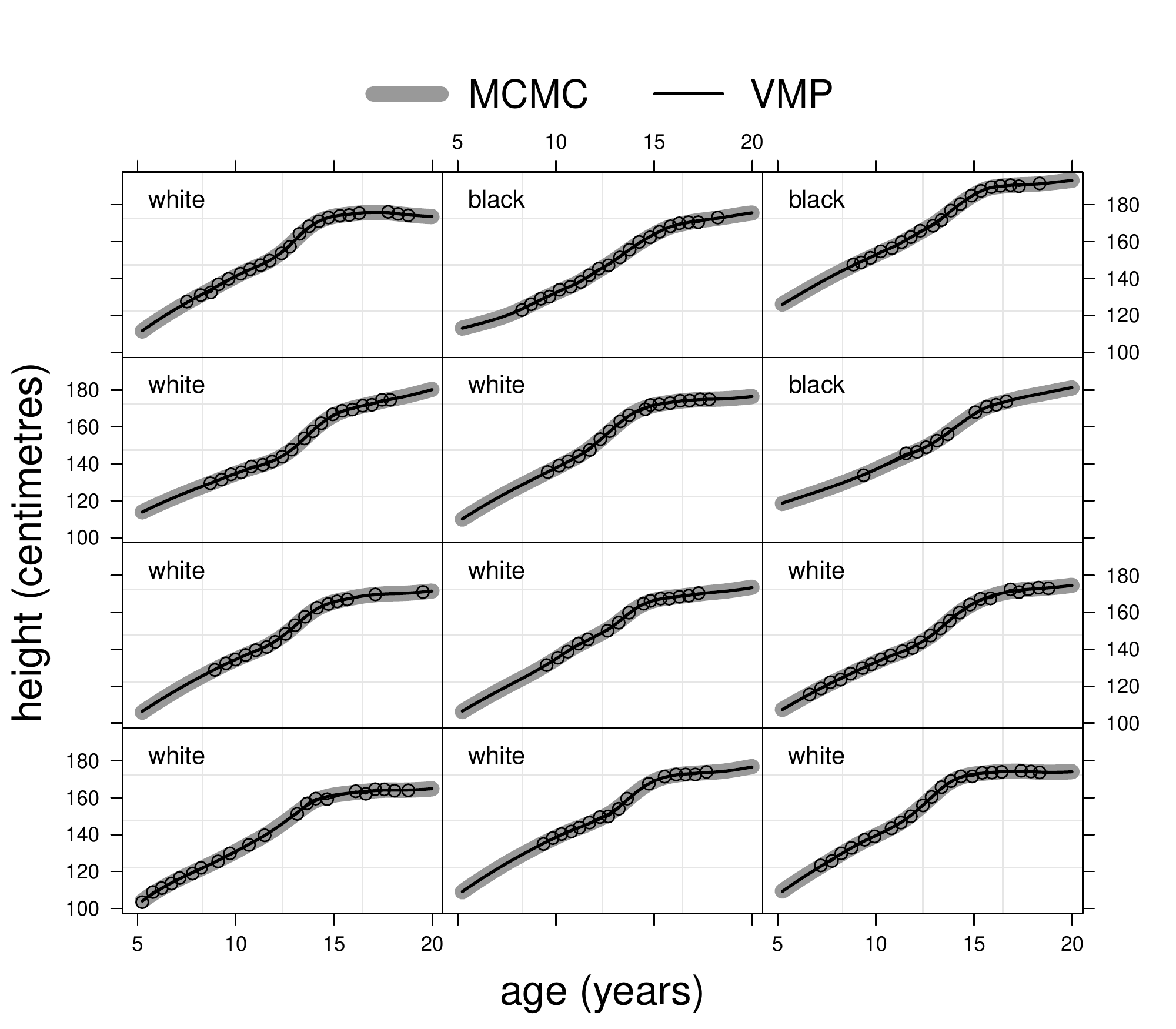}}
&\null{\includegraphics[width=0.45\textwidth]{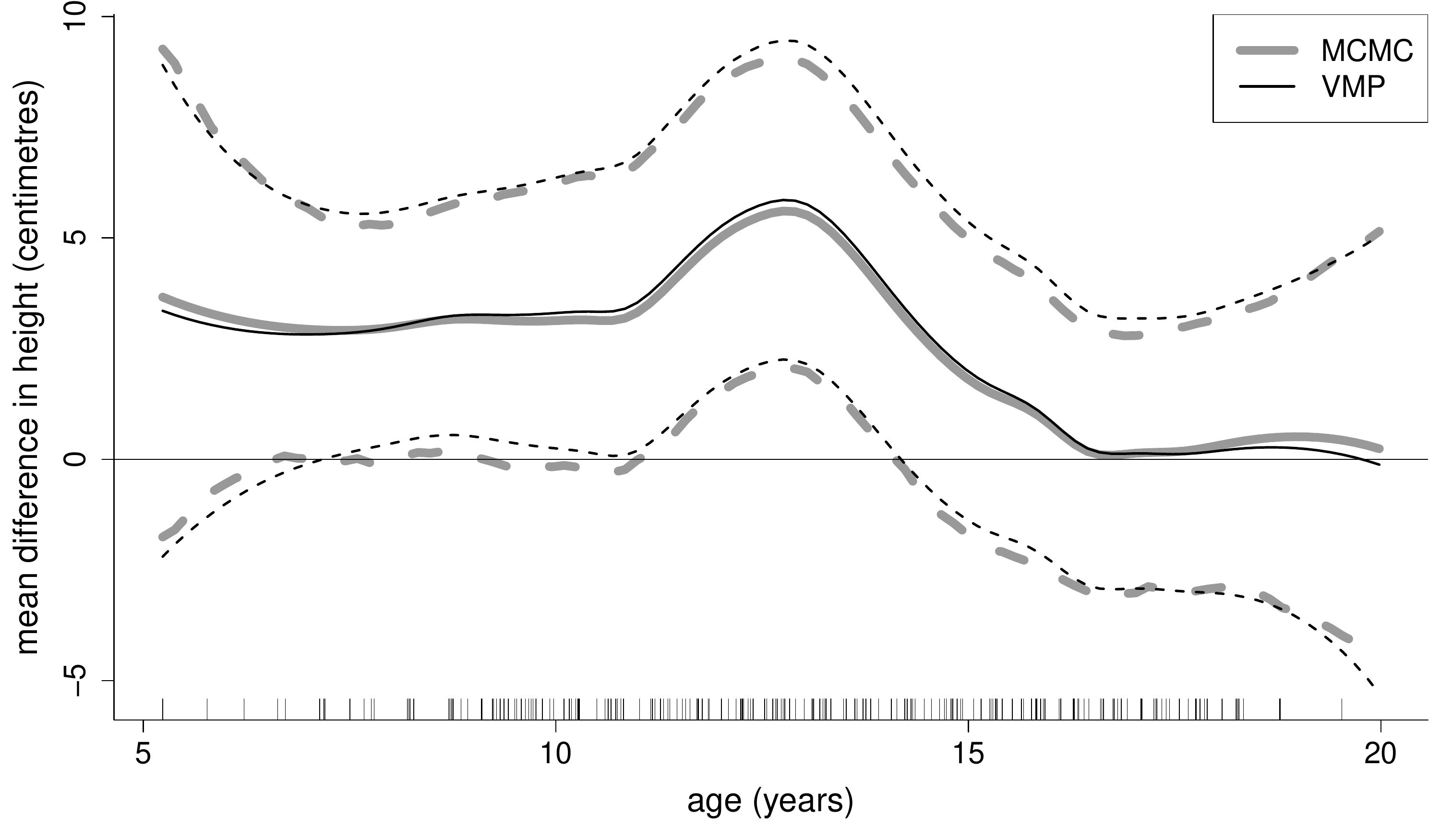}}
\end{tabular}
\caption{\it Left panel: Comparison of MCMC-based and VMP-based 
fitted group-specific curves for 12 randomly chosen subjects 
from the data on adolescent somatic growth (Pratt \textit{et al.}, 1989).
The legend in each panel signifies the subject's ethnicity.
Right panel: Similar to the left panel but for the
estimated contrast curve. The dashed curves 
correspond to approximate pointwise 95\% credible intervals.
The tick marks at the base of the plot show the age data.}
\label{fig:maleGrowIndContVMP} 
\end{figure}
}
{\vskip3mm
\thickboxit{\bf \centerline{Figure here.}}
\vskip3mm
}

\section{Extension to Generalized Semiparametric Regression}\label{sec:generalized}

Now we turn to the situation where the response data are not Gaussian
and, in particular, are binary or counts. This corresponds to the 
\emph{generalized} extension of linear models. In the same vein,
\emph{generalized linear mixed models} and \emph{generalized additive models}
are extensions of models treated in Section \ref{sec:linReg} that fall
under the umbrella of \emph{generalized semiparametric regression}.
Viable VMP algorithms for generalized semiparametric regression need
to be developed on a case-to-case basis. In this section we treat
binary response semiparametric regression, with both logistic and probit
link functions, and Poisson semiparametric regression. 

The logistic case is handled here using the variational lower bound of 
Jaakkola \myand Jordan (2000). In the probit case, a rather different
approach is used based on the auxiliary variable representation
of Albert \myand Chib (1993). 
Girolami \myand Rogers (2006) and Consonni \myand Marin (2007) 
show how the Albert-Chib device results in tractable MFVB
algorithms for probit models. The Poisson case uses yet another
approach based on the non-conjugate VMP infrastructure laid
out in Knowles \myand Minka (2011) and the fully simplified
Multivariate Normal updates derived in Wand (2014).
Knowles \myand Minka (2011) and Tan \myand Nott (2013) also 
propose quadrature-based approaches for handling the logistic
case, but are not investigated here.

The beauty of the VMP approach is that only messages passed
between fragments near the likelihood part of the factor
graph are affected by a change from the Gaussian response 
situation to each of these generalized response situations.
Figure \ref{fig:BernPoisFrags} shows the fragments involved.
The left panel diagram of Figure \ref{fig:BernPoisFrags}  
is appropriate for both logistic models handled via 
the Jaakkola \myand Jordan (2000) 
approach and Poisson response models handled via the
Knowles \myand Minka (2011) approach with the Wand (2014)
updates. The fragment is called the \emph{Jaakkola-Jordan logistic fragment}
or the \emph{Knowles-Minka-Wand fragment} depending on the
response type. In Sections \ref{sec:JJfragment} and \ref{sec:KMWfragment} we 
provide analytic updating formulae for the sufficient statistic of 
$m_{p(\by|\btheta)\to\btheta}(\btheta)$
assuming that Multivariate Normal messages are being passed 
to and from the $\btheta$ stochastic node.

\ifthenelse{\boolean{ShowFigures}}
{
\begin{figure}[!ht]
\centering
{\includegraphics[width=0.45\textwidth]{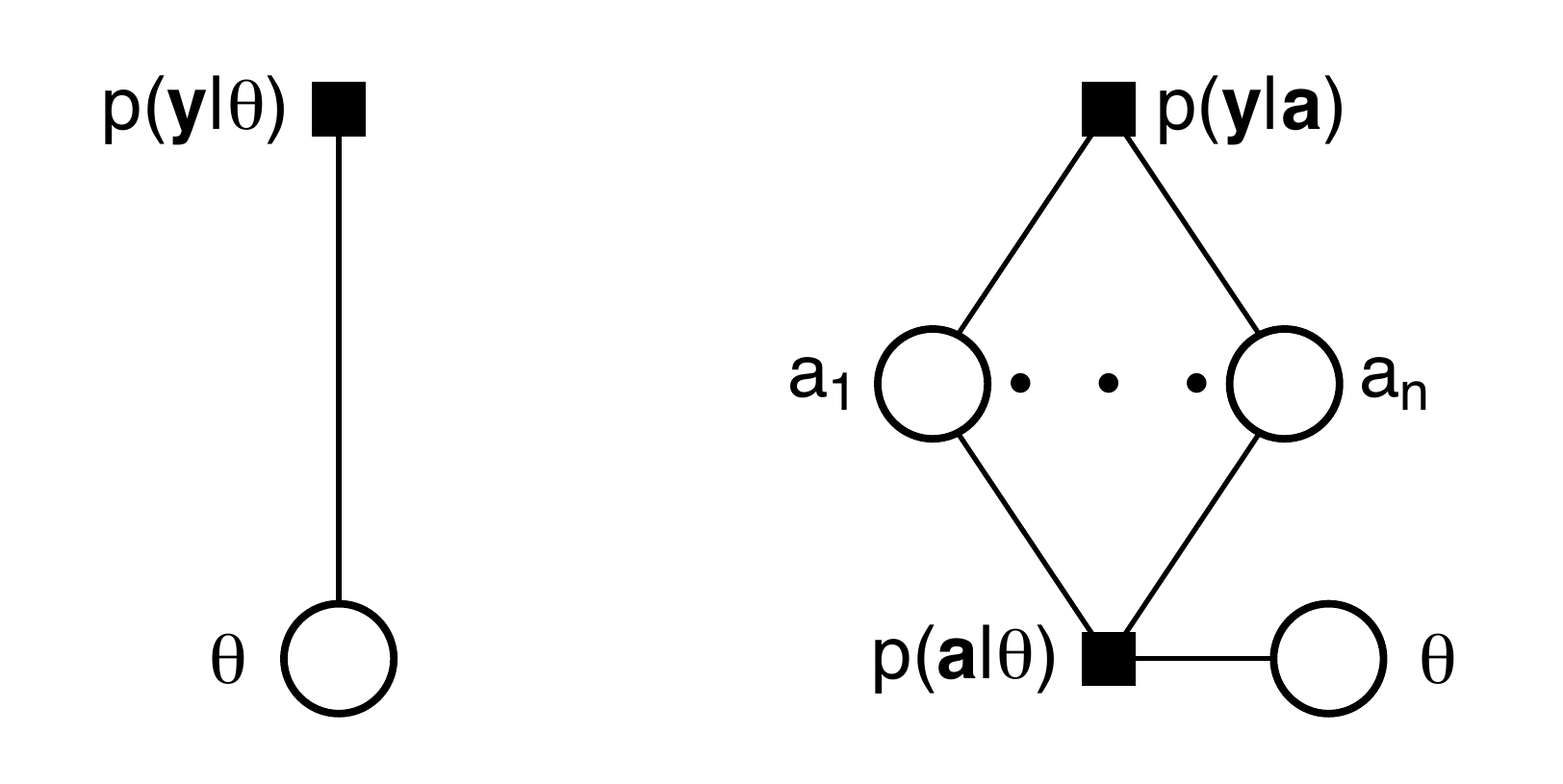}}
\caption{\it Left panel: Diagram for fragment corresponding to 
the likelihood for logistic and Poisson regression models.
Right panel: Fragments for the likelihood for probit
regression models with auxiliary variables $a_1,\ldots,a_n$
corresponding to the Albert-Chib device.
}
\label{fig:BernPoisFrags} 
\end{figure}
}
{\vskip3mm
\thickboxit{\bf \centerline{Figure here.}}
\vskip3mm
}

Throughout this section $\btheta$ denotes an $d\times 1$ random vector
and $\bA$ denotes an $n\times d$ design matrix.

\subsection{Jaakkola-Jordan Updates for the Logistic Likelihood Fragment}\label{sec:JJfragment}

The \emph{logistic fragment} is concerned with the
logistic likelihood specification
$$
y_i|\btheta\simind\mbox{Bernoulli}\big(1/[1+\exp\{-(\bA\btheta)_i\}]\big),
\quad 1\le i\le n.
$$
The factor of this fragment is 
\begin{equation}
p(\by|\,\btheta)=\exp\Big[\by^T\bA\btheta
                    -\bone^T\log\{\bone+\exp(\bA\btheta)\}\Big].
\label{eq:logistLogLik}
\end{equation}
Based on inputs
$$\etaSUBpythetaTOtheta\quad\mbox{and}\quad\etaSUBthetaTOpytheta$$
the variational and natural parameter vectors have the following updates:
\begin{equation}
\begin{array}{l}
\bXi\thickarrow\quarter
\Big\{\vecof^{-1}\Big(\big(\etaSUBpythetaCONNtheta\big)_2\Big)\Big\}^{-1}\\
\qquad\qquad
\times\Big[\big(\etaSUBpythetaCONNtheta\big)_1\big(\etaSUBpythetaCONNtheta\big)_1^T  
\Big\{\vecof^{-1}\Big(\big(\etaSUBpythetaCONNtheta\big)_2\Big)\Big\}^{-1}
-2\bI\Big],\\
\null\\
\bxi\thickarrow\sqrt{\mbox{diagonal}(\bA\,\bXi\bA^T)},\\
\null\\
\etaSUBpythetaTOtheta\thickarrow
\left[
\begin{array}{c}
\bA^T(\by-\smhalf\bone)\\[2ex]
-\vecof\left(\bA^T\diag\left\{\displaystyle{\frac{\tanh(\bxi/2)}{4\bxi}}\right\}\bA\right)
\end{array}
\right].
\end{array}
\label{eq:JJupdates}
\end{equation}

Justification for these updates is given in Section \ref{sec:JJderiv}
of the online supplement.

\subsection{Updates for the Albert-Chib Probit Likelihood Fragments}\label{sec:ACfragment}

The \emph{Albert-Chib probit fragments} deal with the probit likelihood
specification
\begin{equation}
y_i|\btheta\sim\mbox{Bernoulli}\Big(\Phi\{(\bA\btheta)_i\}\Big),\quad 1\le i\le n.
\label{eq:probitFrag}
\end{equation}
where $\Phi$ is the $N(0,1)$ cumulative distribution function.
Following Albert \myand Chib (1993) we re-write (\ref{eq:probitFrag}) as
\begin{equation}
y_i|a_i\sim\mbox{Bernoulli}\Big(I(a_i\ge0)\Big),\quad 1\le i\le n,
\qquad
\ba|\btheta\sim N(\bA\btheta,\bI)
\label{eq:ACtrick}
\end{equation}
and work with the factor graph fragments shown in the right panel of
Figure \ref{fig:BernPoisFrags}.

Based on the inputs
$\etaSUBpathetaTOtheta$ and $\etaSUBthetaTOpatheta$,
the updates for the Albert-Chib probit fragments are
\begin{equation}
\begin{array}{rcl}
\bnu&\thickarrow& -\smhalf\bA
\Big\{\vecof^{-1}\Big(\big(\etaSUBpathetaCONNtheta\big)_2\Big)\Big\}^{-1}
\big(\etaSUBpathetaCONNtheta\big)_1,\\[3ex]
\etaSUBpathetaTOtheta&\thickarrow&
\left[
\begin{array}{c}
\bA^T\left\{\bnu + (2\by-1)\odot\recipMills\Big((2\by-1)\odot\bnu\Big)\right\}\\[2ex]
-\smhalf\vecof(\bA^T\bA)
\end{array}
\right]
\end{array}
\label{eq:AlbertChibUpdate}
\end{equation}
where
$$
\zeta(x)\equiv\log\{2\Phi(x)\}
\quad\mbox{implying that}\quad 
\zeta'(x)=\frac{(2\pi)^{-1/2}e^{-x^2/2}}{\Phi(x)}.
$$
Working with $\zeta'$ has the advantage that software,
such as the function $\texttt{zeta()}$ in the package 
\textsf{sn} (Azzalini, 2015) within the \textsf{R} 
computing environment (R Core Team, 2015),
that facilitates numerically stable computation of 
(\ref{eq:AlbertChibUpdate}). 

Justification for these updates is given in Section \ref{sec:ACderiv}
of the online supplement.

\subsection{Knowles-Minka-Wand Updates for the Poisson Likelihood Fragment}\label{sec:KMWfragment}

The generic Poisson regression likelihood is
$$y_i|\btheta\sim\mbox{Poisson}\Big(\exp\{(\bA\btheta)_i\}\Big),\quad 1\le i\le n.$$
The message passed from $p(\by|\btheta)$ to $\btheta$ is
\begin{equation}
\mSUBpythetaTOtheta=\exp\Big\{\by^T\bA\btheta-\bone^T\exp(\bA\btheta)\Big\}.
\label{eq:PoissonLikMsg}
\end{equation}
Based on inputs 
$$\etaSUBpythetaTOtheta\quad\mbox{and}\quad\etaSUBthetaTOpytheta,$$
the update of $\etaSUBpythetaTOtheta$ involves the steps
\begin{equation}
{\setlength\arraycolsep{3pt}
\begin{array}{rcl}
\bomega&\thickarrow&\exp\Bigg(               
-\smhalf\bA\Big\{\vecof^{-1}\Big(\big(\etaSUBpythetaCONNtheta\big)_2\Big)\Big\}^{-1}
\big(\etaSUBpythetaCONNtheta\big)_1\\[2ex]
\qquad\qquad\qquad&&\qquad\quad-\quarter\diagonal\Big[\bA \Big\{\vecof^{-1}
        \Big(\big(\etaSUBpythetaCONNtheta\big)_2\Big)\Big\}^{-1}\bA^T\Big]\Bigg)\\[4ex]
\etaSUBpythetaTOtheta&\thickarrow&\left[
\begin{array}{l}
\bA^T\Big[\by-\bomega\\[1ex]
-\smhalf\diag(\bomega)\bA\Big\{\vecof^{-1}
\Big(\big(\etaSUBpythetaCONNtheta\big)_2\Big)\Big\}^{-1}
\big(\etaSUBpythetaCONNtheta\big)_1\Big]\\[4ex]
-\smhalf\vecof\big(\bA^T\diag(\bomega)\bA\big)
\end{array}
\right].
\end{array}
}
\label{eq:KMWupdate}
\end{equation}

Full justification of (\ref{eq:KMWupdate}) is given in
Section \ref{sec:KMWderiv} of the online supplement. 
Note that, despite their involved form, the manipulations required to update 
the factor to stochastic node message are purely algebraic.
Again we point out that, according to the message passing approach, 
(\ref{eq:KMWupdate}) only needs to be implemented once when developing 
a suite of computer programs for VMP-based semiparametric regression.

\subsection{Illustration for Generalized Response Nonparametric Regression}

We now provide brief illustration of the fragments presented in
this section for nonparametric regression via 
mixed model-based penalized splines with synthetic data.
Accuracy compared with MCMC-based inference is also addressed.
A timing comparison is given in Section \ref{sec:speed}.

A sample of size $500$ was generated from the Uniform
distribution on $(0,1)$, which we denote by $x_1,\ldots,x_{500}$
and then binary and count responses were generated according to
$$y^b_i|x_i\simind\mbox{Bernoulli}\{\ftrue(x_i)\}\quad\mbox{and}
\quad y^c_i|x_i\simind\mbox{Poisson}\{10\,\ftrue(x_i)\},\quad 1\le i\le 500,$$
where 
$\ftrue(x)\equiv \{1.05-1.02\,x+0.018\,x^2+0.4\,\phi(x;0.38,0.08)
                     +0.08\,\phi(x;0.75,0.03)\}/2.7$
and $\phi(x;\mu,\sigma)\equiv(2\pi\sigma^2)^{-1/2}
\exp\{-\smhalf(x-\mu)/\sigma^2\}$. The logistic, probit and Poisson
penalized spline models for the mean functions take the forms
$$
H\Big(\beta_0+\beta_1\,x+\sum_{k=1}^K\,u_k\,z_k(x)\Big),\quad
u_k\simind N(0,\sigma_u^2)
$$
where, respectively, $H(x)=1/(1+e^{-x})$, $H(x)=\Phi^{-1}(x)$ 
and $H(x)=e^x$ and the $z_k$ are spline basis
functions as defined just after (\ref{eq:penSplineFirst}).
The priors $\beta_0,\beta_1\simind N(0,10^{10})$ and
$\sigma_u\sim\mbox{Half-Cauchy}(10^5)$ were imposed.
A canonical cubic O'Sullivan spline basis with $K=25$ was
used for the $z_k$, formed by placing the interior knots at quantiles
of the $x_i$s.  MCMC samples from the posterior distributions of
the coefficients of size 1000, after a warm-up of 1000,
were obtained using the \textsf{R} package \textsf{rstan}
(Stan Development Team, 2016). VMP fitting is similar to
the updating scheme described in Section \ref{sec:altVMP}
but with likelihood fragment updating steps described
in Sections \ref{sec:JJfragment} to \ref{sec:KMWfragment} rather than those
for the Gaussian likelihood fragment and was iterated
200 times for each model.

Figure \ref{fig:genRespMCMCvsVMP} displays the  
true mean function and the MCMC and VMP fits.
In the logistic and Poisson models it is difficult
to discern a difference between the posterior means
and pointwise 95\% credible intervals.
The probit fits are such that VMP gives credible
sets that are slightly too narrow. This 
shortcoming of MFVB/VMP for the Albert-Chib probit 
approach is attributable to posterior
correlations between the entries $\ba$ and those of 
$(\beta_0,\bu)$ conveniently being set to zero in the mean field 
approximation even though these correlations
are significantly non-zero (e.g. Holmes \myand Held, 2006).

\ifthenelse{\boolean{ShowFigures}}
{
\begin{figure}[!ht]
\centering
{\includegraphics[width=0.85\textwidth]{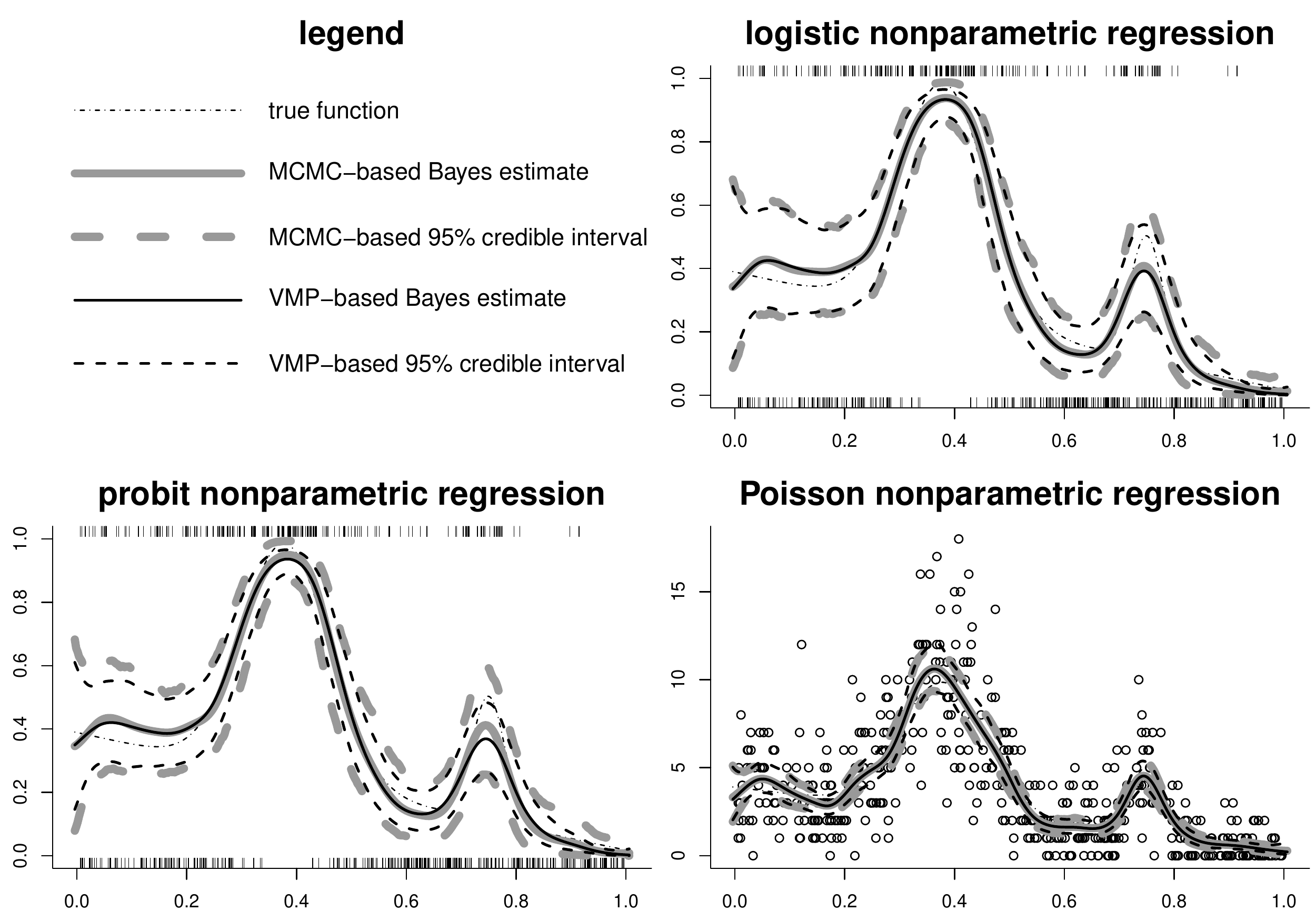}}
\caption{\it Comparison of MCMC- and VMP-based inference
for simulated Bernoulli and Poisson response nonparametric
regression data. The solid curves are Bayes estimates whilst
the dash curves correspond to pointwise 95\% credible intervals.
The tick marks at the top and base of the binary response plots 
show the data.
}
\label{fig:genRespMCMCvsVMP} 
\end{figure}
}
{\vskip3mm
\thickboxit{\bf \centerline{Figure here.}}
\vskip3mm
}

\section{Speed Considerations}\label{sec:speed}

As the title of this article indicates, 
the MFVB/VMP approach offers fast approximate
inference. For models of reasonable size, fits
can be achieved in in a few seconds or less
on ordinary desktop and laptop computers.
Seven of the author's previously published
MFVB articles contain speed comparisons with MCMC 
including Faes \textit{et al.} (2011), Lee \myand Wand (2016b) 
and Luts \myand Wand (2015) out of those which
are referenced earlier. The speed advantages for
the VMP alternative also apply, although some
qualification is necessary due to whether or not
matrix algebraic streamlining is employed.
In Table 1 of Lee \myand Wand (2016a) it is shown that
MFVB/VMP fitting of large semiparametric longitudinal and multilevel models 
with matrix algebraic streamlining and low-level programming language
implementation can be achieved in seconds even when there are tens of 
thousands of groups. A similar story is told by Table 1 of Lee \myand Wand (2016b)
for large to very large group-specific curve models. The MFVB/VMP computing
times range from minutes to tens of minutes for the largest models considered,
although this is without low-level programming language implementation. 
It is stated that MCMC fitting for the same models is expected to take days 
to weeks to run.

Table \ref{tab:timeComp} shows the average and standard deviation of computing times in 
seconds for replications of Figure \ref{fig:genRespMCMCvsVMP} simulation example.
All computations were performed on a laptop computer with 8 gigabytes of random access
memory and a 1.7 gigahertz processor.
There are a number of caveats connected with Table \ref{tab:timeComp}: (a) the computing
times depend on the MCMC sample sizes and the number of VMP iterations, (b) the MCMC
iterations are performed by \textsf{Stan} in the faster low-level \textsf{C++} programming 
language whereas the VMP iterations are performed in the slower high-level \textsf{R}
programming language, (c) the VMP approach was implemented na\"{\i}vely 
using the formulae of Section \ref{sec:GaussPenFrag} without any matrix algebraic streamlining.
Each of these caveats disadvantage the VMP approach in the speed comparison.
Nevertheless, Table \ref{tab:timeComp} shows that VMP takes 1.5 seconds or less
to perform approximate Bayesian inference for the Figure \ref{fig:genRespMCMCvsVMP}
scatterplots, whereas close to a minute is needed for MCMC via \textsf{Stan}.
%
\begin{table}[!ht]
\begin{center}
\begin{tabular}{l|ccc}
\hline\\[-1.3ex]
method & logistic nonpar. reg'n & probit nonpar. reg'n & Poisson nonpar. reg'n\\[0.1ex]
\hline\\[-0.9ex]
MCMC  &  49.50 (6.85)  & 56.200 (7.56)  & 48.60 (4.340) \\
VMP   &  1.36  (0.117) &  0.327 (0.0307)& 1.52  (0.101) \\[0.8ex]
\hline
\end{tabular}
\end{center}
\caption{\textit{Average (standard deviation) of computational times in seconds
over 100 replications of the Figure \ref{fig:genRespMCMCvsVMP} simulated data example.}}
\label{tab:timeComp}
\end{table}
%
\section{Conclusion}\label{sec:conclusion}

We have demonstrated that approximate inference for 
particular classes of arbitrarily large semiparametric regression 
models can be implemented with relatively few computer code
compartments. Moreover, many of these compartments involve 
straightforward matrix algebraic manipulations. Our exposition
transcends ongoing software projects that make use of
MFVB/VMP. Extensions to more elaborate models within the VMP framework 
is elucidated.

Accuracy considerations aside, the algebraic infrastructure that we 
have laid out in this article has far-reaching implications for the analysis of
big data sets via large semiparametric models as both continue
to grow in size. It is also beneficial for other classes 
of statistical models. In situations where inferential accuracy
is paramount, variational message passing algorithms may 
still play important roles in design and model selection 
phases with final reporting based on a more accurate method.


\ifthenelse{\boolean{UnBlinded}}{
\section*{Acknowledgments}

This research was partially supported by the
Australian Research Council Centre of Excellence
for Mathematical and Statistical Frontiers. The author
thanks Ray Carroll, Peter Forrester, Andy Kim, 
Cathy Lee, Matt McLean, Marianne Menictas, Tui Nolan, 
Chris Oates and Donald Richards for their comments on 
this research. Comments from an associate editor and
three referees are also gratefully acknowledged.
}
{\null}


\section*{References}

\begin{small}
\bibmpa
Albert, J.H. \myand Chib, S. (1993). Bayesian analysis of
binary and polychotomous response data.  
{\it Journal of the American Statistical Association}, 
{\bf 88}, 669--679.

\bibmpa
Atay-Kayis, A. \myand Massam, H. (2005).
A Monte Carlo method for computing marginal likelihood
in nondecomposable Gaussian graphical models.
\textit{Biometrika}, {\bf 92}, 317--335.

\bibmpa
Azzalini, A. (2015). The R package 'sn': The skew-normal and skew-t
  distributions (version 1.2). \texttt{http://azzalini.stat.unipd.it/SN}

\bibmpa
Bishop, C.M., Spiegelhalter, D.J. \myand Winn, J. (2003).
\textsf{VIBES}: A variational inference engine for Bayesian
networks. In S. Becker, S. Thrun and K. Obermayer, editors,
\textit{Advances in Neural Information Processing
Systems}, pp. 793--800,
Cambridge, Massachusetts: MIT Press.

\bibmpa
Bishop, C.M. (2006). {\it Pattern Recognition and Machine Learning.}
New York: Springer.

\bibmpa
Consonni, G. \myand Marin, J.-M. (2007).
Mean-field variational approximate Bayesian inference for 
latent variable models. {\it Computational Statistics
and Data Analysis}, {\bf 52}, 790--798.

\bibmpa
Coull, B.A., Ruppert, D. \myand Wand, M.P. (2001).
Simple incorporation of interactions into additive models.
{\it Biometrics}, {\bf 57}, 539--545.

\bibmpa
Diggle, P., Heagerty, P., Liang, K.-L. \myand Zeger, S. (2002).
{\it Analysis of Longitudinal Data (Second Edition)}.
Oxford: Oxford University Press.

\bibmpa
Donnelly, C.A., Laird, N.M. and Ware, J.H. (1995).
Prediction and creation of smooth curves for
temporally correlated longitudinal data.
{\it Journal of the American Statistical Association},
{\bf 90}, 984--989.

\bibmpa
Durban, M., Harezlak, J., Wand, M.P. \myand Carroll, R.J. (2005).
Simple fitting of subject-specific curves for longitudinal data.
{\it Statistics in Medicine}, {\bf 24}, 1153--1167.

\bibmpa
Faes, C., Ormerod, J.T. \myand Wand, M.P. (2011).
Variational Bayesian inference for parametric
and nonparametric regression with missing data.
{\it Journal of the American Statistical Association},
{\bf 106}, 959--971.

\bibmpa
Fitzmaurice, G., Davidian, M., Verbeke,G. \myand Molenberghs, G.
(eds.) (2008). \textit{Longitudinal Data Analysis: A Handbook of
Modern Statistical Methods}. Boca Raton, Florida: CRC Press.


\bibmpa
Frey, B.J., Kschischang, F.R., Loeliger, H.A. \myand Wiberg, N. (1998).
Factor graphs and algorithms. In \textit{Proceedings
of the 35th Allerton Conference on Communication, Control
and Computing 1997}.

\bibmpa
Gelman, A. (2006).  Prior distributions for variance parameters 
in hierarchical models. {\it Bayesian Analysis}, {\bf 1}, 515--533.

\bibmpa
Gelman, A. \myand Hill, J. (2007).
\textit{Data Analysis using Regression and Multilevel/Hierarchical Models},
New York: Cambridge University Press. 

\bibmpa
Gelman, A., Carlin, J.B., Stern, H.S., Dunson, D.B.,
Vehtari, A. \myand Rubin, D.B. (2014).
\textit{Bayesian Data Analysis, Third Edition},
Boca Raton, Florida: CRC Press.

\bibmpa
Ghosh, S. (2015).
\textit{Distributed Systems: An Algorithmic Approach,
Second Edition}. Boca Raton, Florida: CRC Press. 

\bibmpa
Girolami, M. \myand Rogers, S. (2006). Variational Bayesian multinomial probit
regression. {\it Neural Computation}, {\bf 18}, 1790--1817.

\bibmpa
Goldstein, H. (2010). \textit{Multilevel Statistical Models, 4th Edition},
Chichester UK: Wiley. 

\bibmpa
Gopal, V., Matthaiou, M. \myand Zhong, C. (2012).
Performance analysis of distributed MIMO systems
in Rayleigh/Inverse-Gaussian fading channels.
In \textit{Proceedings of
the Global Communications Conference (GLOBECOM) 2012}, 
pp. 2468--2474. IEEE Xplore Digital Library.

\bibmpa
Gurrin, L.C., Scurrah, K.J. \myand Hazelton, M.L. (2005). 
Tutorial in biostatistics: spline smoothing with linear mixed models.
{\it Statistics in Medicine}, {\bf 24}, 3361--3381.

\bibmpa
Hankin, R.K.S. (2007). The R package \textsf{gsl}:
Wrapper for the Gnu Scientific Library (version 2.10).
\texttt{http://cran.r-project.org}

\bibmpa
Hodges, J.S. (2013). \textit{Richly Parameterized Linear Models:
Additive, Time Series, and Spatial Models Using Random Effects}.
Boca Raton, Florida: CRC Press.

\bibmpa
Holmes, C.C. \myand Held, L. (2006).
Bayesian auxiliary variable models for binary and multinomial
regression. \textit{Bayesian Analysis}, {\bf 1}, 145--168.

\bibmpa
Huang, A. \myand Wand, M.P. (2013).
Simple marginally noninformative prior distributions 
for covariance matrices. \textit{Bayesian Analysis}, 
{\bf 8}, 439--452.

\bibmpa
Jaakkola, T.S. \myand Jordan, M.I. (2000). Bayesian
parameter estimation via variational methods.
{\it Statistics and Computing}, {\bf 10}, 25--37.

\bibmpa
Jordan, M.I. (2004). Graphical models.
{\it Statistical Science}, {\bf 19}, 140--155.

\bibmpa
Kammann, E.E. \myand Wand, M.P. (2003). Geoadditive models.
{\it Journal of the Royal Statistical Society, Series C}, {\bf 52}, 1--18.

\bibmpa
Knowles, D.A. \myand Minka, T.P. (2011), 
Non-conjugate message passing for multinomial and binary
regression. In J. Shawe-Taylor, R.S. Zamel, P. Bartlett, F. Pereira 
and K.Q. Weinberger, editors, \textit{Advances in Neural 
Information Processing Systems 24}, pp. 1701--1709.

\bibmpa
Kschischang, F.R., Frey, B.J. \myand Loeliger, H.A. (2001).
Factor graphs and the sum-product algorithm.
\textit{IEEE Transactions of Information Theory}, {\bf 47},
498--519.

\bibmpa
Kucukelbir, A., Tran, D., Ranganath, R., Gelman, A. \myand Blei, D.M. (2016).
Automatic variational inference in \textsf{Stan}.
Unpublished manuscript (\textit{arXiv:1603.00788v1}).

\bibmpa
Lee, C.Y.Y. \myand Wand, M.P. (2016a). Streamlined mean field
variational Bayes for longitudinal and multilevel data analysis.
\textit{Biometrical Journal}, in press.

\bibmpa
Lee, C.Y.Y. and Wand, M.P. (2016b).
Variational inference for fitting complex Bayesian mixed 
effects models to health data. \textit{Statistics in Medicine},
{\bf 35}, 165--188.

\bibmpa
Lock, R.H. (1993). 1993 new car data. \textit{Journal of
Statistics Education}, \textbf{1}.\\
\texttt{http://www.amstat.org/publications/jse/}

\bibmpa
Lunn, D., Jackson, C., Best, N., Thomas, A. \myand Spiegelhalter, D. (2012).
\textit{The \textsf{BUGS} Book -- A Practical Introduction to 
Bayesian Analysis}. Boca Raton, Florida: CRC Press.

\bibmpa
Luts, J. (2015). Real-time semiparametric regression for distributed
data sets. \textit{IEEE Transactions on Knowledge and Data Engineering},
{\bf 27}, 545--557.

\bibmpa
Luts, J., Broderick, T. \myand Wand, M.P. (2014).
Real-time semiparametric regression.
\textit{Journal of Computational and Graphical Statistics},
{\bf 23}, 589--615.

\bibmpa
Luts, J. \myand Wand, M.P. (2015).
Variational inference for count response semiparametric
regression. \textit{Bayesian Analysis}, {\bf 10}, 991--1023.

\bibmpa
Luts, J., Wang, S.S.J., Ormerod, J.T. \myand Wand, M.P. (2015).
Semiparametric regression analysis via \textsf{Infer.NET}. 
Under revision for \textit{Journal of Statistical Software}.

\bibmpa
Marley, J.K. \myand Wand, M.P. (2010).
Non-standard semiparametric regression via \textsf{BRugs}.
\textit{ Journal of Statistical Software}, {\bf 37}, Issue 5,
1--30.

\bibmpa
Minka, T. (2005).
Divergence measures and message passing.
\textit{Microsoft Research Technical Report Series}, 
{\bf MSR-TR-2005-173}, 1--17.

\bibmpa
Minka, T. \myand Winn, J. (2008).
Gates: A graphical notation for mixture models.
\textit{Microsoft Research Technical Report Series}, 
{\bf MSR-TR-2008-185}, 1--16.

\bibmpa
Minka, T., Winn, J., Guiver, J., Webster, S., Zaykov, Y., 
Yangel, B., Spengler, A. \myand Bronskill, J. (2014).
\textsf{Infer.NET 2.6}, Microsoft Research Cambridge.\\
\texttt{http://research.microsoft.com/infernet}

\bibmpa
Ormerod, J.T. \myand Wand, M.P. (2010).
Explaining variational approximations.
{\it The American Statistician},
{\bf 64}, 140--153.

\bibmpa
Park, T. \myand Casella, G. (2008). The Bayesian Lasso.
\textit{Journal of the American Statistical Association}, 
{\bf 103}, 681--686.

\bibmpa
Pratt, J.H., Jones, J.J, Miller, J.Z., Wagner, M.A. \myand
Fineberg, N.S. (1989). Racial differences in aldosterone
excretion and plasma aldosterone concentrations in
children. \textit{New England Journal of Medicine}, 
\textbf{321}, 1152--1157.

\bibmpa
R Core Team (2015). \textsf{R}: A language and environment
for statistical computing. R Foundation for Statistical
Computing, Vienna, Austria. 
\texttt{http://www.R-project.org}

\bibmpa
Ruppert, D., Wand, M.P. \myand Carroll, R.J. (2003).
{\it Semiparametric Regression}.
New York: Cambridge University Press.

\bibmpa
Ruppert, D., Wand, M.P. \myand Carroll, R.J. (2009).
Semiparametric regression during 2003-2007.
\textit{Electronic Journal of Statistics}, {\bf 3}, 1193--1256.

\bibmpa
Stan Development Team (2016). \textsf{Stan}: A C++ Library for 
Probability and Sampling, Version 2.9.0.
\texttt{http://mc-stan.org}.

\bibmpa
Tan, L.S.L. \myand Nott, D.J. (2013).
Variational inference for generalized linear mixed models using
partially noncentered parametrizations. 
\textit{Statistical Science}, \textbf{28}, 168--188.

\bibmpa
The Mathworks Incorporated (2015). Natick, Massachusetts, U.S.A.

\bibmpa
Uhler, C., Lenkoski, A. \myand Richards, D. (2014).
Exact formulas for the normalizing constants of Wishart
distributions for graphical models. Unpublished manuscript
(\textit{arXiv:1406.490}).

\bibmpa
Verbyla, A.P., Cullis, B.R., Kenward, M.G. and Welham, S.J. (1999).
The analysis of designed experiments and longitudinal data
by using smoothing splines (with discussion).
{\it Applied Statistics}, {\bf 48}, 269--312.

\bibmpa
Wainwright, M.J. \myand Jordan, M.I. (2008).
Graphical models, exponential families and variational 
inference. \textit{Foundations and Trends in 
Machine Learning}, {\bf 1}, 1--305.

\bibmpa
Wand, M.P. (2009).
Semiparametric regression and graphical models.
\textit{ Australian and New Zealand Journal
of Statistics}, {\bf 51}, 9--41.

\bibmpa
Wand, M.P. (2014).
Fully simplified Multivariate Normal updates
in non-conjugate variational message passing.
\textit{Journal of Machine Learning Research},
{\bf 15}, 1351--1369.

\bibmpa
Wand, M.P. \myand Ormerod, J.T. (2008).
On semiparametric regression with O'Sullivan penalized splines.
{\it Australian and New Zealand Journal of Statistics},
{\bf 50}, 179--198.

\bibmpa
Wand, M.P. \myand Ormerod, J.T. (2011).
Penalized wavelets: embedding wavelets
into semiparametric regression.
\textit{Electronic Journal of Statistics},
{\bf 5}, 1654--1717.


\bibmpa
Wang, S.S.J. \myand Wand, M.P. (2011).
Using \textsf{Infer.NET} for statistical analyses.
{\it The American Statistician}, {\bf 65}, 115--126.

\bibmpa
Winn, J. \myand Bishop, C.M. (2005).
Variational message passing. 
\textit{Journal of Machine Learning Research},
{\bf 6}, 661--694.

\bibmpa
Wood, S.N. (2006).
\textit{Generalized Additive Models: An Introduction with \textsf{R}}.
Boca Raton, Florida: CRC Press.

\bibmpa
Wood, S.N., F. Scheipl \myand J.J. Faraway (2013). Straightforward
intermediate rank tensor product smoothing in mixed models. 
\textit{Statistics and Computing}, {\bf 23}, 341--3601.

\end{small}

\vfill\eject
%
%
%
%
\renewcommand{\theequation}{S.\arabic{equation}}
\renewcommand{\thesection}{S.\arabic{section}}
\renewcommand{\thetable}{S.\arabic{table}}
\setcounter{equation}{0}
\setcounter{table}{0}
\setcounter{section}{0}
\setcounter{page}{1}
\setcounter{footnote}{0}

\begin{center}

\ifthenelse{\boolean{DoubleSpaced}}{\setstretch{1.5}}{}

{\Large Supplement for:}

\vskip7mm
{\Large\bf Fast Approximate Inference for Arbitrarily Large}
\vskip2mm
{\Large\bf Semiparametric Regression Models via Message Passing}

\vskip6mm
\ifthenelse{\boolean{UnBlinded}}{
{\sc By M.P. Wand}\footnote{M.P. Wand is Distinguished Professor, 
School of Mathematical and Physical Sciences,
University of Technology Sydney, P.O. Box 123, Broadway 2007, Australia, 
and Chief Investigator, Australian Research Council Centre of Excellence
for Mathematical and Statistical Frontiers.}}
{\null}
\end{center}


\section{Exponential Family Theory and Results}\label{sec:expFam}

The sufficient statistic and log-partition function are linked
by the results
\begin{equation}
E\{\bT(\bx)\}=\Diff_{\bdeta}A(\bdeta)^T\quad\mbox{and}\quad
\mbox{Cov}\{\bT(\bx)\}=\Diff_{\bdeta}\{\Diff_{\bdeta}A(\bdeta)^T\}
\label{eq:DrvResult}
\end{equation}
where $\mbox{Cov}\{\bT(\bx)\}$ is the covariance matrix of $\bT(\bx)$, 
and for $\bdf$ a $\real^p$-valued function with argument 
$\bx\in\real^d$, $\Diff_{\bx}f(\bx)$ is the $p\times d$ matrix
whose $(i,j)$ entry is $\partial{\bdf(\bx)_i}/\partial x_j$. 
The first expression in (\ref{eq:DrvResult}) is particularly important 
for variational message passing since the messages from factors to stochastic
nodes in conjugate factor graphs reduce to sufficient statistic expectations.

The digamma function, denoted by $\psi$, is
$$
\psi(x)\equiv \frac{d}{dx}\log\Gamma(x).
$$
Evaluation of $\psi$ is supported in the 
\textsf{MATLAB} computing environment (The Mathworks Incorporated, 2015)
via the function \texttt{psi()} and in the \textsf{R} computing environment 
(R Core Team, 2015) via the function \texttt{digamma()}.

The exponential integral function is
\begin{equation}
\mbox{Ei}(x)\equiv\,-\int_{-x}^{\infty}\frac{\exp(-t)}{t}\,dt,\quad x\in\real\backslash\{0\}.
\label{eq:EiDefn}
\end{equation}
Evaluation of $\mbox{Ei}$ is supported in the
\textsf{MATLAB} via the function \texttt{expint()},
which returns values of $-\mbox{Ei}(-x)$ for an input $x$, 
and in \textsf{R} via the function \texttt{expint{\textunderscore}Ei()} 
within the package \textsf{gsl} (Hankin, 2007).

\subsection{Bernoulli Distribution}

The probability mass function of the Bernoulli distribution
with probability of success $\wp\in(0,1)$ is
$$p(x)=\wp^x(1-\wp)^{1-x},\quad x\in\{0,1\}.$$
The sufficient statistic and base measure are
$$T(x)=x
\quad\mbox{and}\quad
h(x)=I(x\in\{0,1\}).
$$
The natural parameter vector and its inverse mapping are 
$$
\eta=\log\{\wp/(1-\wp)\}\quad\mbox{and}\quad 
\wp=e^{\eta}/(1+e^{\eta})
$$
and the log-partition function is
$$A(\eta)=\log(1+e^{\eta}).$$

\subsection{Univariate Normal Distribution}

The density function of the Univariate Normal distribution
with mean $\mu\in\real$ and variance $\sigma^2>0$ is
$$p(x)=(2\pi\sigma^2)^{-1/2}\exp\{-(x-\mu)^2/(2\sigma^2)\},\quad x\in\real.$$
The sufficient statistic and base measure are
$$
\bT(x)=\left[
\begin{array}{c}
x\\
x^2
\end{array}
\right]
\quad\mbox{and}\quad
h(x)=(2\pi)^{-1/2}.
$$
The natural parameter vector and its inverse mapping are 
$$
\bdeta=
\left[
\begin{array}{c}
\eta_1\\
\eta_2
\end{array}
\right]=\left[
\begin{array}{c}
\mu/\sigma^2\\[1ex]
-1/(2\sigma^2)
\end{array}
\right]
\quad\mbox{and}\quad
\left[
\begin{array}{c}
\mu\\
\sigma^2
\end{array}
\right]=\left[
\begin{array}{c}
-\eta_1/(2\eta_2)\\[1ex]
-1/(2\eta_2)
\end{array}
\right]
$$
and the log-partition function is 
$$A(\bdeta)=-\quarter\,(\eta_1^2/\eta_2)-\smhalf\log(-2\eta_2).$$

\subsection{Inverse Chi-Squared and Inverse Gamma Distributions}

The random variable $x$ has an \textit{Inverse Chi-Squared} distribution
with shape parameter $\kappa>0$ and scale parameter $\lambda>0$,
written $x\sim\mbox{Inverse-$\chi^2$}(\kappa,\lambda)$,
if the density function of $x$ is 
$$p(x)=\{(\lambda/2)^{\kappa/2}/\Gamma(\kappa/2)\}\,x^{-(\kappa/2)-1}
\exp\{-(\lambda/2)/x\},\quad x>0.
$$
The random variable $x$ has an \textit{Inverse Gamma} distribution
with shape parameter $\kappaIG>0$ and scale parameter $\lambdaIG>0$,
written $x\sim\mbox{Inverse-Gamma}(\kappaIG,\lambdaIG)$
if the density function of $x$ is 
$$p(x)=\{\lambdaIG^{\kappaIG}/\Gamma(\kappaIG)\}\,x^{-\kappaIG-1}
\exp(-\lambdaIG/x),\quad x>0.
$$
The Inverse Chi-Squared and Inverse Gamma distributions are simple
reparametrizations of each other in that
$$x\sim\mbox{Inverse-$\chi^2$}(\kappa,\lambda)\quad 
\mbox{if and only if}
\quad 
x\sim\mbox{Inverse-Gamma}(\kappa/2,\lambda/2).
$$
As explained in Section \ref{sec:invWishDist}, 
the Inverse Wishart distribution for random 
matrices reduces to the Inverse Chi-Squared 
distribution in the $1\times1$ case. 

The sufficient statistic and base measure are 
$$
\bT(x)=\left[
\begin{array}{c}
\log(x)\\
1/x
\end{array}
\right]
\quad\mbox{and}\quad
h(x)=I(x>0).
$$
The natural parameter vector and its inverse mappings are
\begin{equation}
\begin{array}{l}
\bdeta=
\left[
\begin{array}{c}
\eta_1\\
\eta_2
\end{array}
\right]=\left[
\begin{array}{c}
-\smhalf(\kappa+2)\\[1ex]
-\smhalf\lambda
\end{array}
\right]
=\left[
\begin{array}{c}
-(\kappatilde+1)\\[1ex]
-\lambdatilde
\end{array}
\right],
\\[3ex]
\left[
\begin{array}{c}
\kappa\\
\lambda
\end{array}
\right]=\left[
\begin{array}{c}
-2-2\eta_1\\[1ex]
-2\eta_2\end{array}
\right]
\quad\mbox{and}\quad
\left[
\begin{array}{c}
\kappatilde\\
\lambdatilde
\end{array}
\right]=\left[
\begin{array}{c}
-1-\eta_1\\[1ex]
-\eta_2\end{array}
\right]
\end{array}
\label{eq:InvChiSqNatVsComm}
\end{equation}
and the log-partition function is
$$A(\bdeta)=(\eta_1+1)\log(-\eta_2)+\log\Gamma(-\eta_1-1).$$

\subsection{Beta Distribution}\label{sec:BetaDistn}

The density function of the Beta distribution
with shape parameters $\shpOne>0$ and $\shpTwo>0$ is
$$p(x)=\frac{\Gamma(\shpOne+\shpTwo)\,
             x^{\shpOne-1}(1-x)^{\shpTwo-1}}
             {\Gamma(\shpOne)\Gamma(\shpTwo)},\quad 0<x<1.$$
The sufficient statistic and base measure are
$$
\bT(x)=\left[
\begin{array}{c}
\log(x)\\
\log(1-x)
\end{array}
\right]
\quad\mbox{and}\quad
h(x)=I(0<x<1).
$$
The natural parameter vector and its inverse mapping are 
$$
\bdeta=
\left[
\begin{array}{c}
\eta_1\\
\eta_2
\end{array}
\right]=\left[
\begin{array}{c}
\shpOne-1\\[1ex]
\shpTwo-1
\end{array}
\right]
\quad\mbox{and}\quad
\left[
\begin{array}{c}
\shpOne\\
\shpTwo
\end{array}
\right]=\left[
\begin{array}{c}
\eta_1+1\\[1ex]
\eta_2+1\end{array}
\right]
$$
and the log-partition function is
$$A(\bdeta)=\log\Gamma(\eta_1+1)+\log\Gamma(\eta_2+1)
-\log\Gamma(\eta_1+\eta_2+2).$$

\subsection{Inverse Gaussian Distribution}\label{sec:InvGaussDistn}

The random variable $x$ has an Inverse Gaussian distribution
with parameters $\mu>0$ and $\lambdaInvGau>0$, written 
$x\sim\mbox{Inverse-Gaussian}(\mu,\lambdaInvGau)$, if the
density function of $x$ is
$$p(x)=\lambdaInvGau^{1/2}(2\pi\,x^3)^{-1/2}\exp\left\{-\frac{\lambdaInvGau(x-\mu)^2}
{2\mu^2\,x}\right\},\quad x>0.$$
The sufficient statistic and base measure are
$$
\bT(x)=\left[
\begin{array}{c}
x\\
1/x
\end{array}
\right]
\quad\mbox{and}\quad
h(x)=(2\pi\,x^3)^{-1/2}\,I(x>0).
$$
The natural parameter vector and its inverse mapping are 
$$
\bdeta=
\left[
\begin{array}{c}
\eta_1\\
\eta_2
\end{array}
\right]=\left[
\begin{array}{c}
-\lambdaInvGau/(2\mu^2)\\[1ex]
-\lambdaInvGau/2
\end{array}
\right]
\quad\mbox{and}\quad
\left[
\begin{array}{c}
\mu\\
\lambdaInvGau
\end{array}
\right]=\left[
\begin{array}{c}
(\eta_2/\eta_1)^{1/2}\\[1ex]
-2\eta_2\end{array}
\right]
$$
and the log-partition function is
$$A(\bdeta)=\,-2(\eta_1\eta_2)^{1/2}-\smhalf\log(-2\eta_2).$$
The Inverse Gaussian distribution is the only exponential family
distribution in Section \ref{sec:expFam} with a non-constant base measure.
This implies that the entropy contribution from $h$,
$E\{-\log h(x)\}$ where $x\sim\mbox{Inverse-Gaussian}(\mu,\lambdaInvGau)$,
is not trivial and so we list it here. Using, for example, 
Lemma 1 of Gopal \textit{et al.} (2012) we obtain  
$$E\{-\log h(x)\}=\quarter\log(4\pi^2\eta_2^3/\eta_1^3)+\frac{3}{2}\exp\Big(4(\eta_1\eta_2)^{1/2}\Big)
                      \mbox{Ei}\Big(-4(\eta_1\eta_2)^{1/2}\Big).$$
where the function $\mbox{Ei}$ is defined in (\ref{eq:EiDefn}).

\subsection{Multivariate Normal Distribution}\label{sec:MultNormDistn}

The $d\times 1$ random vector $\bx$ has a \textit{Multivariate Normal} 
distribution
with mean $\bmu$ and covariance matrix $\bSigma$,
a symmetric positive definite $d\times d$ matrix, 
written $\bx\sim N(\bmu,\bSigma)$, 
if the density function of $\bx$ is
$$p(\bx)=(2\pi)^{-d/2}|\bSigma|^{-1/2}\exp\{-\smhalf(\bx-\bmu)^T
\bSigma^{-1}(\bx-\bmu)\},\quad\bx\in\real^d.$$
The sufficient statistic and base measure are 
$$\bT(\bx)=\left[
\begin{array}{c}
\bx\\[1ex]
\vecof(\bx\bx^T)
\end{array}
\right]
\quad\mbox{and}\quad
h(\bx)=(2\pi)^{-d/2}.
$$
The natural parameter vector and inverse mapping are
\begin{equation}
\bdeta=
\left[
\begin{array}{c}
\bdeta_1\\
\bdeta_2
\end{array}
\right]=\left[
\begin{array}{c}
\bSigma^{-1}\bmu\\[1ex]
-\smhalf\vecof(\bSigma^{-1})
\end{array}
\right]
\quad\mbox{and}\quad
\left\{
\begin{array}{l}
\bmu\,=\,-\smhalf\{\vecof^{-1}(\bdeta_2)\}^{-1}\bdeta_1\\[1ex]
\bSigma\,=\,-\smhalf\{\vecof^{-1}(\bdeta_2)\}^{-1}
\end{array}
\right.
\label{eq:MVNnatVScomm}
\end{equation}
and log-partition function is 
$$A(\bdeta)=-\quarter\bdeta_1^T\{\vecof^{-1}(\bdeta_2)\}\bdeta_1
-\smhalf\log\Big|-2\vecof^{-1}(\bdeta_2)\Big|.
$$

\subsection{Inverse Wishart Distribution}\label{sec:invWishDist}

The $d\times d$ random matrix $\bX$ has an 
\textit{Inverse Wishart} distribution
with shape parameter $\kappa>d-1$ and scale matrix $\bLambda$,
a symmetric positive definite $d\times d$ matrix, written 
$\bX\sim\mbox{Inverse-Wishart}(\kappa,\bLambda)$, 
if the density function of $\bX$ is
{\setlength\arraycolsep{3pt}
\begin{eqnarray*}
p(\bX)&=&\frac{|\bLambda|^{\kappa/2}}{2^{d\kappa/2}\pi^{d(d-1)/4}
\prod_{j=1}^d\Gamma(\frac{\kappa+1-j}{2})}\,
|\bX|^{-(\kappa+d+1)/2}\exp\{-\smhalf\tr(\bLambda\bX^{-1} )\}\\[1ex]
&&\quad\times 
I(\bX\ \mbox{a symmetric and positive definite $d\times d$ matrix}).
\end{eqnarray*}
}
The special case of $d=1$ coincides with the Inverse Chi-Squared
distribution. The sufficient statistic
and base measure are 
\begin{equation}
\bT(\bX)=\left[
\begin{array}{c}
\log|\bX|\\[1ex]
\vecof(\bX^{-1})
\end{array}
\right]
\quad\mbox{and}\quad
h(\bX)=\frac{I(\mbox{$\bX$ is symmetric and positive definite})}
{\pi^{d(d-1)/4}}.
\label{eq:natStatInvWish}
\end{equation}
The natural parameter vector and inverse mapping are
\begin{equation}
\bdeta=
\left[
\begin{array}{c}
\eta_1\\[1ex]
\bdeta_2
\end{array}
\right]=\left[
\begin{array}{c}
-\smhalf(\kappa+d+1)\\[1ex]
-\smhalf\vecof(\bLambda)
\end{array}
\right]
\quad\mbox{and}\quad
\left\{
\begin{array}{l}
\kappa\,=\,-\,d-1-2\eta_1\\
\bLambda\,=\,
-2\vecof^{-1}(\bdeta_2)
\end{array}
\right.
\label{eq:natParmInvWish}
\end{equation}
and log-partition function is 
$$A(\bdeta)=\{\eta_1+\smhalf(d+1)\}\log\big|-\vecof^{-1}(\bdeta_2)\big|
+\sum_{j=1}^d\log\Gamma\{-\eta_1-\smhalf(d+j)\}.$$

\subsubsection{Inverse G-Wishart Extension}\label{sec:GinvWishDist}

Now consider the extension of the Inverse Wishart distribution 
corresponding to the inverse of the $d\times d$ random matrix $\bX$
having some off-diagonal entries forced to equal zero. Such structure can
be represented using undirected graphs and, following the 
nomenclature of Atay-Kayis \myand Massam (2005), is referred to 
as the \emph{Inverse G-Wishart} distribution. Let $G$ be an
undirected graph with $d$ nodes labeled $1,\ldots,d$ and
set $E$ consisting of sets of pairs of nodes that are connected by an edge. 
We say that the $d\times d$ matrix $\bM$ 
\emph{respects} $G$ if 
$$\bM_{ij}=0\quad\mbox{for all}\quad \{i,j\}\notin E.$$
Then the $d\times d$ random matrix $\bX$ has an \emph{Inverse G-Wishart}
distribution with $d$-node undirected graph $G$, 
shape parameter $\kappa>d-1$ and scale matrix $\bLambda$, a
symmetric positive definite $d\times d$ matrix that respects $G$,
written $\bX\sim\mbox{Inverse-G-Wishart}(G,\kappa,\bLambda)$,
if the density function of $\bX$ is 
{\setlength\arraycolsep{3pt}
\begin{eqnarray*}
p(\bX)&\propto&|\bX|^{-(\kappa+d+1)/2}\exp\{-\smhalf\tr(\bLambda\bX^{-1} )\}
\,I(\bX \mbox{is symmetric and positive definite})\\
&&\qquad\times I(\bX^{-1}\ \mbox{respects\ \ $G$}).
\end{eqnarray*}
}
The normalizing factor follows from the formulae of 
Uhler \textit{et al.} (2014),
although it is quite complicated for general $G$.

The sufficient statistic $\bT(\bX)$ and natural parameter
vector take the same form as for the ordinary Inverse Wishart
distribution, given at (\ref{eq:natStatInvWish})
and (\ref{eq:natParmInvWish}).

The special case of diagonal matrices coincides with $G$
such that $E=\emptyset$, meaning that $G$ is a totally
disconnected graph. We denote such $G$ by $\Gdiag$. 
Note that
$$\bX\sim\mbox{Inverse-G-Wishart}(\Gdiag,\kappa,\bLambda)$$
if and only if 
{\setlength\arraycolsep{3pt}
\begin{eqnarray*}
p(\bX)&=&\frac{1}{2^{d(\kappa+d-1)/2}\Gamma(\frac{\kappa+d-1}{2})^d}
\left\{\prod_{i=1}^d\bLambda_{ii}^{(\kappa+d-1)/2}\bX_{ii}^{-(\kappa+d+1)/2}
\exp(-\smhalf\bLambda_{ii}/\bX_{ii})I(\bX_{ii}>0)\right\}
\end{eqnarray*}
}
and is simply a product of Inverse Chi-Squared density functions.


\subsection{Table  of Sufficient Statistic Expectations}\label{sec:tabETxAndEntropies}

Table \ref{tab:ETx} lists the sufficient statistic
expectations for each of the exponential family distributions
covered in Section \ref{sec:expFam}. All expressions are in terms of natural
parameters.

\vfill\eject
\vskip15mm

\begin{table}[ht]
\begin{center}
\begin{tabular}{lcc}
\hline\\[-1.3ex]
Distribution       & $T(x),\bT(\bx),\bT(\bX)$   &$E\{T(x)\},E\{\bT(\bx)\},E\{\bT(\bX)\}$\\[0.1ex]
\hline\\[-0.9ex]
Bernoulli          & $x$   &$1/(1+e^{-\eta})$               \\[2ex]
Univariate Normal  & $\left[\begin{array}{c}
                            x\\[1ex]
                            x^2
                       \end{array}\right]$
    &$\left[\begin{array}{c}
                        -\eta_1/(2\eta_2)\\[1ex]
                         (\eta_1^2-2\eta_2)/(4\eta_2^2)
                       \end{array}\right]$                   \\[4ex]
Inverse Chi-Squared      &    $\left[\begin{array}{c}
                            \log(x)\\[0.5ex]
                              1/x
                       \end{array}\right]$
&$\left[\begin{array}{c}
                    \log(-\eta_2)-\psi(-\eta_1-1)\\[0.5ex]
                         (\eta_1+1)/\eta_2
                       \end{array}\right]$                   \\[4ex]
Beta                    &    $\left[\begin{array}{c}
                            \log(x)\\[0.5ex]
                            \log(1-x)
                       \end{array}\right]$
&$\left[\begin{array}{c}
                     \psi(\eta_1+1)-\psi(\eta_1+\eta_2+2)\\[0.5ex]
                     \psi(\eta_2+1)-\psi(\eta_1+\eta_2+2)
                       \end{array}\right]$                   \\[4ex]
Inverse Gaussian      &    $\left[\begin{array}{c}
                              x\\[0.5ex]
                              1/x
                       \end{array}\right]$
&$\left[\begin{array}{c}
                         (\eta_2/\eta_1)^{1/2}\\[0.5ex]
                         (\eta_1/\eta_2)^{1/2}-1/(2\eta_2)
                       \end{array}\right]$                   \\[4ex]
Multivariate Normal  & $\left[\begin{array}{c}
                              \bx\\[1ex]
                             \vecof(\bx\bx^T)
                       \end{array}\right]$
&
$\left[\begin{array}{l}
           -\smhalf\{\vecof^{-1}(\bdeta_2)\}^{-1}\bdeta_1\\[4ex]
   \quarter\vecof\Big(\{\vecof^{-1}(\bdeta_2)\}^{-1}\\
             \qquad\times[\bdeta_1\bdeta_1^T\{\vecof^{-1}(\bdeta_2)\}^{-1}
                         -2\,\bI]\Big)\end{array}\right]$\\[10ex]
Inverse Wishart  & $\left[\begin{array}{l}
                              \log|\bX|\\[1ex]
                             \vecof(\bX^{-1})
                       \end{array}\right]$
&$\left[\begin{array}{l}
           \log|-\vecof^{-1}(\bdeta_2)|\\
     \qquad-\displaystyle{\sum_{j=1}^d}\psi\{-\eta_1-\smhalf(d+j)\}\\[6ex]
           \{\eta_1+\smhalf(d+1)\}\vecof[\{\vecof^{-1}(\bdeta_2)\}^{-1}]
        \end{array}\right]$                   \\[8ex]
\hline
\end{tabular}
\end{center}
\caption{\textit{Expressions for sufficient statistics and their expectations
in terms of natural parameters for some common exponential family
distributions.}}
\label{tab:ETx}
\end{table}

\section{Derivational Details}\label{sec:derivDetails}

Here we provide details on various derivations appearing
throughout the article.

\subsection{Derivation of  Message Functional Forms Given by (\ref{eq:linModMsgForms})}\label{sec:msgFormsDeriv}

With `const' denoting terms that do not depend on the function argument,
the logarithms of each of the factors can be expressed as follows:

{\setlength\arraycolsep{3pt}
\begin{eqnarray*}
\log\,p(\bbeta)&=&
\left[
\begin{array}{c}   
\bbeta\\[1ex]
\vecof(\bbeta\bbeta^T)
\end{array}
\right]^T
\left[
\begin{array}{c}   
\bSigma_{\bbeta}^{-1}\bmu_{\bbeta}\\[1ex]
-\smhalf\vecof(\bSigma_{\bbeta}^{-1})
\end{array}
\right]+\mbox{const},\\[2ex]
\log\,p(\by\,|\,\bbeta,\sigma^2)&=&
\left\{
\begin{array}{l}
\left[
\begin{array}{c}   
\bbeta\\[1ex]
\vecof(\bbeta\bbeta^T)
\end{array}
\right]^T
\left[
\begin{array}{c}   
\bX^T\by\\[1ex]
-\smhalf\vecof(\bX^T\bX) 
\end{array}
\right]\left(\frac{1}{\sigma^2}\right)
\mbox{+ const, as a function of $\bbeta$,}\\[4ex]
\left[
\begin{array}{c}   
\log(\sigma^2)\\[1ex]
1/\sigma^2
\end{array}
\right]^T
\left[
\begin{array}{c}   
-\smhalf\,n\\[1ex]
-\smhalf\Vert\by-\bX\bbeta\Vert^2 
\end{array}
\right]\mbox{+ const, as a function of $\sigma^2$,}
\end{array}
\right.\\[1ex]
\log\,p(\sigma^2\,|\,a)&=&
\left\{
\begin{array}{l}
\left[
\begin{array}{c}   
\log(\sigma^2)\\[1ex]
1/\sigma^2
\end{array}
\right]^T
\left[
\begin{array}{c}   
-\threehalves\\[1ex]
-1/(2a)
\end{array}
\right]
\mbox{+const, as a function of $\sigma^2$,}\\[4ex]
\left[
\begin{array}{c}   
\log(a)\\[1ex]
1/a
\end{array}
\right]^T
\left[
\begin{array}{c}   
-\smhalf\\[1ex]
-1/(2\sigma^2)
\end{array}
\right]
\mbox{+ const, as a function of $a$, and}
\end{array}
\right.\\[3ex]
\log\,p(a)&=&
\left[
\begin{array}{c}   
\log(a)\\[1ex]
1/a
\end{array}
\right]^T
\left[
\begin{array}{c}   
-\threehalves\\[1ex]
-1/(2\,A^2)
\end{array}
\right]+\mbox{const}.
\end{eqnarray*}
}
Then, since the only neighbor of $p(\bbeta)$ in Figure \ref{fig:linRegFacGraphMsgs}
is $\bbeta$, the expectation in (\ref{eq:facToStochOne}) disappears and we 
immediately get 
$$
\mSUBpbetaTObeta\thickarrow\exp\left\{
\left[
\begin{array}{c}   
\bbeta\\
\vecof(\bbeta\bbeta^T)
\end{array}
\right]^T
\left[
\begin{array}{c}   
\bSigma_{\bbeta}^{-1}\bmu_{\bbeta}\\[1ex]
-\smhalf\vecof(\bSigma_{\bbeta}^{-1})
\end{array}
\right]
\right\}
$$
which confirms the first part of (\ref{eq:linModMsgForms}).
The factor $p(\by\,|\,\bbeta,\sigma^2)$ has
both stochastic nodes $\bbeta$ and $\sigma^2$ 
as neighbors so
$$\mSUBpybetasigsqTObeta\thickarrow\exp\left\{
\left[
\begin{array}{c}   
\bbeta\\[1ex]
\vecof(\bbeta\bbeta^T)
\end{array}
\right]^T
\left[
\begin{array}{c}   
\bX^T\by\\[1ex]
-\smhalf\vecof(\bX^T\bX) 
\end{array}
\right]\biggerE_{p(\by|\bbeta,\sigma^2)\to\bbeta}
\left(\frac{1}{\sigma^2}\right)
\right\}
$$
and
$$\mSUBpybetasigsqTOsigsq\thickarrow\exp\left\{
\left[
\begin{array}{c}   
\log(\sigsq)\\[1ex]
1/\sigsq
\end{array}
\right]^T
\left[
\begin{array}{c}   
-\smhalf\,n\\[1ex]
-\smhalf\biggerE_{p(\by|\bbeta,\sigma^2)\to\sigsq}\Vert\by-\bX\bbeta\Vert^2
\end{array}
\right]
\right\}
$$
which are also of the forms given in (\ref{eq:linModMsgForms}).
Similar arguments show that $\mSUBpsigsqaTOsigsq$,
$\mSUBpsigsqaTOa$ and $\mSUBpaTOa$ have the stated
Inverse-$\chi^2$ forms after the first iteration
of VMP.

\subsection{Derivation of the Jaakkola-Jordan Updates}\label{sec:JJderiv}

According to (\ref{eq:facToStochOne}) and (\ref{eq:facToStochTwo}), 
the message passed from the factor $p(\by|\btheta)$, given at (\ref{eq:logistLogLik}),
is
\begin{equation}
\mSUBpythetaTOtheta=p(\by|\,\btheta)=\exp\Big[\by^T\bA\btheta
                    -\bone^T\log\{\bone+\exp(\bA\btheta)\}\Big].
\label{eq:logistLikMsg}
\end{equation}
This, however, is not conjugate with the Multivariate Normal messages typically passed
to $\btheta$ from other neighboring factors. 
The Jaakkola-Jordan device (Jaakkola \myand Jordan, 2000)
is based on the following variational representation of the troublesome function in
(\ref{eq:logistLikMsg}):
\begin{equation}
-\log(1+e^x)=\maxsub{\xi\in\real}\left\{A(\xi)x^2-\smhalf\,x+C(\xi)\right\}
\ \mbox{for all $x\in\real$}
\label{eq:JJbound}
\end{equation}
where $A(\xi)\equiv\,-\tanh(\xi/2)/(4\xi)$ and 
$C(\xi)\equiv\xi/2-\log(1+e^{\xi})+\xi\tanh(\xi/2)/4$.
Representation (\ref{eq:JJbound}) leads to the following 
family of variational lower bounds on the logarithm of 
(\ref{eq:logistLogLik}):
$$
\log\,p(\by|\,\btheta)=-\btheta^T\bA^T
\diag\left\{\displaystyle{\frac{\tanh(\bxi/2)}{4\bxi}}\right\}\bA\,\btheta
+(\by-\smhalf\bone)^T\bA\btheta+\bone^TC(\bxi)
$$
and corresponding family of conjugate messages
$$
\munderSUBpythetaTOthetaxi\equiv
\exp\left\{
\left[
\begin{array}{c}
\btheta\\[1ex]
\vecof(\btheta\btheta^T)
\end{array}
\right]^T
\left[
\begin{array}{c}
\bA^T(\by-\smhalf\,\bone)\\[1ex]
-\vecof\left(\bA^T\diag\left\{\displaystyle{\frac{\tanh(\bxi/2)}{4\bxi}}\right\}\bA\right)
\end{array}
\right]^T
\right\}
$$
where $\bxi$ is an $n\times 1$ vector of variational parameters. 

The updates (\ref{eq:JJupdates}) are driven by the goal
of maximizing the following $\btheta$-\emph{localized} approximate marginal log-likelihood:
$$
\log\punder(\by;q)^{[\btheta]}=\Entropy\{q(\btheta)\}+E_q\{\log\,p(\by|\btheta)\}
                  +E_q(\mbox{other log-factors neighboring $\btheta$}).
$$
Rohde \myand Wand (2015) contains further details on localized 
approximate marginal log-likelihoods. Application of the Jaakkola-Jordan device
leads to the family of approximate marginal log-likelihoods:
\begin{equation}
{\setlength\arraycolsep{1pt}
\begin{array}{rcl}
\log\punder(\by;q,\bxi)^{[\btheta]}&=&\Entropy\{q(\btheta;\bxi)\}
-E_{q(\btheta;\bxi)}\left[\btheta^T\bA^T
\diag\left\{\displaystyle{\frac{\tanh(\bxi/2)}{4\bxi}}\right\}\bA\,\btheta\right]
\\[2ex]
&&\qquad +E_{q(\btheta;\bxi)}\{(\by-\smhalf\bone)^T\bA\btheta\}+\bone^TC(\bxi)\\[2ex]
&&\qquad +E_{q(\btheta;\bxi)}\big(\mbox{other log-factors neighboring $\btheta$}\big)
\end{array}
}
\label{eq:logMLxi}
\end{equation}
where the function $C$ is defined in Section \ref{sec:JJfragment}
and, courtesy of (\ref{eq:qStarVMP}), the current $q(\btheta;\bxi)$ 
density function satisfies
\begin{equation}
q(\btheta;\bxi)
\propto\munderSUBpythetaTOthetaxi\times
(\mbox{product of messages to $\btheta$ from its other neighbors}).
\label{eq:qJJ}
\end{equation}
Note that update (\ref{eq:stochToFac}) allows us to replace (\ref{eq:qJJ}) by
\begin{equation}
q(\btheta;\bxi)\propto\munderSUBpythetaTOthetaxi\,\mSUBthetaTOpytheta.
\label{eq:qthetaxi}
\end{equation}
As explained in, for example, Section 21.8 of Murphy (2012), a
practical approach to optimizing the $\bxi$ vector is coordinate ascent
applied to (\ref{eq:logMLxi}) with the moments of $q(\btheta;\bxi)$
held fixed. This approach also has an Expectation-Maximization algorithm
representation (Jaakkola \myand Jordan, 2000). Under this strategy
\begin{equation}
{\setlength\arraycolsep{1pt}
\begin{array}{rcl}
\log\punder(\by;q,\bxi)^{[\btheta]}&=&
\bone^TC(\bxi)-E_{q(\btheta;\bxi)}\left[\btheta^T\bA^T
\diag\left\{\displaystyle{\frac{\tanh(\bxi/2)}{4\bxi}}\right\}\bA\,\btheta\right]
\\[2ex]
&&\qquad+\mbox{terms not involving $\bxi$, excluding moments of $q(\btheta;\bxi)$.}\\[2ex]
\end{array}
}
\label{eq:logMLxiMain}
\end{equation}
The first line of (\ref{eq:logMLxiMain}) is maximized over $\bxi$ 
by
$$\bxi=\sqrt{\mbox{diagonal}\big\{\bA\,E_{q(\btheta;\bxi)}(\btheta\btheta^T)\bA\big\}}.$$
(e.g. Murphy, 2012, Section 21.8.3). From (\ref{eq:qthetaxi}),
$$q(\btheta;\bxi)\propto\exp\left\{\left[
\begin{array}{c}
\btheta\\
\vecof(\btheta\btheta^T)
\end{array}
\right]^T\big(\etaSUBpythetaTOtheta+\etaSUBthetaTOpytheta\big)\right\}
$$
and, so from Table \ref{tab:ETx},
\begin{eqnarray*}
&&E_{q(\btheta;\bxi)}(\btheta\btheta^T)=
\Big\{\vecof^{-1}\Big(\big(\etaSUBpythetaCONNtheta\big)_2\Big)\Big\}^{-1}\\
&&\qquad\qquad
\times\Big[\big(\etaSUBpythetaCONNtheta\big)_1\big(\etaSUBpythetaCONNtheta\big)_1^T  
\Big\{\vecof^{-1}\Big(\big(\etaSUBpythetaCONNtheta\big)_2\Big)\Big\}^{-1}
-2\bI\Big]
\end{eqnarray*}
and the updates (\ref{eq:JJupdates}) follow immediately.

\subsection{Derivation of the Albert-Chib Updates}\label{sec:ACderiv}

The relevant factor graph fragments are displayed in the right
panel of Figure \ref{fig:BernPoisFrags}. The factors are
$$p(\by|\,\ba)=\prod_{i=1}^n\left\{y_i\,I(a_i\ge0)+(1-y_i)\,I(a_i<0)\right\}$$
and
$$p(\ba|\,\btheta)=(2\pi)^{-n/2}\exp\{-\smhalf\Vert\ba-\bA\btheta\Vert^2\}.$$

According to (\ref{eq:facToStochOne}) and (\ref{eq:facToStochTwo}), the messages 
from $p(\by|\,\ba)$ to each $a_i$ are 
$$\mSUBpyaTOai=y_i\,I(a_i\ge0)+(1-y_i)\,I(a_i<0),\quad 1\le i\le n,$$
and the messages from $p(\ba|\btheta)$ to each $a_i$ are
$$
\mSUBpathetaTOai=\exp\left[-\smhalf\{a_i-\big(\bA\,E_{p(\ba|\btheta)\to a_i}(\btheta)\big)_i\}^2\right],
\quad 1\le i\le n
$$
where, with the assistance of Table \ref{tab:ETx},
$$
E_{p(\ba|\btheta)\to a_i}(\btheta)=-\smhalf
\Big\{\vecof^{-1}\Big(\big(\etaSUBpathetaCONNtheta\big)_2\Big)\Big\}^{-1}
\big(\etaSUBpathetaCONNtheta\big)_1,\quad 1\le i\le n.
$$
Since, from (\ref{eq:stochToFac}), 
$\mSUBaiTOpatheta\thickarrow\mSUBpyaTOai$ we have
\begin{eqnarray*}
&&\mSUBpathetaTOai\mSUBaiTOpatheta\propto\\
&&\qquad\left\{
\begin{array}{l}
\mbox{the $N((\bA\,E_{p(\ba|\btheta)\to a_i}(\btheta))_i,1)$ 
density function truncated to $(-\infty,0)$ if $y_i=0$}\\[1ex]
\mbox{the $N((\bA\,E_{p(\ba|\btheta)\to a_i}(\btheta))_i,1)$ 
density function truncated to $[0,\infty)$ if $y_i=1$}.\
\end{array}
\right.
\end{eqnarray*}
Standard manipulations then lead to the
mean of the normalized 
$$\mSUBpathetaTOai\mSUBaiTOpatheta$$ 
equaling
\begin{equation}
\muSUBpathetaCONNai\equiv \nu_i+(2\,y_i-1)\recipMills\big((2\,y_i-1)\nu_i\big)
\label{eq:muWithMills}
\end{equation}
%
%
%
where
\begin{equation}
\nu_i\equiv (\bA\,E_{p(\ba|\btheta)\to a_i}(\btheta))_i
=-\smhalf\Big(\bA
\Big\{\vecof^{-1}\Big(\big(\etaSUBpathetaCONNtheta\big)_2\Big)\Big\}^{-1}
\big(\etaSUBpathetaCONNtheta\big)_1\Big)_i.
\label{eq:nuiExprn}
\end{equation}
Lastly, the message from $p(\ba|\btheta)$ to $\btheta$ is 
$$\mSUBpathetaTOtheta=
\exp\left\{
\left[
\begin{array}{c}
\btheta\\
\vecof(\btheta\btheta^T)
\end{array}
\right]^T\etaSUBpathetaTOtheta
\right\}
$$
where
\begin{equation}
\etaSUBpathetaTOtheta\thickarrow
\left[
\begin{array}{c}
\bA^T\bmuSUBpathetaCONNa\\[1ex]
-\smhalf\vecof(\bA^T\bA)
\end{array}
\right]
\label{eq:ACpenul}
\end{equation}
and $\bmuSUBpathetaCONNa$ is the $n\times 1$ vector containing the 
$\muSUBpathetaCONNai$. The updates in (\ref{eq:AlbertChibUpdate})
arise from substitution of (\ref{eq:muWithMills}) and (\ref{eq:nuiExprn})
into (\ref{eq:ACpenul}).

\subsection{Derivation of the  Knowles-Minka-Wand Updates}\label{sec:KMWderiv}

The message passed from $p(\by|\btheta)$ to $\btheta$, 
given by  (\ref{eq:PoissonLikMsg}), is not conjugate with 
Multivariate Normal messages passed to $\btheta$ from
other factors. A remedy proposed by Knowles \myand Minka (2011) and dubbed
\emph{non-conjugate VMP} involves, in this case, replacement of 
(\ref{eq:PoissonLikMsg}) by 
$$\mtildeSUBpythetaTOtheta\equiv\exp\left\{
\left[
\begin{array}{c}
\btheta \\[1ex]
\vecof(\btheta\btheta^T)
\end{array}
\right]^T
\etaSUBpythetaTOtheta
\right\}
$$
to enforce conjugacy with Multivariate Normal messages. 

Knowles \myand Minka (2011) propose that $\etaSUBpythetaTOtheta$ 
be updated according to maximization of a localized Kullback-Leibler divergence criterion,
summarized in their Algorithm 1. For the Poisson regression likelihood this 
criterion can be expressed in closed form. However, expressions in 
Algorithm 1 of Knowles \myand Minka (2011) involve inversion of a 
matrix that is quartic in the length of $\btheta$. Wand (2014)
derived fully simplified updates for non-conjugate VMP in the special 
case of Multivariate Normal message approximation.

As with Section \ref{sec:JJderiv}, the derivation starts with the
$\btheta$-localized approximate marginal log-likelihood for the 
Poisson likelihood fragment:
\begin{equation}
{\setlength\arraycolsep{1pt}
\begin{array}{rcl}
\log\punder(\by;q)^{[\btheta]}&=&\Entropy\{q(\btheta)\}+E_q\{\log\,p(\by|\btheta)\}\\[1ex]
                  &&\quad+E_q(\mbox{other log-factors neighboring $\btheta$})
\end{array}
}
\label{eq:logMLlocal}
\end{equation}
but now the logarithm of the likelihood factor is 
$$\log\,p(\by|\btheta)=\by^T\bA\btheta-\bone^T\exp(\bA\btheta)-\bone^T\log(\by!).$$
As explained in Section \ref{sec:KMWfragment} we enforce conjugacy with Multivariate
Normal messages sent from other factors neighboring $\btheta$ by simply working
with
$$\mtildeSUBpythetaTOtheta\equiv\exp\left\{
\left[
\begin{array}{c}
\btheta \\[1ex]
\vecof(\btheta\btheta^T)
\end{array}
\right]^T
\etaSUBpythetaTOtheta
\right\}
$$
instead of (\ref{eq:PoissonLikMsg}). Using the same argument that led to 
(\ref{eq:qthetaxi}), the current $q(\btheta)$ density function is the
Multivariate Normal density function with natural parameter vector 
$\etaSUBpythetaCONNtheta$. Let $\bmu_{q(\btheta)}$ and 
$\bSigma_{q(\btheta)}$ be the mean vector and covariance matrix
of $q(\btheta)$ . Then, because of (\ref{eq:MVNnatVScomm}), the
natural parameters and common parameters are the following
functions of one another:
\begin{equation}
\begin{array}{l}
\etaSUBpythetaCONNtheta=
\left[
\begin{array}{c}
(\etaSUBpythetaCONNtheta)_1\\
(\etaSUBpythetaCONNtheta)_2
\end{array}
\right]=\left[
\begin{array}{c}
\bSigma_{q(\btheta)}^{-1}\bmu_{q(\btheta)}\\[1ex]
-\smhalf\vecof(\bSigma_{q(\btheta)}^{-1})
\end{array}
\right]\\[5ex]
\quad\mbox{and}\quad
\left\{
\begin{array}{l}
\bmu_{q(\btheta)}\,=\,-\smhalf\{\vecof^{-1}
\big((\etaSUBpythetaCONNtheta)_2\big)\}^{-1}(\etaSUBpythetaCONNtheta)_1\\[1ex]
\bSigma_{q(\btheta)}\,=\,-\smhalf\{\vecof^{-1}\big((\etaSUBpythetaCONNtheta)_2\big)\}^{-1}.
\end{array}
\right.
\end{array}
\label{eq:MVNnatVScommKMW}
\end{equation}
For the next part of the derivation with work with the common parameters 
to make use of a key result in Wand (2014) (see also Rohde \myand Wand, 2015),
and then transform to natural parameter vectors after that.

Under conjugacy, the non-entropy 
component of (\ref{eq:logMLlocal}) is 
$$\mbox{NonEntropy}(q)=E_q\{\by^T\bA\btheta-\bone^T\exp(\bA\btheta)\}
-\bone^T\log(\by!)+E_q\left\{\left[
\begin{array}{c}   
\btheta \\
\vecof(\btheta\btheta^T)
\end{array}\right]\right\}^T\bdeta^{\dagger}
$$
where $\bdeta^{\dagger}$ is the sum of the 
sufficient statistics of messages passed to $\btheta$
other than the message from $p(\by|\btheta)$.
But, because of (\ref{eq:stochToFac}),
$$\etaSUBthetaTOpytheta\thickarrow\bdeta^{\dagger}$$
and so we get the following explicit form depending only
on the messages passed between the nodes of the Poisson
likelihood fragment:
{\setlength\arraycolsep{1pt}
\begin{eqnarray*}
\mbox{NonEntropy}(q;\bmu_{q(\btheta)},\bSigma_{q(\btheta)})&=&\by^T\bA\bmu_{q(\btheta)}
-\bone^T\exp\{\bA\bmu_{q(\btheta)}
+\smhalf\mbox{diagonal}(\bA\bSigma_{q(\btheta)}\bA^T)\}\\[2ex]
&&\quad+\bmu_{q(\btheta)}^T(\etaSUBthetaTOpytheta)_1
+\bmu_{q(\btheta)}^T
\vecof^{-1}\big((\etaSUBthetaTOpytheta)_2\big)\bmu_{q(\btheta)}\\[2ex]
&&\quad+\tr\Big\{\bSigma_{q(\btheta)}\vecof^{-1}\big((\etaSUBthetaTOpytheta)_2\big)\Big\}
-\bone^T\log(\by!).
\end{eqnarray*}
}

From equation (7) of Wand (2014) and Result 2 of Rohde \myand Wand (2015), 
fixed-point iteration with respect to the natural parameter vector 
for maximization of (\ref{eq:logMLlocal}) reduces to 
\begin{equation}
\left\{
{\setlength\arraycolsep{3pt}
\begin{array}{rcl}
\bv_{q(\btheta)}&\thickarrow&\,\Diff_{\bmu_{q(\btheta)}}
\NonEntropy(q;\bmu_{q(\btheta)},\bSigma_{q(\btheta)})^T\\[2ex]
\bSigma_{q(\btheta)}&\thickarrow&-\{\Hess_{\bmu_{q(\btheta)}}
\NonEntropy(q;\bmu_{q(\btheta)},\bSigma_{q(\btheta)})\}^{-1}\\[2ex]
\mu_{q(\btheta)}&\thickarrow&\bmu_{q(\btheta)}+\bSigma_{q(\btheta)}\bv_{q(\btheta)}
\end{array}
}
\right.
\label{eq:RWupdates}
\end{equation}
where $\Diff_{\bmu_{q(\btheta)}}$ and $\Hess_{\bmu_{q(\btheta)}}$ 
denote, respectively, the derivative vector and Hessian matrix
with respect to $\bmu_{q(\btheta)}$. Formal definitions are given
in Wand (2014). Arguments analogous to those given in Appendix A.4 of
Menictas \myand Wand (2015) lead to the explicit forms for
the non-entropy component of (\ref{eq:logMLlocal}):
{\setlength\arraycolsep{1pt}
\begin{eqnarray*}
\Diff_{\bmu_{q(\btheta)}}\NonEntropy(q;\bmu_{q(\btheta)},\bSigma_{q(\btheta)})^T
&=&\bA^T(\by-\bomega)+(\etaSUBthetaTOpytheta)_1\\[2ex]
&&\qquad+2\,\vecof^{-1}\big((\etaSUBthetaTOpytheta)_2\big)\bmu_{q(\btheta)}
\end{eqnarray*}
}
and
{\setlength\arraycolsep{1pt}
\begin{eqnarray*}
\Hess_{\bmu_{q(\btheta)}}\NonEntropy(q;\bmu_{q(\btheta)},\bSigma_{q(\btheta)})
&=&-\bA^T\diag(\bomega)\bA
+2\,\vecof^{-1}\big((\etaSUBthetaTOpytheta)_2\big)
\end{eqnarray*}
}
where
$$\bomega\equiv \exp\{\bA\bmu_{q(\btheta)}
+\smhalf\mbox{diagonal}(\bA\bSigma_{q(\btheta)}\bA^T)\}.$$
Substitution into (\ref{eq:RWupdates}) then gives the updating
scheme
\begin{equation}
{\setlength\arraycolsep{3pt}
\begin{array}{rcl}
\bomega&\thickarrow&\exp\{\bA\bmu_{q(\btheta)}
+\smhalf\mbox{diagonal}(\bA\bSigma_{q(\btheta)}\bA^T)\}\\[2ex]
\bSigma_{q(\btheta)}&\thickarrow&\Big\{\bA^T\diag(\bomega)\bA
-2\,\vecof^{-1}\big((\etaSUBthetaTOpytheta)_2\big)\Big\}^{-1}\\[2ex]
\mu_{q(\btheta)}&\thickarrow&\bmu_{q(\btheta)}+\bSigma_{q(\btheta)}
\big\{\bA^T(\by-\bomega)+(\etaSUBthetaTOpytheta)_1\\[2ex]
&&\qquad\qquad+2\,\vecof^{-1}\big((\etaSUBthetaTOpytheta)_2\big)\bmu_{q(\btheta)}\big\}.
\end{array}
}
\label{eq:WandResult}
\end{equation}
Using (\ref{eq:MVNnatVScommKMW}) the update for $\bomega$ can be expressed
in terms of the natural parameter vectors as
{\setlength\arraycolsep{1pt}
\begin{eqnarray*}
\bomega&\thickarrow&\exp\Bigg(               
-\smhalf\bA\Big\{\vecof^{-1}\Big(\big(\etaSUBpythetaCONNtheta\big)_2\Big)\Big\}^{-1}
\big(\etaSUBpythetaCONNtheta\big)_1\\
&&\qquad-\quarter\diagonal\Big[\bA \Big\{\vecof^{-1}
        \Big(\big(\etaSUBpythetaCONNtheta\big)_2\Big)\Big\}^{-1}\bA^T\Big]\Bigg).
\end{eqnarray*}
}
Again using (\ref{eq:MVNnatVScommKMW}), the $\bSigma_{q(\btheta)}$ update can 
be written as 
$$-\smhalf\{\vecof^{-1}\big((\etaSUBpythetaCONNtheta)_2\big)\}^{-1}
\thickarrow
\Big\{\bA^T\diag(\bomega)\bA
-2\,\vecof^{-1}\big((\etaSUBthetaTOpytheta)_2\big)\Big\}^{-1}
$$
which is equivalent to 
$$(\etaSUBpythetaTOtheta)_2+(\etaSUBthetaTOpytheta)_2\thickarrow
\,-\smhalf\vecof(\bA^T\diag(\bomega)\bA)+(\etaSUBthetaTOpytheta)_2$$
which, in turn, is equivalent to the second component of $\etaSUBpythetaTOtheta$
being updated according to 
\begin{equation}
(\etaSUBpythetaTOtheta)_2\thickarrow\, -\smhalf\vecof(\bA^T\diag(\bomega)\bA).
\label{eq:etaTwoUpdate}
\end{equation}
For the update of the first component of $\etaSUBpythetaTOtheta$ we note that
the last update of (\ref{eq:WandResult}) is equivalent to 
\begin{equation}
{\setlength\arraycolsep{3pt}
\begin{array}{rcl}
\bSigma_{q(\btheta)}^{-1}\mu_{q(\btheta)}&\thickarrow&\bSigma_{q(\btheta)}^{-1}\bmu_{q(\btheta)}+
\bA^T(\by-\bomega)+(\etaSUBthetaTOpytheta)_1\\[2ex]
&&\qquad\qquad\qquad+2\,\vecof^{-1}\big((\etaSUBthetaTOpytheta)_2\big)\bmu_{q(\btheta)}
\end{array}
}
\label{eq:sickOfLabelling}
\end{equation}
where, on the right-hand side,
\begin{equation}
\bSigma_{q(\btheta)}^{-1}=\bA^T\diag(\bomega)\bA
-2\,\vecof^{-1}\big((\etaSUBthetaTOpytheta)_2\big)
\label{eq:rustBucket}
\end{equation}
according to its updated value and 
\begin{equation}
\mu_{q(\btheta)}=-\smhalf\{\vecof^{-1}
\big((\etaSUBpythetaCONNtheta)_2\big)\}^{-1}(\etaSUBpythetaCONNtheta)_1
\label{eq:figTree}
\end{equation}
is the terms of the sufficient statistics from the previous iteration
before (\ref{eq:etaTwoUpdate}) has taken place. Substitution of 
(\ref{eq:rustBucket}) and (\ref{eq:figTree}) into (\ref{eq:sickOfLabelling})
we get
{\setlength\arraycolsep{3pt}
\begin{eqnarray*}
&&(\etaSUBpythetaTOtheta)_1+(\etaSUBthetaTOpytheta)_1\thickarrow\\[1ex]
&&\ \Big\{-\smhalf\bA^T\diag(\bomega)\bA+\,\vecof^{-1}\big((\etaSUBthetaTOpytheta)_2\big)\Big\}
\{\vecof^{-1}\big((\etaSUBpythetaCONNtheta)_2\big)\}^{-1}(\etaSUBpythetaCONNtheta)_1\\[1ex]
&&\quad+\bA^T(\by-\bomega)+(\etaSUBthetaTOpytheta)_1\\[1ex]
&&\quad-\vecof^{-1}\big((\etaSUBthetaTOpytheta)_2\big)
\{\vecof^{-1}\big((\etaSUBpythetaCONNtheta)_2\big)\}^{-1}(\etaSUBpythetaCONNtheta)_1
\end{eqnarray*}
}
which is equivalent to 
$$(\etaSUBpythetaTOtheta)_1\thickarrow-\smhalf\bA^T\diag(\bomega)
\bA\{\vecof^{-1}\big((\etaSUBpythetaCONNtheta)_2\big)\}^{-1}(\etaSUBpythetaCONNtheta)_1
+\bA^T(\by-\bomega).$$
Scheme (\ref{eq:KMWupdate}) follows immediately.

\subsection{Streamlined Derivation of the Approximate Marginal Log-Likelihood}\label{sec:logMLderiv}

When performing MFVB-based inference the variational lower bound
on the marginal log-likelihood, given by (\ref{eq:logMLfirst}),  
is commonly used to assess convergence. However, the algebra required 
to obtain the lower bound expression is demanding for large models. 
The VMP approach offers efficiencies for its calculation, which we now summarize.

In Section \ref{sec:VMP} we described VMP for a general statistical
model with observed data $\bD$ in terms of factors $f_j$, $1\le j\le M$,
such that $p(\btheta,\bD)=\prod_{j=1}^N f_j$
where each $f_j$ is a function of a sub-vector of $\btheta$.
The mean field approximation to the posterior density function
takes the form 
$$p(\btheta|\bD)\approx\prod_{i=1}^M\,q(\btheta_i)$$
for some partition $\{\btheta_1,\ldots,\btheta_M\}$
of $\btheta$. The expressions in Winn \myand Bishop (2005) 
and Minka \myand Winn (2008) give rise to 
\begin{equation}
\log\punder(q;\bD)=\sum_{i=1}^M\Entropy\{q(\btheta_i)\}
+\sum_{j=1}^NE_q\{\log(f_j)\}
\label{eq:logMLgeneral}
\end{equation}
where 
$$\Entropy\{q(\btheta_i)\}\equiv E_{q(\btheta_i)}\{-\log\,q(\btheta_i)\}$$
is the \emph{entropy} (also known as the \emph{differential entropy}) of $q$.

For models such that the optimal $q(\btheta_i)$ are exponential density functions,
which includes each of the models treated in Sections \ref{sec:GaussResp} and 
\ref{sec:generalized}, the value of $\Entropy\{q(\btheta_i)\}$ can be looked
up in a table. Table \ref{tab:Entropies} lists the entropies for each of 
the exponential family distributions covered in Section \ref{sec:expFam}. 
All expressions are in terms of natural parameters.

\begin{table}[ht]
\begin{center}
\begin{tabular}{lc}
\hline\\[-1.3ex]
Distribution       &Entropy\\[0.1ex]
\hline\\[-0.9ex]
Bernoulli        & $\log(1+e^{\eta})-\eta\,e^{\eta}/(1+e^{\eta})$   \\[2ex]
Univariate Normal &$\smhalf\{1+\log(2\pi)\}+\smhalf\log\left(\displaystyle{\frac{-1}{2\eta_2}}\right)$ \\[2ex]
Inverse Chi-Squared    &$\log\Gamma(-\eta_1-1)+\eta_1\psi(-\eta_1-1)
                       +\log(-\eta_2)-\eta_1-1$\\[4ex]
Beta                   &$\log\Gamma(\eta_1+1)+\log\Gamma(\eta_2+1)-\log\Gamma(\eta_1+\eta_2+2)$\\[1ex]
                       &$-\eta_1\psi(\eta_1+1)-\eta_2\psi(\eta_2+1)
                         +(\eta_1+\eta_2)\psi(\eta_1+\eta_2+2)$\\[2ex]
Inverse Gaussian      & $\smhalf+\quarter\log(\pi^2\eta_2/\eta_1^3)+\frac{3}{2}\exp\Big(4(\eta_1\eta_2)^{1/2}\Big)
                      \mbox{Ei}\Big(-4(\eta_1\eta_2)^{1/2}\Big)$      \\[4ex]
Multivariate Normal  & $\frac{d}{2}\{1+\log(2\pi)\}+\smhalf\log\Big|-\smhalf\{\vecof^{-1}(\bdeta_2)\}^{-1}\Big|$\\[2ex]
Inverse Wishart      & $\displaystyle{\sum_{j=1}^d}\Big[\log\Gamma\{-\eta_1-\smhalf(d+j)\}
                       +\eta_1\psi\{-\eta_1-\smhalf(d+j)\}\Big]$  \\[1ex]
                     &$+\smhalf(d+1)\log\Big|-\vecof^{-1}(\bdeta_2)\Big|
                    -d\eta_1-\smhalf\,d(d+1)+\quarter d(d-1)\log(\pi)$\\[2ex]
\hline
\end{tabular}
\end{center}
\caption{\textit{Expressions for entropies
in terms of natural parameters for some common exponential family
distributions.}}
\label{tab:Entropies}
\end{table}

As an example, consider VMP fitting of the linear regression model described
in Section \ref{sec:linReg} and the updates of the stochastic node natural
parameters given by (\ref{eq:qetaupdates}). From Table \ref{tab:Entropies},
the entropy contributions to $\log\punder(q;\by)$ are 
\begin{equation}
{\setlength\arraycolsep{3pt}
\begin{array}{rcl}
\Entropy\{q(\bbeta)\}&=&\textstyle{\frac{d}{2}}\{1+\log(2\pi)\}
+\smhalf\log\Big|-\smhalf\Big\{\vecof^{-1}
 \Big(\big(\bdeta_{q(\bbeta)}\big)_2\Big)\Big\}^{-1}\Big|,\\[1ex]
\Entropy\{q(\sigma^2)\}&=&\log\Gamma\Big(-\big(\eta_{q(\sigma^2)}\big)_1-1\Big)
   +\big(\eta_{q(\sigma^2)}\big)_1\psi\Big(-\big(\eta_{q(\sigma^2)}\big)_1-1\Big)\\[1ex]
     &&\quad+\log\Big(-\big(\eta_{q(\sigma^2)}\big)_2\Big)
        -\big(\eta_{q(\sigma^2)}\big)_1-1,\\[1ex]
\mbox{and}\quad\Entropy\{q(a)\}&=&\log\Gamma\Big(-\big(\eta_{q(a)}\big)_1-1\Big)
   +\big(\eta_{q(a)}\big)_1\psi\Big(-\big(\eta_{q(a)}\big)_1-1\Big)\\[1ex]
  &&\quad+\log\Big(-\big(\eta_{q(a)}\big)_2\Big)-\big(\eta_{q(a)}\big)_1-1.
\end{array}
}
\label{eq:EntropyForLM}
\end{equation}

For conjugate models with exponential family stochastic nodes, 
the factor contributions reduce to linear combinations of 
expected values of sufficient statistics. Their formulae in
terms of natural parameters can be looked up in tables such as 
Table \ref{tab:ETx} in Section \ref{sec:tabETxAndEntropies}. 
For the linear regression model of Section \ref{sec:linReg}
the $q$-density expectation of the logarithm of the likelihood
factor is 
{\setlength\arraycolsep{1pt}
\begin{eqnarray*}
&&E_{q(\bbeta,\sigma^2)}\{\log\,p(\by|\bbeta,\sigma^2)\}=\\[1ex]
&&\quad E_{q(\sigma^2)}(1/\sigma^2)\left\{
\left[  
\begin{array}{c}
E_{q(\bbeta)}(\bbeta)\\[1ex]
E_{q(\bbeta)}\{\vecof(\bbeta\bbeta^T)\}
\end{array}
\right]^T
\left[  
\begin{array}{c}
\bX^T\by\\[1ex]
-\smhalf\vecof(\bX^T\bX)
\end{array}
\right]
-\smhalf\by^T\by
\right\}\\[1ex]
&&\qquad\qquad-\frac{n}{2}E_{q(\sigma^2)}\{\log(\sigma^2)\}-\frac{n}{2}\log(2\pi)\\[2ex]
\end{eqnarray*}
\begin{eqnarray*}
&&=\left\{\frac{\big(\bdeta_{q(\sigma^2)}\big)_1+1}{\big(\bdeta_{q(\sigma^2)}\big)_2}\right\}\\[1ex]
&&\ \times\left\{
\left[\begin{array}{l}
           -\smhalf\{\vecof^{-1}\Big(\big(\bdeta_{q(\bbeta)}\big)_2\Big)\}^{-1}
              \big(\bdeta_{q(\bbeta)}\big)_1\\[4ex]
   \quarter\vecof\Bigg(\left\{\vecof^{-1}\Big(\big(\bdeta_{q(\bbeta)}\big)_2\Big)\right\}^{-1}\\[2ex]
       \ \times\Big[\big(\bdeta_{q(\bbeta)}\big)_1\big(\bdeta_{q(\bbeta)}\big)_1^T
             \left\{\vecof^{-1}\Big(\big(\bdeta_{q(\bbeta)}\big)_2\Big)\right\}^{-1}
              -2\,\bI\Big]\Bigg)\end{array}\right]^T
\left[  
\begin{array}{c}
\bX^T\by\\[1ex]
-\smhalf\vecof(\bX^T\bX)
\end{array}
\right]
-\smhalf\by^T\by
\right\}\\[1ex]
&&\qquad\qquad-\frac{n}{2}\left\{\log\Big(-\big(\bdeta_{q(\sigma^2)}\big)_2\Big)
-\psi\Big(-\big(\bdeta_{q(\sigma^2)}\big)_1-1\Big)\right\}-\frac{n}{2}\log(2\pi).
\end{eqnarray*}
}
The contributions from the remaining three factors in Figure \ref{fig:linRegFacGraphMsgs}
can be handled using similar algebra. These expressions can then be added to 
the $E_{q(\bbeta,\sigma^2)}\{\log\,p(\by|\bbeta,\sigma^2)\}$ expression
and the entropy expressions given in (\ref{eq:EntropyForLM}) to give the full
$\log\punder(q;\by)$ expression.

For the classes of semiparametric regression models treated in Sections \ref{sec:GaussResp}
and \ref{sec:generalized} the $E_q\{\log(f_j)\}$ terms in (\ref{eq:logMLgeneral}) can be handled
efficiently via fragment categorization. The marginal log-likelihood lower bound contributions
of each of the fragments identified in Sections \ref{sec:GaussResp} and \ref{sec:generalized} 
only need to be worked out once and can be tabulated and looked up. 

Next we derive the $E_q\{\log(f_j)\}$-type 
contributions from each of the Section \ref{sec:fiveFundFrags} 
fragment factors. Illustration is then provided for the penalized spline regression
model introduced in Section \ref{sec:arbit}.
Other fragments, such as the generalized response fragments 
of Section \ref{sec:generalized}, can be handled similarly.

\subsubsection{Contribution from an Gaussian Prior Fragment Factor}

In the notation of Section \ref{sec:GaussPriorFrag} the logarithm of the
factor in the Gaussian prior fragment is
$$
\log\,p(\btheta)=\,
\left[
\begin{array}{c}
\btheta\\[1ex]
\vecof(\btheta\btheta^T)
\end{array}
\right]^T
\left[
\begin{array}{c}
\bSigma_{\btheta}^{-1}\bmu_{\btheta}\\[1ex]
-\smhalf\vecof(\bSigma_{\btheta}^{-1})
\end{array}
\right]
-\smhalf\dtheta\log(2\pi)-\smhalf\log|\bSigma_{\btheta}|.$$
Hence, using Table \ref{tab:ETx},
{\setlength\arraycolsep{1pt}
\begin{eqnarray*}
E_q\{\log\,p(\btheta)\}
&=&
\left[\begin{array}{l}
           -\smhalf\{\vecof^{-1}\Big(\big(\bdeta_{q(\btheta)}\big)_2\Big)\}^{-1}
              \big(\bdeta_{q(\btheta)}\big)_1\\[4ex]
\quarter\vecof\Bigg(\left\{\vecof^{-1}\Big(\big(\bdeta_{q(\btheta)}\big)_2\Big)\right\}^{-1}\\[2ex]
       \ \times\Big[\big(\bdeta_{q(\btheta)}\big)_1\big(\bdeta_{q(\btheta)}\big)_1^T
             \left\{\vecof^{-1}\Big(\big(\bdeta_{q(\btheta)}\big)_2\Big)\right\}^{-1}
      -2\,\bI\Big]\Bigg)\end{array}\right]^T
\left[
\begin{array}{c}
\bSigma_{\btheta}^{-1}\bmu_{\btheta}\\[1ex]
-\smhalf\vecof(\bSigma_{\btheta}^{-1})
\end{array}
\right]\\[2ex]
&&\qquad-\smhalf\dtheta\log(2\pi)-\smhalf\log|\bSigma_{\btheta}|.
\end{eqnarray*}
}

\subsubsection{Contribution from an Inverse Wishart Prior Fragment Factor}

In the notation of Section \ref{sec:InvWishPriorFrag} the logarithm of the
factor in the Inverse Wishart prior fragment is
$$
\log\,p(\bTheta)=
\left[
\begin{array}{c}
\log|\bTheta|\\[1ex]
\vecof(\bTheta^{-1})
\end{array}
\right]^T
\left[
\begin{array}{c}
-(\kappaTheta+\dTheta+1)/2\\[1ex]
-\smhalf\vecof(\LambdaTheta)
\end{array}
\right]-\log(\Csc_{\dTheta,\kappaTheta})+\smhalf\kappaTheta\log|\LambdaTheta|.
$$
Table \ref{tab:ETx} then gives 
\begin{equation}
{\setlength\arraycolsep{1pt}
\begin{array}{l}
E_q\{\log\,p(\bTheta)\}=\\[1ex]
\qquad\left[
\begin{array}{l}
           \log\Big|-\vecof^{-1}\big((\bdeta_{q(\bTheta)})_2\big)\Big|\\
     \qquad-\displaystyle{\sum_{j=1}^{\dTheta}}\psi\{-(\eta_{q(\bTheta)})_1
          -\smhalf(\dTheta+j)\}\\[6ex]
           \{(\eta_{q(\bTheta)})_1+\smhalf(\dTheta+1)\}
           \vecof[\{\vecof^{-1}\big((\bdeta_{q(\bTheta)})_2\big)\}^{-1}]
\end{array}
\right]^T
\left[
\begin{array}{c}
-\smhalf(\kappaTheta+\dTheta+1)\\[1ex]
-\smhalf\vecof(\LambdaTheta)
\end{array}
\right]\\[0ex]
\\
\qquad\qquad\qquad-\log(\Csc_{\dTheta,\kappaTheta})+\smhalf\kappaTheta\log|\LambdaTheta|.
\end{array}
}
\label{eq:logMLinvWishPrior}
\end{equation}

\subsubsection{Contribution from an Iterated Inverse G-Wishart Fragment Factor}

As in Section \ref{sec:IterInvGWish} we first treat the scalar case before dealing
with the more delicate matrix case.

\subsubsubsection{The Case of $\dTheta=1$}

When $\dTheta=1$ the covariance matrices $\bTheta_1$ and $\bTheta_2$ reduce
to scalars $\btheta_1$ and $\btheta_2$ 
and the logarithm of the fragment factor is
$$\log\,p(\theta_1|\theta_2)=
\left[
\begin{array}{c}
\log(\theta_1)\\[1ex]
1/\theta_1
\end{array}
\right]^T
\left[
\begin{array}{c}
-\smhalf(\kappa+2)\\[1ex]
-\smhalf(1/\theta_2)
\end{array}
\right]
-\smhalf\kappa\log(\theta_1)-\smhalf\kappa\log(2)-\log\Gamma(\smhalf\kappa)$$
so using Table \ref{tab:ETx} we get 
\begin{equation}
{\setlength\arraycolsep{1pt}
\begin{array}{l}
E_q\{\log\,p(\theta_1|\theta_2)\}=\\[1ex]
\qquad\qquad\left[
\begin{array}{c}
\log\big(-(\eta_{q(\theta_1)})_2\big)
-\psi\big(-(\eta_{q(\theta_1)})_1-1\big)\\[1ex]
\big((\eta_{q(\theta_1)})_1+1\big)/(\eta_{q(\theta_1)})_2
\end{array}
\right]^T
\left[
\begin{array}{c}
-\smhalf(\kappa+2)\\[1ex]
-\smhalf\big((\eta_{q(\theta_2)})_1+1\big)/(\eta_{q(\theta_2)})_2
\end{array}
\right]\\[4ex]
\qquad\qquad
-\smhalf\kappa\Big\{\log\big(-(\eta_{q(\theta_2)})_2\big)
-\psi\big(-(\eta_{q(\theta_2)})_1-1\big)\Big\}
-\smhalf\kappa\log(2)-\log\Gamma(\smhalf\kappa).
\end{array}
}
\label{eq:logMLIterInvGWishartscalar}
\end{equation}

\subsubsubsection{The Case of $\dTheta>1$ and $G$ Totally Connected or Totally Disconnected}

If $\bTheta_1|\bTheta_2\sim\mbox{Inverse-G-Wishart}(G,\kappa,\bTheta_2^{-1})$
where $G$ is totally connected then 
$$\log\,p(\bTheta_1|\bTheta_2)=
\left[
\begin{array}{c}
\log|\bTheta_1|\\[1ex]
\vecof(\bTheta_1^{-1})
\end{array}
\right]^T
\left[
\begin{array}{c}
-\smhalf(\kappa+\dTheta+1)\\[1ex]
-\smhalf\vecof(\bTheta_2^{-1})
\end{array}
\right]-\smhalf\kappa\,\log|\bTheta_2|-\log(\Csc_{\dTheta,\kappa}).
$$
Table \ref{tab:ETx} immediately gives
{\setlength\arraycolsep{1pt}
\begin{eqnarray*}
&&E_q\{\log\,p(\bTheta_1|\bTheta_2)\}=\\[1ex]
&&\quad\left[\begin{array}{l}
           \log|-\vecof^{-1}\big((\bdeta_{q(\bTheta_1)})_2\big)|\\
     \qquad-\displaystyle{\sum_{j=1}^d}\psi\{-(\bdeta_{q(\bTheta_1)})_1-\smhalf(\dTheta+j)\}\\[6ex]
           \{(\bdeta_{q(\bTheta_1)})_1+\smhalf(\dTheta+1)\}\vecof[\{\vecof^{-1}\big((\bdeta_{q(\bTheta_1)})_2\big)\}^{-1}]
        \end{array}\right]^T
\left[
\begin{array}{c}
-\smhalf(\kappa+\dTheta+1)\\[1ex]
-\smhalf\vecof(E_{q(\bTheta_2)}(\bTheta_2^{-1}))
\end{array}
\right]\\[1ex]
&&\quad-\smhalf\kappa\,E_{q(\bTheta_2)}\{\log|\bTheta_2|\}-\log(\Csc_{\dTheta,\kappa}).
\end{eqnarray*}
}
If $\bTheta_2$ has a totally disconnected Inverse G-Wishart distribution then
$$E_{q(\bTheta_2)}\{\log|\bTheta_2|\}=\log\Big|-\vecof^{-1}\big(\bdeta_{q(\bTheta_2)}\big)_2\Big|
     -\displaystyle{\sum_{j=1}^{\dTheta}}
      \psi\big\{-\big(\bdeta_{q(\bTheta_2)}\big)_1-\smhalf(\dTheta+j)\big\}$$
and
$$E_{q(\bTheta_2)}(\bTheta_2^{-1})=\left\{\big(\bdeta_{q(\bTheta_2)}\big)_1+\smhalf(\dTheta+1)\right\}
\left\{\vecof^{-1}\big(\bdeta_{q(\bTheta_2)}\big)_2\right\}^{-1}.
$$
If $\bTheta_2$ has an totally disconnected Inverse G-Wishart distribution, which is the case for
the auxiliary variable representation of the covariance matrix prior of Huang \myand Wand (2013), then
$$E_{q(\bTheta_2)}\{\log|\bTheta_2|\}=\sum_{j=1}^{\dTheta}
\left\{\log\Big(-\big(\bdeta_{q((\bTheta_2)_{jj})}\big)_2\Big)
-\psi\Big(-\big(\bdeta_{q((\bTheta_2)_{jj})}\big)_1-1\Big)\right\}
$$
and
$$E_{q(\bTheta_2)}(\bTheta_2^{-1})=\diagarg{1\le j\le\dTheta}
\left(\frac{\big(\bdeta_{q((\bTheta_2)_{jj})}\big)_1+1}{\big(\bdeta_{q((\bTheta_2)_{jj})}\big)_2}\right).
$$

\subsubsubsection{Other Cases}

The other cases such as $\bTheta_1$ having a Inverse G-Wishart distribution
with $G$ partially connected or totally disconnected are not
common in Bayesian semiparametric regression analysis and are left aside here.

\subsubsection{Contribution from a Gaussian Penalization Factor}

For this fragment, the logarithm of the factor is
{\setlength\arraycolsep{1pt}
\begin{eqnarray*}
&&\log\,p(\btheta_0,\ldots,\btheta_L|\bTheta_1,\ldots,\bTheta_L)=
\left[  
\begin{array}{c}
\btheta_0\\[1ex]
\vecof(\btheta_0\btheta_0^T)
\end{array}
\right]^T
\left[  
\begin{array}{c}
\bSigma_{\btheta_0}^{-1}\bmu_{\btheta_0}\\[1ex]
-\smhalf\vecof(\bSigma_{\btheta_0}^{-1})
\end{array}
\right]
-\smhalf\dtheta_0\log(2\pi)-\smhalf\log|\bSigma_{\btheta_0}|
\\[2ex]
&&\qquad+\sum_{\ell=1}^L\left\{\left[  
\begin{array}{c}
\btheta_{\ell}\\[1ex]
\vecof(\btheta_{\ell}\btheta_{\ell}^T)
\end{array}
\right]^T
\left[  
\begin{array}{c}
\bzero\\[1ex]
-\smhalf\vecof\big(\bI_{\mR_{\ell}}\otimes\bTheta_{\ell}^{-1}\big)
\end{array}
\right]-\smhalf\mR_{\ell}\dTheta_{\ell}\log(2\pi)
-\smhalf\mR_{\ell}\log|\bTheta_{\ell}|\right\}.
\end{eqnarray*}
}
Application of results in Table \ref{tab:ETx} then gives
\begin{equation}
{\setlength\arraycolsep{1pt}
\begin{array}{rcl}
&&E_q\{\log\,p(\btheta_0,\ldots,\btheta_L|\bTheta_1,\ldots,\bTheta_L)\}\\[1ex]
&&=
\left[\begin{array}{l}
           -\smhalf\left\{\vecof^{-1}\Big(\big(\bdeta_{q(\btheta_0)}\big)_2\Big)\right\}^{-1}
              \big(\bdeta_{q(\btheta_0)}\big)_1\\[4ex]
   \quarter\vecof\Bigg(\left\{\vecof^{-1}\Big(\big(\bdeta_{q(\btheta_0)}\big)_2\Big)\right\}^{-1}\\[2ex]
       \ \times\Big[\big(\bdeta_{q(\btheta_0)}\big)_1\big(\bdeta_{q(\btheta_0)}\big)_1^T
             \left\{\vecof^{-1}\Big(\big(\bdeta_{q(\btheta_0)}\big)_2\Big)\right\}^{-1}
              -2\,\bI\Big]\Bigg)\end{array}\right]^T
\left[  
\begin{array}{c}
\bSigma_{\btheta_0}^{-1}\bmu_{\btheta_0}\\[1ex]
-\smhalf\vecof(\bSigma_{\btheta_0}^{-1})
\end{array}
\right]\\[2ex]
&&\qquad
-\smhalf\dtheta_0\log(2\pi)-\smhalf\log|\bSigma_{\btheta_0}|\\[2ex]
&&\qquad+\sum_{\ell=1}^L\Bigg\{\left[  
\begin{array}{c}
\quarter\vecof\Bigg(\left\{\vecof^{-1}\Big(\big(\bdeta_{q(\btheta_{\ell})}\big)_2\Big)\right\}^{-1}\\[2ex]
       \ \times\Big[\big(\bdeta_{q(\btheta_{\ell})}\big)_1\big(\bdeta_{q(\btheta_{\ell})}\big)_1^T
             \left\{\vecof^{-1}\Big(\big(\bdeta_{q(\btheta_{\ell})}\big)_2\Big)\right\}^{-1}
              -2\,\bI\Big]\Bigg)
\end{array}
\right]^T\\
&&\quad\times\left[  
\begin{array}{c}
-\smhalf\vecof\Big(\bI_{\mR_{\ell}}\otimes
\left[\{(\eta_{q(\bTheta_{\ell})})_1+\smhalf(\dTheta_{\ell}+1)\}
\{\vecof^{-1}((\bdeta_{q(\bTheta_{\ell})})_2)\}^{-1}\right]\Big)
\end{array}
\right]\\[1ex]
&&\qquad-\smhalf\mR_{\ell}\dTheta_{\ell}\log(2\pi)\\[1ex]
&&\qquad-\smhalf\mR_{\ell}\log\Big|-\vecof^{-1}((\bdeta_{q(\bTheta_{\ell})})_2)\Big|
+\smhalf\mR_{\ell}\sum_{j=1}^{\dTheta_{\ell}}\psi\{-(\eta_{q(\bTheta_{\ell})})_1-\smhalf(\dTheta_{\ell}+j)\}
\Bigg\}.
\end{array}
}
\label{eq:logMLfromGauPen}
\end{equation}

\subsubsection{Contribution from a Gaussian Likelihood Factor}

The logarithm of the factor is
{\setlength\arraycolsep{1pt}
\begin{eqnarray*}
\log\,p(\by|\,\btheta_1,\theta_2)&=&\frac{1}{\theta_2}\left\{\left[  
\begin{array}{c}
\btheta_1\\[1ex]
\vecof(\btheta_1\btheta_1^T)
\end{array}
\right]^T
\left[  
\begin{array}{c}
\bA^T\by\\[1ex]
-\smhalf\vecof(\bA^T\bA)
\end{array}
\right]
-\smhalf\by^T\by\right\}-\frac{n}{2}\log(\theta_2)-\frac{n}{2}\log(2\pi).
\end{eqnarray*}
}
Then, from Table \ref{tab:ETx} we have
\begin{equation}
{\setlength\arraycolsep{1pt}
\begin{array}{l}
E_q\{\log\,p(\by|\btheta_1,\theta_2)\}=\\[1ex]
=\left\{\frac{\big(\bdeta_{q(\theta_2)}\big)_1+1}{\big(\bdeta_{q(\theta_2)}\big)_2}\right\}\\[1ex]
\ \times\left\{
\left[\begin{array}{l}
           -\smhalf\left\{\vecof^{-1}\Big(\big(\bdeta_{q(\btheta_1)}\big)_2\Big)\right\}^{-1}
              \big(\bdeta_{q(\btheta_1)}\big)_1\\[4ex]
   \quarter\vecof\Bigg(\left\{\vecof^{-1}\Big(\big(\bdeta_{q(\btheta_1)}\big)_2\Big)\right\}^{-1}\\[2ex]
       \ \times\Big[\big(\bdeta_{q(\btheta_1)}\big)_1\big(\bdeta_{q(\btheta_1)}\big)_1^T
             \left\{\vecof^{-1}\Big(\big(\bdeta_{q(\btheta_1)}\big)_2\Big)\right\}^{-1}
              -2\,\bI\Big]\Bigg)\end{array}\right]^T
\left[  
\begin{array}{c}
\bA^T\by\\[1ex]
-\smhalf\vecof(\bA^T\bA)
\end{array}
\right]
-\smhalf\by^T\by
\right\}\\[2ex]
\qquad\qquad-\frac{n}{2}\left\{\log\Big(-\big(\bdeta_{q(\theta_2)}\big)_2\Big)
-\psi\Big(-\big(\bdeta_{q(\theta_2)}\big)_1-1\Big)\right\}-\frac{n}{2}\log(2\pi).
\end{array}
}
\label{eq:logMLGaussLikContrib}
\end{equation}

\subsubsection{Illustration for Penalized Spline Nonparametric Regression}

We now illustrate approximate marginal log-likelihood calculation
for penalized spline regression, corresponding to the factor graph
shown in Figure \ref{fig:penSplFacGraph}. Using Table \ref{tab:Entropies}, 
the first two entropy contributions to $\log\punder(q;\by)$ are 
\begin{equation}
\Entropy\{q(\bbeta,\bu)\}=\textstyle{\frac{2+K}{2}}\{1+\log(2\pi)\}
+\smhalf\log\Big|-\smhalf\Big\{\vecof^{-1}
 \Big(\big(\bdeta_{q(\bbeta,\bu)}\big)_2\Big)\Big\}^{-1}\Big|
\label{eq:EntropiesForPenSplineOne}
\end{equation}
and
\begin{equation}
{\setlength\arraycolsep{3pt}
\begin{array}{rcl}
\Entropy\{q(\sigeps^2)\}&=&\log\Gamma\Big(-\big(\eta_{q(\sigeps^2)}\big)_1-1\Big)
   +\big(\eta_{q(\sigeps^2)}\big)_1\psi\Big(-\big(\eta_{q(\sigeps^2)}\big)_1-1\Big)\\[1ex]
     &&\quad+\log\Big(-\big(\eta_{q(\sigeps^2)}\big)_2\Big)
        -\big(\eta_{q(\sigeps^2)}\big)_1-1.
\end{array}
}
\label{eq:EntropiesForPenSplineTwo}
\end{equation}
The entropy contributions 
\begin{equation}
\Entropy\{q(\sigma_u^2)\},\quad 
\Entropy\{q(a_{\varepsilon})\}\quad \mbox{and}\quad \Entropy\{q(a_u)\}
\label{eq:EntropiesForPenSplineThree}
\end{equation}
take exactly the same form as (\ref{eq:EntropiesForPenSplineTwo})
but as functions of the natural parameter vectors $\bdeta_{q(\sigma_u^2)}$,
$\bdeta_{q(a_{\varepsilon})}$ and $\bdeta_{q(a_u)}$.

The factor contributions are each special cases of 
(\ref{eq:logMLinvWishPrior})--(\ref{eq:logMLGaussLikContrib}).
The contribution from the factor $p(\aeps)$ is 
\begin{equation}
\mbox{the right-hand side of (\ref{eq:logMLinvWishPrior}) with}\ \bTheta=\aeps,\ \dTheta=1,\ \kappaTheta=1
\ \mbox{and}\ \LambdaTheta=1/\Aeps^2.
\label{eq:logMLpaepsContrib}
\end{equation}
The contribution from the factor $p(a_u)$ is 
\begin{equation}
\mbox{the right-hand side of (\ref{eq:logMLinvWishPrior}) with}\ \bTheta=a_u,\ \dTheta=1,\ \kappaTheta=1
\ \mbox{and}\ \LambdaTheta=1/\Au^2.
\label{eq:logMLpauContrib}
\end{equation}
The contribution from the factor $p(\sigeps^2|\,\aeps)$ is
\begin{equation}
\mbox{the right-hand side of (\ref{eq:logMLIterInvGWishartscalar}) with}\ 
\theta_1=\sigeps^2,\ \theta_2=\aeps\ \mbox{and}\ \kappa=1.
\label{eq:logMLpsigsqepsaepsContrib}
\end{equation}
The contribution from the factor $p(\sigma_u^2|\,a_u)$ is
\begin{equation}
\mbox{the right-hand side of (\ref{eq:logMLIterInvGWishartscalar}) with}\ 
\theta_1=\sigma_u^2,\ \theta_2=a_u\ \mbox{and}\ \kappa=1.
\label{eq:logMLpsigsquauContrib}
\end{equation}
The contribution from the factor $p(\bbeta,\bu|\,\sigma_u^2)$ is
\begin{equation}
\begin{array}{l}
\mbox{the right-hand side of (\ref{eq:logMLfromGauPen}) with}\
L=1,\ \dTheta=1,\ \mR_1=K,\ \btheta_0=\bbeta,\\[1ex]
\btheta_1=\bu\ \mbox{and}\ \bTheta_1=\sigma_u^2.
\end{array}
\label{eq:logMLpbetausigsquContrib}
\end{equation}
The contribution from the factor $p(\by|\,\bbeta,\bu,\sigeps^2)$ is
\begin{equation}
\mbox{the right-hand side of (\ref{eq:logMLGaussLikContrib}) with}\ 
\btheta_1=(\bbeta,\bu),\ \theta_2=\sigeps^2\ \mbox{and}\ \bA=[\bX\ \bZ].
\label{eq:logMLpybetausigsqepsContrib}
\end{equation}
During the VMP iterations for fitting (\ref{eq:penSplRegMod}), the approximate
marginal log-likelihood $\log\punder(\by;q)$ can be computed by
obtaining 
$$
\begin{array}{rcl}
\bdeta_{q(\bbeta,\bu)}&\thickarrow&\etaSUBpbetausigsquTObetau+\etaSUBpybetausigsqepsTObetau,\\[2ex]
\bdeta_{q(\sigeps^2)}&\thickarrow&\etaSUBpybetausigsqepsTOsigsqeps+\etaSUBpsigsqepsaepsTOsigsqeps,\\[2ex]
\bdeta_{q(\sigma_u^2)}&\thickarrow&\etaSUBpbetausigsquTOsigsqu+\etaSUBpsigsquauTOsigsqu,\\[2ex]
\bdeta_{q(a_{\varepsilon})}&\thickarrow&\etaSUBpsigsqepsaepsTOaeps+\etaSUBpaepsTOaeps\\[2ex]
\mbox{and}\quad\bdeta_{q(a_u)}&\thickarrow&\etaSUBpsigsquauTOau+\etaSUBpauTOau
\end{array}
$$
and summing up the entropy contributions 
(\ref{eq:EntropiesForPenSplineOne}--\ref{eq:EntropiesForPenSplineThree}) and the
factor contributions (\ref{eq:logMLpaepsContrib})--(\ref{eq:logMLpybetausigsqepsContrib}).


\section*{Additional References}

\begin{small}

\bibmpa
Murphy, K.P. (2012). \textit{Machine Learning: A Probabilistic Perspective}.
Cambridge, Massachusetts: The MIT Press.

\bibmpa
Menictas, M. and Wand, M.P. (2015).
Variational inference for heteroscedastic semiparametric
regression. \textit{Australian and New Zealand Journal of Statistics}, 
{\bf 57}, 119--138.

\bibmpa
Rohde, D. \myand Wand, M.P. (2015).
Semiparametric mean field variational Bayes:
general principles and numerical issues.
Under revision for \textit{Journal of Machine Learning Research}.

\end{small}

\end{document}